\documentclass[12pt]{article}

\usepackage{amsfonts,amssymb,subeqnarray,eqsection,indent,eepicemu}

\date{September 15, 2003 \\[1mm] revised September 8, 2004}

\begin{document}

\title{\vspace*{-2cm}The Repulsive Lattice Gas, \\
                        the Independent-Set Polynomial, \\
                        and the Lov\'asz Local Lemma}

\author{
     \\
     {\small Alexander D.~Scott}
\\[-2mm]
     {\small\it Department of Mathematics}  \\[-2mm]
     {\small\it University College London} \\[-2mm]
     {\small\it London WC1E 6BT, England}                         \\[-2mm]
     {\small\tt scott@math.ucl.ac.uk}                        \\[5mm]
     {\small Alan D.~Sokal}                  \\[-2mm]
     {\small\it Department of Physics}       \\[-2mm]
     {\small\it New York University}         \\[-2mm]
     {\small\it 4 Washington Place}          \\[-2mm]
     {\small\it New York, NY 10003 USA}      \\[-2mm]
     {\small\tt sokal@nyu.edu}               \\[-2mm]
     {\protect\makebox[5in]{\quad}}  
     \\
}

\maketitle
\thispagestyle{empty}   

\begin{abstract}
We elucidate the close connection between the repulsive lattice gas
in equilibrium statistical mechanics
and the Lov\'asz local lemma in probabilistic combinatorics.
We show that the conclusion of the Lov\'asz local lemma
holds for dependency graph $G$ and probabilities $\{p_x\}$
if and only if the independent-set polynomial for $G$ is nonvanishing
in the polydisc of radii $\{p_x\}$.
Furthermore, we show that the usual proof of the Lov\'asz local lemma
--- which provides a sufficient condition for this to occur ---
corresponds to a simple inductive argument for the nonvanishing
of the independent-set polynomial in a polydisc,
which was discovered implicitly by Shearer \cite{Shearer_85}
and explicitly by Dobrushin \cite{Dobrushin_96a,Dobrushin_96b}.
We also present some refinements and extensions of both arguments,
including a generalization of the Lov\'asz local lemma
that allows for ``soft'' dependencies.
In addition, we prove some general properties of the partition function
of a repulsive lattice gas, most of which are consequences of
the alternating-sign property for the Mayer coefficients.
We conclude with a brief discussion of the repulsive lattice gas
on countably infinite graphs.
\end{abstract}


\clearpage

\newtheorem{theorem}{Theorem}[section]
\newtheorem{proposition}[theorem]{Proposition}
\newtheorem{lemma}[theorem]{Lemma}
\newtheorem{corollary}[theorem]{Corollary}
\newtheorem{definition}[theorem]{Definition}
\newtheorem{conjecture}[theorem]{Conjecture}
\newtheorem{question}[theorem]{Question}
\newtheorem{example}[theorem]{Example}

\renewcommand{\theenumi}{\alph{enumi}}
\renewcommand{\labelenumi}{(\theenumi)}
\def\eop{\hbox{\kern1pt\vrule height6pt width4pt
depth1pt\kern1pt}\medskip}
\def\prf{\par\noindent{\bf Proof.\enspace}\rm}
\def\rmk{\par\medskip\noindent{\bf Remark\enspace}\rm}

\newcommand{\be}{\begin{equation}}
\newcommand{\ee}{\end{equation}}
\newcommand{\<}{\langle}
\renewcommand{\>}{\rangle}
\newcommand{\widebar}{\overline}
\def\reff#1{(\protect\ref{#1})}
\def\spose#1{\hbox to 0pt{#1\hss}}
\def\ltapprox{\mathrel{\spose{\lower 3pt\hbox{$\mathchar"218$}}
    \raise 2.0pt\hbox{$\mathchar"13C$}}}
\def\gtapprox{\mathrel{\spose{\lower 3pt\hbox{$\mathchar"218$}}
    \raise 2.0pt\hbox{$\mathchar"13E$}}}
\def\textprime{${}^\prime$}
\def\proof{\par\medskip\noindent{\sc Proof.\ }}
\def\qed{\hbox{\hskip 6pt\vrule width6pt height7pt depth1pt
\hskip1pt}\bigskip}
\def\proofof#1{\bigskip\noindent{\sc Proof of #1.\ }}
\def\half{ {1 \over 2} }
\def\third{ {1 \over 3} }
\def\twothird{ {2 \over 3} }
\def\smfrac#1#2{\textstyle{#1\over #2}}
\def\smhalf{ \smfrac{1}{2} }
\newcommand{\real}{\mathop{\rm Re}\nolimits}
\renewcommand{\Re}{\mathop{\rm Re}\nolimits}
\newcommand{\imag}{\mathop{\rm Im}\nolimits}
\renewcommand{\Im}{\mathop{\rm Im}\nolimits}
\newcommand{\sgn}{\mathop{\rm sgn}\nolimits}
\newcommand{\tr}{\mathop{\rm tr}\nolimits}
\newcommand{\supp}{\mathop{\rm supp}\nolimits}
\def\hboxscript#1{ {\hbox{\scriptsize\em #1}} }
\renewcommand{\emptyset}{\varnothing}

\newcommand{\restrict}{\upharpoonright}
\newcommand{\implies}{\;\Longrightarrow\;}

\newcommand{\scra}{{\mathcal{A}}}
\newcommand{\scrb}{{\mathcal{B}}}
\newcommand{\scrc}{{\mathcal{C}}}
\newcommand{\scrf}{{\mathcal{F}}}
\newcommand{\scrg}{{\mathcal{G}}}
\newcommand{\scrh}{{\mathcal{H}}}
\newcommand{\scrk}{{\mathcal{K}}}
\newcommand{\scrl}{{\mathcal{L}}}
\newcommand{\scro}{{\mathcal{O}}}
\newcommand{\scrp}{{\mathcal{P}}}
\newcommand{\scrr}{{\mathcal{R}}}
\newcommand{\scrs}{{\mathcal{S}}}
\newcommand{\scrt}{{\mathcal{T}}}
\newcommand{\scrv}{{\mathcal{V}}}
\newcommand{\scrw}{{\mathcal{W}}}
\newcommand{\scrz}{{\mathcal{Z}}}

\newcommand{\w}{\mathbf{w}}
\newcommand{\wtilde}{{\widetilde{\bf w}}}
\newcommand{\what}{{\widehat{\bf w}}}
\newcommand{\z}{{\bf z}}
\newcommand{\R}{{\bf R}}  
\newcommand{\Rtilde}{{\widetilde{\bf R}}}
\newcommand{\Rhat}{{\widehat{\bf R}}}
\newcommand{\K}{{\bf K}}
\newcommand{\p}{{\bf p}}
\renewcommand{\k}{{\bf k}}
\newcommand{\n}{{\bf n}}
\renewcommand{\a}{{\bf a}}
\renewcommand{\b}{{\bf b}}
\renewcommand{\r}{{\bf r}}
\newcommand{\smalln}{\mbox{\scriptsize\bf n}}
\newcommand{\blambda}{\mbox{\boldmath $\lambda$}}
\newcommand{\smallblambda}{\mbox{\scriptsize\boldmath $\lambda$}}
\newcommand{\btheta}{\mbox{\boldmath $\theta$}}
\newcommand{\balpha}{\mbox{\boldmath $\alpha$}}
\newcommand{\bdelta}{\mbox{\boldmath $\delta$}}
\newcommand{\smallbalpha}{\mbox{\scriptsize\boldmath $\alpha$}}
\def\twotilde#1{{\widetilde{\widetilde{#1}}}}
\def\twohat#1{{\widehat{\widehat{#1}}}}

\newcommand{\C}{{\mathbb C}}
\newcommand{\Z}{{\mathbb Z}}
\newcommand{\N}{{\mathbb N}}
\newcommand{\RR}{{\mathbb R}}


\newenvironment{sarray}{
             \textfont0=\scriptfont0
             \scriptfont0=\scriptscriptfont0
             \textfont1=\scriptfont1
             \scriptfont1=\scriptscriptfont1
             \textfont2=\scriptfont2
             \scriptfont2=\scriptscriptfont2
             \textfont3=\scriptfont3
             \scriptfont3=\scriptscriptfont3
           \renewcommand{\arraystretch}{0.7}
           \begin{array}{l}}{\end{array}}

\newenvironment{scarray}{
             \textfont0=\scriptfont0
             \scriptfont0=\scriptscriptfont0
             \textfont1=\scriptfont1
             \scriptfont1=\scriptscriptfont1
             \textfont2=\scriptfont2
             \scriptfont2=\scriptscriptfont2
             \textfont3=\scriptfont3
             \scriptfont3=\scriptscriptfont3
           \renewcommand{\arraystretch}{0.7}
           \begin{array}{c}}{\end{array}}

\tableofcontents

\clearpage

\section{Introduction}

The lattice gas with repulsive pair interactions is
an important model in equilibrium statistical mechanics
\cite{Dobrushin_68,Ginibre_70,%
Heilmann_73,Heilmann_74a,Heilmann_74b,Runnels_75,%
Baxter_81,Baxter_82,vandenBerg_94,Haggstrom_97,Brightwell_99,%
Todo_99,Heringa_00,Guo_02,Sokal_cergy,Kahn_01,Kahn_02,Galvin_04}.
In the special case of a hard-core self-repulsion and
hard-core nearest-neighbor exclusion
(i.e.\ no site can be multiply occupied and
no pair of adjacent sites can be simultaneously occupied),
the partition function of the lattice gas coincides with the
independent-set polynomial in combinatorics
(also known as the independence polynomial or the stable-set polynomial)
\cite{Arocha_84,Fisher_90a,Fisher_90b,Hamidoune_90,Gutman_91,Gutman_92a,%
Gutman_92b,Hoede_94,Brown_00,Brown_01,Horrocks_02,Levit_03,Chudnovsky_04}.
Moreover, the hard-core lattice gas
is the {\em universal}\/ statistical-mechanical model
in the sense that any statistical-mechanical model
living on a vertex set $V_0$ can be mapped onto a gas
of nonoverlapping ``polymers'' on $V_0$,
i.e.\ a hard-core lattice gas on the intersection graph of $V_0$
\cite[Section~5.7]{Simon_93}.
This construction, which is termed the
``polymer expansion'' or ``cluster expansion'',
is an important tool in mathematical statistical mechanics
\cite{Seiler_82,Brydges_86,Glimm_87,Brydges_99,Borgs_lectures},
and much effort has been devoted to finding complex polydiscs
in which the polymer expansion is convergent,
i.e.\ in which the (polymer-)lattice-gas partition function is nonvanishing
\cite{Penrose_67,Cammarota_82,Seiler_82,Brydges_86,Kotecky_86,%
Brydges_87,Brydges_88,Simon_93,Dobrushin_96a,Dobrushin_96b,%
Procacci_98,Procacci_99,Brydges_99,Miracle_00,Bovier_00,%
Sokal_chromatic_bounds,Sokal_cergy,Sokal_Mayer_in_prep}.
One goal of this paper is to make a modest contribution
to this line of development.

The Lov\'asz local lemma \cite{EL,ES,Spencer_75,Spencer_77}
is an important tool
in probabilistic existence proofs in combinatorics.
It provides a lower bound on the probability that none of a
collection of ``bad'' events occurs, when those events are subject to
a set of ``local'' dependencies, controlled by a ``dependency graph''.
The Lov\'asz local lemma (and its algorithmic versions \cite{Beck_91,Alon_91})
has found a significant variety of applications,
especially in graph coloring, Ramsey theory,
and related algorithmic problems \cite{AS,Bollobas_01,Molloy_02}.

In this paper we would like to elucidate the close relation
between these two apparently disparate subjects.
Following a seminal (but apparently little-known) paper
of Shearer \cite{Shearer_85},
we shall show that the conclusion of the Lov\'asz local lemma
holds for dependency graph $G$ and probabilities $\{p_x\}$
if and only if the independent-set polynomial for $G$ is nonvanishing
in the polydisc of radii $\{p_x\}$.
Moreover, we shall show that the usual proof of the Lov\'asz local lemma
--- which provides a {\em sufficient}\/ condition for this to occur ---
corresponds to a simple inductive argument for the nonvanishing
of the independent-set polynomial in a polydisc,
which was discovered implicitly by Shearer \cite{Shearer_85}
and explicitly by Dobrushin \cite{Dobrushin_96a,Dobrushin_96b}.
Finally, we shall present some refinements and extensions
of both arguments.

We have tried hard to make this paper comprehensible
to the union (not the intersection!)\
of combinatorialists, probabilists,
classical analysts and mathematical physicists.
We apologize in advance to experts in each of these fields
for boring them every now and then with overly detailed
explanations of elementary facts.

\subsection{The repulsive lattice gas}

In statistical mechanics, a ``grand-canonical gas''
is defined by a {\em single-particle state space}\/ $X$
(here a nonempty finite set),
a {\em fugacity vector}\/ $\w = \{w_x\}_{x \in X} \in \C^X$,
and a {\em two-particle Boltzmann factor}\/
$W \colon\, X \times X \to \C$ with $W(x,y) = W(y,x)$.
The (grand) partition function $Z_W(\w)$ is then defined to be the sum
over ways of placing $n \ge 0$ ``particles''
on ``sites'' $x_1, \ldots, x_n \in X$,
with each configuration assigned a ``Boltzmann weight''
given by the product of the corresponding factors
$w_{x_i}$ and $W(x_i,x_j)$:
\begin{subeqnarray}
      Z_W(\w)  & = & \sum\limits_{n=0}^\infty {1 \over n!}
                      \sum_{x_1,\ldots,x_n \in X}
                      \left( \prod\limits_{i=1}^n w_{x_i} \right)
                      \left( \prod\limits_{1 \le i < j \le n} W(x_i,x_j)
\right)
        \slabel{defZ_1}   \\[2mm]
      & = & \sum\limits_{\n}
             \left( \prod\limits_{x \in X}
                    {w_x^{n_x} \, W(x,x)^{n_x (n_x-1)/2}  \over  n_x!}
             \right)
             \left( \prod\limits_{\{x,y\} \subseteq X}
                    W(x,y)^{n_x n_y}
             \right)
        \slabel{defZ_2}
        \label{defZ}
\end{subeqnarray}
where in \reff{defZ_2} the sum runs over all multi-indices
$\n = \{n_x\}_{x \in X}$ of nonnegative integers,
and the product runs over all two-element subsets $\{x,y\} \subseteq X$
($x \neq y$).
In this paper we shall limit attention to the {\em repulsive}\/ case
in which $0 \le W(x,y) \le 1$ for all $x,y$.
{}From this assumption it follows immediately that $Z_W(\w)$ is
an entire analytic function of $\w$ satisfying
$|Z_W(\w)| \le \exp( \sum_{x \in X} |w_x|)$.

If $W(x,x) = 0$ for all $x \in X$ --- in statistical mechanics this is
called a {\em hard-core self-repulsion}\/ ---
then the only nonvanishing terms in \reff{defZ_2}
have $n_x = 0$ or 1 for all $x$
(i.e.\ each site can be occupied by at most one particle),
so that $Z_W(\w)$ can be written as a sum over subsets:
\be
      Z_W(\w)   \;=\;  \sum_{X' \subseteq X}
                       \left( \prod\limits_{x \in X'} w_x \right)
                       \left( \prod\limits_{\{x,y\} \subseteq X'}  W(x,y)
                       \right)
      \;.
    \label{eq1.2}
\ee
In this case $Z_W(\w)$ is a multiaffine polynomial,
i.e.\ of degree 1 in each $w_x$ separately.
Combinatorially, $Z_W(\w)$ is the generating polynomial
for {\em induced subgraphs}\/ of the complete graph,
in which each vertex $x$ gets weight $w_x$ and
each edge $xy$ gets weight $W(x,y)$.

If, in addition to hard-core self-repulsion,
we have $W(x,y) = 0$ or 1 for each pair $x \neq y$ ---
in statistical mechanics this is
called a {\em hard-core pair interaction}\/ ---
then we can define a (simple loopless) graph $G=(X,E)$
by setting $xy \in E$ whenever $W(x,y)=0$ and $x \neq y$,
so that $Z_W(\w)$ is precisely the {\em independent-set polynomial}\/
for $G$:
\be
      Z_G(\w)   \;=\;  \!\!\!\sum_{\begin{scarray}
                                      X' \subseteq X \\
                                      X' \, {\rm independent}
                                   \end{scarray}}
                       \prod\limits_{x \in X'} w_x
      \;.
  \label{eq1.3}
\ee
Traditionally the independent-set polynomial is defined as a
univariate polynomial $Z_G(w)$ in which $w_x$ is set equal to
the same value $w$ for all vertices $x$
\cite{Arocha_84,Fisher_90a,Fisher_90b,Hamidoune_90,Gutman_91,Gutman_92a,%
Gutman_92b,Hoede_94,Brown_00,Brown_01,Horrocks_02,Levit_03}.
But one of our main contentions in this paper
is that $Z_G$ is more naturally understood
as a multivariate polynomial;
this allows us, in particular, to exploit the fact that $Z_G$
is multiaffine.\footnote{
    More generally, let $G=(X,E)$ be a graph,
    and suppose that for $x \neq y$ we have $W(x,y) = 0$ or 1
    according as $xy \in E$ or $xy \notin E$;
    but let $0 \le W(x,x) \le 1$ be arbitrary.
    Then we have the amusing identity $Z_W(\w) = Z_G(\widetilde{\w})$ where
    $$ \widetilde{w}_x  \;=\;
       \sum\limits_{n=1}^\infty {W(x,x)^{n(n-1)/2} \over n!} \, w_x^n  \;. $$
}

More generally, given any $W$
satisfying $0 \le W(x,y) \le 1$ for all $x,y$,
let us define a simple loopless graph $G = G_W$
(the {\em support graph} of $W$)
by setting $xy \in E(G)$
if and only if $W(x,y) \neq 1$ and $x \neq y$.\footnote{
    Strictly speaking, $G_W$ ought to be called the
    (simple) support graph of $1-W$,
    since $xy \in E(G)$ if and only if $1 - W(x,y) \neq 0$ and $x \neq y$.
    In particular, in the case of a hard-core self-repulsion and
    hard-core pair interaction,
    $W$ is the adjacency matrix of the complementary graph $\bar{G}$.
    Please note also that $G_W$ does not contain any information
    about the diagonal weights $W(x,x)$.
}
The partition function $Z_W(\w)$ can be thought of
as a ``soft'' version of the independent-set polynomial for $G$,
in which an edge $xy \in E(G)$ has ``strength'' $1-W(x,y) \in (0,1]$.

In summary, we shall consider in this paper three levels of generality:
\begin{itemize}
    \item[(a)]  The general repulsive lattice gas \reff{defZ},
       in which multiple occupation of a site may be permitted.
    \item[(b)]  The lattice gas with hard-core self-repulsion \reff{eq1.2},
       in which multiple occupation of a site is forbidden,
       but in which the interactions between adjacent sites may be ``soft''.
    \item[(c)]  The lattice gas with hard-core self-repulsion and
       hard-core pair interaction \reff{eq1.3},
       which is simply the independent-set polynomial for the graph $G$.
\end{itemize}

Let us now return to the general case of a repulsive lattice gas \reff{defZ}.
Since $Z_W(\w)$ is an entire function of $\w$ satisfying $Z_W({\mathbf 0})
= 1$,
its logarithm is analytic in some neighborhood of $\w = {\mathbf 0}$,
and so can be expanded in a convergent Taylor series:
\be
      \log Z_W(\w)   \;=\;   \sum_{\n} c_{\n}(W) \, \w^{\n}
      \;,
    \label{logZ}
\ee
where we have used the notation $\w^{\n} = \prod_{x \in X} w_x^{n_x}$,
and of course $c_{\bf 0} = 0$.
In statistical mechanics, \reff{logZ} is called the
{\em Mayer expansion}\/ \cite{Uhlenbeck_62},
and there is a beautiful formula for the coefficients $c_{\n}(W)$,
whose derivation we will review briefly (Proposition~\ref{prop2.1}).
As a corollary of this formula, we will show
(Proposition~\ref{prop_derivs_of_cn})
that the Taylor series \reff{logZ} has alternating signs
whenever the lattice gas is repulsive:
\be
      (-1)^{|\n| - 1} c_{\n}(W)  \;\ge\; 0
\ee
for $0 \le W \le 1$.  (Here $|\n| = \sum_{x \in X} n_x$.)
Moreover, in this interval, $|c_{\n}(W)|=(-1)^{|\n| - 1} c_{\n}(W)$
is a decreasing function of each $W(x,y)$,
i.e.\ an increasing function of the ``interaction strength'' $1-W(x,y)$.

As a simple consequence of the alternating-sign property for $\log Z_W$,
we will then prove (Theorem~\ref{thm2.fund}) the equivalence of a number of
conditions for the nonvanishing of $Z_W$ in a closed polydisc
$\bar{D}_{\R} = \{ \w \colon\; |w_x| \le R_x \hbox{ for all } x \}$.
These equivalent conditions will play a central role in our study of
the Lov\'asz local lemma (Section~\ref{sec.LLL}).

Finally, we will provide (Section~\ref{sec.dob}) some sufficient conditions
for the nonvanishing of $Z_W$ in a closed polydisc $\bar{D}_{\R}$,
based on ``local'' properties of the interaction $W$ (or of the graph $G$).
Results of this type have traditionally been proven
\cite{Penrose_67,Cammarota_82,Seiler_82,Brydges_86,Brydges_87,Brydges_88,%
Simon_93,Brydges_99,Sokal_Mayer_in_prep}
by explicitly bounding the terms in the Mayer expansion \reff{logZ};
this requires some rather nontrivial combinatorics
(for example, Proposition~\ref{prop_penrose} below
together with the counting of trees).
Once this is done, an immediate consequence is that $Z_W$ is nonvanishing
in any polydisc where the series for $\log Z_W$ is convergent.
Dobrushin's brilliant idea \cite{Dobrushin_96a,Dobrushin_96b}
was to prove these two results in the opposite order.
First one proves, by an elementary induction on the cardinality
of the state space, that $Z_W$ is nonvanishing in a suitable polydisc
(Theorem~\ref{thm.dobrushin});
it then follows immediately that $\log Z_W$ is analytic in that polydisc,
and hence that its Taylor series \reff{logZ} is convergent there.
In Section~\ref{sec.dob} we will prove some refinements of this result
and investigate their sharpness;
and in Sections~\ref{sec.tree_interpretation} and \ref{sec.unravel}
we will provide some complementary results that give additional insight
into the nature of these bounds.
Let us remark that the Dobrushin--Shearer inductive method
employed in Section~\ref{sec.dob} is limited, at present,
to models with hard-core self-repulsion \reff{eq1.2},
for which $Z_W$ is a multiaffine polynomial.
It is an interesting open question to know whether this approach
can be made to work without the assumption of hard-core self-repulsion.

\subsection{The Lov\'asz local lemma}

In combinatorics we are frequently faced with a finite family
$(A_x)_{x\in X}$ of ``bad'' events in some probability space,
having probabilities $p_x = {\mathbb P}(A_x)$,
and we want to show that there is a positive probability that none of
the events $A_x$ occurs.
Under what conditions can we do this?
In general all we can say is that
${\mathbb P}(\bigcap_{x\in X}\overline A_x) \ge 1 - \sum_x p_x$,
since in the worst case the events $A_x$ could be disjoint;
so we would need $\sum_x p_x < 1$ to ensure that
${\mathbb P}(\bigcap_{x\in X}\overline A_x) > 0$.
On the other hand, if
the events $(A_x)_{x\in X}$ are independent,
then ${\mathbb P}(\bigcap_{x\in X}\overline A_x) = \prod_x (1-p_x)$,
which is positive as soon as $p_x < 1$ for all $x$.
This suggests that if the $(A_x)_{x\in X}$ are
in some sense ``not too strongly dependent'',
then it might be possible to prove
${\mathbb P}(\bigcap_{x\in X}\overline A_x) > 0$
under relatively mild conditions on the $\{p_x\}$.
This is the situation addressed by the Lov\'asz local lemma:
we allow strong dependence among {\em some}\/ subsets of the $(A_x)_{x\in X}$,
but insist that {\em most}\/ of these events are independent.
Specifically, the local lemma applies to collections of
events in which the dependencies are controlled by a
dependency graph, so that each event is independent from
the events that are not adjacent to it.

More precisely, let $G$ be a graph with vertex set $X$.
We say that $G$ is a {\em dependency graph}\/ for the family
$(A_x)_{x\in X}$
if, for each $x\in X$, the event $A_x$ is independent from the
$\sigma$-algebra generated by the events
$\{A_y \colon\; y \in X \setminus \Gamma^*(x)\}$.
[Here we have used the notation $\Gamma^*(x) = \Gamma(x) \cup \{x\}$,
where $\Gamma(x)$ is the set of vertices of $G$ adjacent to $x$.]
Erd\H os and Lov\'asz \cite{EL} proved the following fundamental result:

\begin{theorem}[Lov\'asz local lemma]
    \label{thm.LLL}
Let $G$ be a dependency graph for the family of events $(A_x)_{x\in X}$,
and suppose that $(r_x)_{x\in X}$ are
real numbers in $[0,1)$ such that, for each $x$,
\be
      {\mathbb P}(A_x) \;\le\;  r_x \prod_{y\in\Gamma(x)} (1-r_y)
      \;.
\label{eq1.6}
\ee
Then
${\mathbb P}(\bigcap_{x\in X}\overline A_x) \ge
    \prod_{x\in X}(1-r_x) > 0$.
\end{theorem}

Erd\H os and Spencer \cite{ES} (see also \cite{AS,Molloy_02})
later noted that the same conclusion holds
even if $A_x$ and $\sigma(A_y \colon\; y \in X \setminus \Gamma^*(x))$
are not independent, provided that the ``harmful'' conditional
probabilities are suitably bounded.  More precisely:

\begin{theorem}[Lopsided Lov\'asz local lemma]
    \label{thm.LLL_ESversion}
Let $(A_x)_{x\in X}$ be a family of events on some probability space,
and let $G$ be a graph with vertex set $X$.
Suppose that $(r_x)_{x\in X}$ are real numbers in $[0,1)$ such that,
for each $x$ and each $Y \subseteq X \setminus \Gamma^*(x)$,
we have
\be
      {\mathbb P}(A_x|\bigcap_{y\in Y}\overline A_y)  \;\le\;
      r_x \prod_{y\in\Gamma(x)} (1-r_y)
      \;.
\label{eq1.7}
\ee
Then
${\mathbb P}(\bigcap_{x\in X}\overline A_x) \ge
    \prod_{x\in X}(1-r_x) > 0$.
\end{theorem}

In fact, the arguments of \cite{EL,ES}
(see also \cite{Spencer_75,Spencer_77})
show that in Theorems~\ref{thm.LLL} and \ref{thm.LLL_ESversion}
a slightly stronger conclusion holds:
for all pairs $Y$, $Z$ of subsets of $X$ we have
\be
{\mathbb P}(\bigcap_{x\in Y}\overline A_x | \bigcap_{x\in Z}\overline A_x)
\;\ge\;  \prod_{x\in Y\setminus Z}(1-r_x)
\;.
\label{eq1.8}
\ee

In this paper we shall (following Shearer \cite{Shearer_85})
approach the problem by dividing our discussion into two parts:
\begin{itemize}
      \item[1)] A best-possible condition to have
${\mathbb P}(\bigcap_{x\in X}\overline A_x) > 0$,
in terms of the independent-set polynomial $Z_G(-{\mathbf p})$;  and
      \item[2)] A {\em sufficient}\/ condition to have
$Z_G(-{\mathbf p}) > 0$, along the lines of
Lov\'asz, Dobrushin and Shearer.
\end{itemize}
We shall treat the first problem in Section~\ref{sec.LLL}
and the second problem in Sections~\ref{sec.dob}--\ref{sec.unravel}.
We shall also prove a generalization of the Lov\'asz local lemma
that allows for ``soft'' dependencies
(Theorems~\ref{thm.softshearer} and \ref{softlopsided}).


Let us remark that we have been able to relate the Lov\'asz local lemma
to a combinatorial polynomial (namely, the independent-set polynomial)
only in the case of an {\em undirected}\/ dependency graph $G$.
Although the local lemma can be formulated quite naturally
for a dependency {\em digraph}\/ \cite{AS,Bollobas_01,Molloy_02},
we do not know whether the digraph Lov\'asz problem can be related
to any combinatorial polynomial.
(Clearly the independent-set polynomial cannot be the right object
 in the digraph context, since exclusion of simultaneous occupation is
 manifestly a symmetric condition.)

\subsection{Plan of this paper}

The plan of this paper is as follows:
In Section~\ref{sec2} we prove some general properties
of the partition function of a repulsive lattice gas.
In Section~\ref{sec3} we study the additional properties
that arise in the case of a hard-core self-repulsion \reff{eq1.2},
for which $Z_W$ is a multiaffine polynomial.
We also calculate some simple examples;
in particular, we show that when the support graph $G = G_W$ is a tree,
the partition function $Z_W$ can easily be calculated
by working upwards from the leaves.
In Section~\ref{sec.LLL} we prove our lower bound on
${\mathbb P}(\bigcap_{x\in X}\overline A_x) > 0$
in terms of the independent-set polynomial $Z_G(-{\mathbf p})$.
We also prove a ``soft-core'' generalization of this result.
In Section~\ref{sec.dob} we give some sufficient conditions
for the nonvanishing of $Z_W$ in a polydisc,
and investigate their sharpness.
In Section~\ref{sec.tree_interpretation} we show how the bounds of
Section~\ref{sec.dob} can be interpreted in terms of a lattice gas
on either the ``self-avoiding-walk tree''
or the ``pruned self-avoiding-walk tree'' of $G$.
In Section~\ref{sec.unravel} we show how this tree bound
can be understood as arising from the repeated application
of a single ``unfolding'' step.
Finally, in Section~\ref{sec.infinite} we study the
repulsive lattice gas on an infinite graph.

The reader primarily interested in the Lov\'asz local lemma
can read the statement of Theorem~\ref{thm2.fund} (skipping the proof)
and Definition~\ref{def.scrr},
quickly read Section~\ref{sec3.1},
and then jump directly to Section~\ref{sec.LLL}.
The reader primarily interested in the lattice gas
can skip Section~\ref{sec.LLL}.

\section{The lattice-gas partition function}  \label{sec2}

In this section we prove some general properties
of the partition function of a repulsive lattice gas.
We begin with some elementary identities (Section~\ref{sec2.identities}).
In Section~\ref{sec2.mayer}
we derive the Mayer expansion and prove some of its properties,
notably the alternating-sign property (Proposition~\ref{prop_derivs_of_cn}).
In Section~\ref{sec2.fundamental} we state and prove
the ``fundamental theorem'' (Theorem~\ref{thm2.fund}),
which sets forth a number of equivalent conditions
defining the set that we shall call $\scrr(W)$ [Definition~\ref{def.scrr}]
and that plays a central role in the remainder of this paper.
In Section~\ref{sec2.scrr} we derive the principal properties of
the set $\scrr(W)$.
In Section~\ref{sec2.alternating} we derive some further consequences
of the alternating-sign property.
In Section~\ref{sec2.irred} (which is a digression from the main thread
of the paper and can be omitted on a first reading)
we discuss the algebraic irreducibility of the multivariate
partition function $Z_W(\w)$.
Finally, in Section~\ref{sec2.convexity}
we make a brief remark concerning the convexity of
$\log Z_W(\w)$ at nonnegative fugacity $\w$.

Nearly all of the results in this section are valid for an
arbitrary repulsive lattice gas \reff{defZ},
in which multiple occupation of a site is permitted.
A few of the results are restricted to the case of a
hard-core self-repulsion \reff{eq1.2},
in which multiple occupation of a site is forbidden.

\subsection{Elementary identities}  \label{sec2.identities}

Consider a general lattice gas \reff{defZ} defined on a finite set $X$.
For any subset $\Lambda \subseteq X$, let us define the restricted
partition function
\be
    Z_{W,\Lambda}(\w)  \;=\;  Z_W(\w \, {\bf 1}_\Lambda)
\ee
where
\be
    (\w \, {\bf 1}_\Lambda)_x  \;=\;
    \cases{ w_x   & if $x \in \Lambda$ \cr
            0     & otherwise \cr
          }
\ee
This simply forces the sites in $X \setminus \Lambda$ to be unoccupied.
Since $W$ will be fixed throughout this subsection,
we shall often omit it from the notation and write simply $Z_\Lambda$.

For any $x \in \Lambda$, we can expand the partition function \reff{defZ_2}
in terms of the occupation number $n_x$:
we obtain the {\em fundamental identity}\/
\be
    Z_\Lambda(\w)  \;=\;
    \sum_{n=0}^\infty {w_x^n \over n!} \, W(x,x)^{n(n-1)/2} \,
        Z_{\Lambda \setminus x}(W(x,\cdot)^n \, \w)
  \label{eq.fund_general}
\ee
where
\be
      [W(x,\cdot)^n \, \w]_y  \;=\;  W(x,y)^n \, w_y  \;.
\ee
In the special case of a hard-core self-repulsion at site $x$
[i.e.\  $W(x,x) = 0$], only the terms $n=0,1$ appear in this sum;
this case will play a key role in this paper starting in Section~\ref{sec3.1}.
(Analogous identities can be derived where we expand in the
  occupation numbers at an arbitrary set of sites $S \subseteq \Lambda$;
  for example, a two-site identity for the hard-core case
  will play a major role in Section~\ref{sec.unravel}.)

{}From \reff{eq.fund_general} we can deduce the
{\em differentiation identity}\/
\begin{subeqnarray}
    {\partial Z_\Lambda(\w) \over \partial w_x}
    & = &
    \sum_{n=1}^\infty {w_x^{n-1} \over (n-1)!} \, W(x,x)^{n(n-1)/2} \,
        Z_{\Lambda \setminus x}(W(x,\cdot)^n \, \w)
      \\ [2mm]
    & = &
    \sum_{n=0}^\infty {w_x^{n} \over n!} \, W(x,x)^{n(n+1)/2} \,
        Z_{\Lambda \setminus x}(W(x,\cdot)^{n+1} \, \w)
      \\ [2mm]
    & = &
    \sum_{n=0}^\infty {[W(x,x) w_x]^n \over n!} \, W(x,x)^{n(n-1)/2} \,
        Z_{\Lambda \setminus x}(W(x,\cdot)^{n} W(x,\cdot) \w)
      \\ [2mm]
    & = &
    Z_\Lambda(W(x,\cdot) \w)  \;.
  \label{eq.differentiation}
\end{subeqnarray}
Repeated application of \reff{eq.differentiation} yields the
{\em multiple differentiation identity}\/
\be
    {\partial^n Z_\Lambda(\w) \over \partial w_{x_1} \cdots \partial w_{x_n}}
    \;=\;
    \left( \prod\limits_{1 \le i < j \le n} W(x_i,x_j) \right)
    Z_\Lambda\Biggl( \prod\limits_{i=1}^n W(x_i, \cdot) \, \w \Biggr)
    \;,
  \label{eq.differentiation.multiple}
\ee
where $(\prod\limits_{i=1}^n 
W(x_i,\cdot)\,\w)_x=w_x\prod\limits_{i=1}^nW(x_i,x)$.

\subsection{The Mayer expansion} \label{sec2.mayer}


We begin by reviewing the derivation of the Mayer expansion \reff{logZ}.
The first step is to trivially rewrite the partition function
\reff{defZ_1} as
\be
      Z_W(\w)  \;=\;  \sum\limits_{n=0}^\infty {1 \over n!}
                      \sum_{x_1,\ldots,x_n \in X}
                      \left( \prod\limits_{i=1}^n w_{x_i} \right)
                      \sum_{G \in \scrg_n}  \prod_{ij \in E(G)} F(x_i,x_j)
      \;,
\ee
where $\scrg_n$ is the set of all (simple loopless undirected) graphs
on the vertex set $\{1,\ldots,n\}$,
and
\be
      F(x,y)  \;=\;  W(x,y) - 1
\ee
is called the {\em two-particle Mayer factor}\/.
This is of the form
\be
      Z_W(\w)  \;=\;  \sum\limits_{n=0}^\infty {1 \over n!}
                      \sum_{G \in \scrg_n}  \scrw(G)
\ee
where the weights
\be
      \scrw(G)  \;=\;  \sum_{x_1,\ldots,x_n \in X}
                       \left( \prod\limits_{i=1}^n w_{x_i} \right)
                       \prod_{ij \in E(G)} F(x_i,x_j)
\ee
satisfy
\begin{itemize}
      \item[(a)]  $\scrw(\emptyset) = 1$;
      \item[(b)]  $\scrw(G) = \scrw(G')$ whenever $G \cong G'$
         (i.e.\ whenever $G$ and $G'$ differ only by a relabelling of 
vertices);
         and
      \item[(c)]  $\scrw(G) = \scrw(G_1) \, \scrw(G_2)$
         whenever $G$ is isomorphic to the disjoint union of $G_1$ and $G_2$.
\end{itemize}
It then follows from the exponential formula
\cite{Uhlenbeck_62,Wilf_94,Bergeron_98,Stanley_99} that
\begin{subeqnarray}
      \log Z_W(\w)  & = & \sum\limits_{n=0}^\infty {1 \over n!}
                           \sum_{G \in \scrc_n}  \scrw(G)          \\[2mm]
      & = &
\sum\limits_{n=0}^\infty {1 \over n!}
                      \sum_{x_1,\ldots,x_n \in X}
                      \left( \prod\limits_{i=1}^n w_{x_i} \right)
                      \sum_{G \in \scrc_n}  \prod_{ij \in E(G)} F(x_i,x_j),
    \label{mayer_exp}
\end{subeqnarray}
at least in the sense of formal power series in $\w$,
where $\scrc_n \subseteq \scrg_n$ is the set of {\em connected}\/ graphs
on $\{1,\ldots,n\}$.
Therefore:

\begin{proposition}
      \label{prop2.1}
The coefficients $c_{\n}(W)$ of the Mayer expansion \reff{logZ}
are given by
\begin{subeqnarray}
      c_{\n}(W)
      & = &
      {1 \over n!} \!\!\!
      \sum\limits_{\begin{scarray}
                     x_1,\ldots,x_{n} \in X \\
                     \#\{i \colon\, x_i=x\} = n_x \, \forall x
                   \end{scarray}}
      \! \sum\limits_{G\in\scrc_n}
      \prod\limits_{ij\in E(G)}   F(x_i,x_j)
           \slabel{eq.mayer_coeff.a}  \\[2mm]
      & = &
      \!\!\!
      \sum_{\begin{scarray}
              (S_x)_{x \in X}  \\
              \biguplus\limits_{x \in X} \!\! S_x = \{1,\ldots,n\} \\
              |S_x| = n_x \,\forall x
            \end{scarray}}
      \! \sum\limits_{G\in\scrc_n}
      \prod_{\{x,y\} \subseteq X} F(x,y)^{e_G(S_x,S_y)}
      \prod_{x \in X} F(x,x)^{e_G(S_x)}
        \qquad\qquad
           \slabel{eq.mayer_coeff.b}
    \label{eq.mayer_coeff}
\end{subeqnarray}
where $n = |\n|$, the first sum in \reff{eq.mayer_coeff.b}
runs over all partitions of $\{1,\ldots,n\}$
into disjoint subsets $(S_x)_{x \in X}$ with the specified cardinalities,
and $e_G(S_x,S_y)$ [resp.\ $e_G(S_x)$] denotes the number of edges of $G$
connecting $S_x$ to $S_y$ [resp.\ within $S_x$].
\end{proposition}

\medskip
\par\noindent
{\bf Remark.}
Even in the simple case $X = \{x\}$ and $W(x,x)=0$, for which
$\log Z_W(w) = \log(1+w) = \sum_{n=1}^\infty {(-1)^{n-1} \over n} w^n$,
the Mayer expansion \reff{mayer_exp} implies the nontrivial identity
\be
      \sum_{G \in \scrc_n} (-1)^{|E(G)|}  \;=\;
      (-1)^{n-1} (n-1)!
    \label{Kn.identity}
\ee
for the generating function of connected spanning subgraphs
of the complete graph $K_n$.
We leave it as an exercise for the reader to find
a direct combinatorial proof of \reff{Kn.identity}.

\bigskip

In order to analyze the Mayer coefficients $c_{\n}(W)$,
it is convenient to proceed in a bit more generality.
So let $H=(V,E)$ be a graph --- we allow loops and multiple edges ---
and let $\z = \{z_e\}_{e \in E}$ be a family of
complex edge weights for $H$.
[Later we will specialize to $H=K_n$.]
Define the generating function of connected spanning subgraphs of $H$
(or ``connected sum'' for short),
\be
      C_H(\z)   \;=\;
      \!\!\sum\limits_{\begin{scarray}
                          E' \subseteq E  \\
                          (V,E') \, {\rm connected}
                       \end{scarray}}
      \prod\limits_{e \in E'} z_e
      \;.
\ee
It is easy to see that $C_H$ satisfies the
deletion-contraction relation
\be
      C_H(\z)   \;=\;   C_{H \setminus e}(\z_{\neq e})  \,+\,
                        z_e C_{H/e}(\z_{\neq e})
    \label{eq.delcon}
\ee
for any edge $e \in E$:
here $H \setminus e$ is the graph $H$ with edge $e$ deleted,
$H/e$ is the graph $H$ with edge $e$ contracted
(note that we do {\em not}\/ delete any loops or multiple edges
    that may be formed),
and $\z_{\neq e}$ denotes the family $\{z_f\}_{f \in E \setminus e}$.
Please note that if $e$ is a loop, then $H/e=H\setminus e$ and hence
\be
    C_H(\z) \;=\;  C_{H \setminus e}(\z_{\neq e})  \,+\,
                        z_e C_{H/e}(\z_{\neq e})
            \;=\;  (1+z_e) \, C_{H\setminus e}(\z_{\neq e})
    \;.
\ee

Let $\scrc$ (resp.\ $\scrt$) be the set of subsets $E' \subseteq E$
such that $(V,E')$ is connected (resp.\ is a tree).
Clearly $\scrc$ is an increasing family of subsets of $E$
with respect to set-theoretic inclusion,
and the minimal elements of $\scrc$ are precisely those of $\scrt$
(i.e.\ the spanning trees).
It is a nontrivial but well-known fact
\cite[Sections 7.2 and 7.3]{Bjorner_92}
\cite[Section 8.3]{Ziegler_95}
that the (anti-)complex $\scrc$ is {\em partitionable}\/:
that is, there exists a map ${\sf R} \colon\, \scrt \to \scrc$
such that ${\sf R}(T) \supseteq T$ for all $T \in \scrt$
and $\scrc = \biguplus_{T \in \scrt} [T, \, {\sf R}(T)]$ (disjoint union),
where $[E_1,E_2]$ denotes the Boolean interval
$\{ E' \colon\; E_1 \subseteq E' \subseteq E_2 \}$.
In fact, many alternative choices of ${\sf R}$ are available
\cite[Sections 7.2 and 7.3]{Bjorner_92}
\cite[Sections 2 and 6]{Gessel_96}
\cite[Proposition 13.7 et seq.]{Biggs_93}
\cite[Proposition 4.1]{Sokal_chromatic_bounds},
and most of our arguments will not depend on any specific choice of ${\sf R}$.
However, at one point (Proposition~\ref{prop_alternating.strict})
we shall need the existence of an ${\sf R}$ with the following
additional property:

\begin{lemma}
    \label{lemma.partitionability}
Let $H=(V,E)$ be a connected graph, and fix any $T_0 \in \scrt$.
Then there exists a map ${\sf R} \colon\, \scrt \to \scrc$ such that
\begin{itemize}
    \item[(a)]  ${\sf R}(T) \supseteq T$ for all $T \in \scrt$;
    \item[(b)]  $\scrc$ is the disjoint union of the Boolean intervals
         $[T, \, {\sf R}(T)]$, $T\in \scrt$; and
    \item[(c)]  ${\sf R}(T_0) = T_0$.
\end{itemize}
\end{lemma}

\proof
If $H$ has one vertex and no edges,
then $\scrt=\scrc=\{\emptyset\}$ and the result holds trivially;
so assume henceforth that $E \neq \emptyset$.
Assign arbitrary weights $w_e > 0$ chosen so that
no two spanning trees have equal weight
(for example, one can choose the $w_e$ to be
linearly independent over the rationals).
For each $E' \in \scrc$, let ${\sf S}(E')$ be the (unique) minimum-weight
spanning tree contained in $E'$.
(This can be constructed by a greedy algorithm,
i.e.\ start from $\emptyset$ and keep adding
the lowest-weight edge in $E'$ that does not create a cycle.
See, for instance, \cite[Section I.2]{Bollobas_98}.)
We then define ${\sf R}(T)$ to be the union of all $E'$
that have ${\sf S}(E') = T$.
To verify that this works, we need to show that if
${\sf S}(E_1) = {\sf S}(E_2) = T$, then ${\sf S}(E_1 \cup E_2) = T$;
but this follows easily from the validity of the greedy algorithm.
[Note that this construction includes, as a special case,
the lexicographically minimum spanning tree for any ordering of $E$:
it suffices to take $w_n = 2^n$.]

By choosing the weights so that $w_e > w_f$ whenever
$e \in T_0$ and $f \in E \setminus T_0$,
we can ensure that for $E' \neq T_0$ the first edge chosen
by the greedy algorithm will not belong to $T_0$.
Therefore, ${\sf R}(T_0) = T_0$.
\qed

\medskip
\par\noindent
{\bf Remark.}
The greedy algorithm works, more generally, for an arbitrary matroid
\cite[Section 1.8]{Oxley_92}.
It follows that the independent-set complex of a matroid
(or, dually, the spanning-set anti-complex)
is partitionable \cite[Sections 7.2 and 7.3]{Bjorner_92}.

\bigskip

Given the existence of ${\sf R}$, we have the following
simple but fundamental identity:

\begin{proposition}[partitionability identity]
Let ${\sf R} \colon\, \scrt \to \scrc$
be any map satisfying ${\sf R}(T) \supseteq T$ for all $T \in \scrt$
and $\scrc$ is the disjoint union of the Boolean intervals
$[T, \, {\sf R}(T)]$, for $T\in \scrt$.
Then
\begin{eqnarray}
      C_H(\z)
      & = &
\sum\limits_{\begin{scarray}
                      T \subseteq E  \\
                      (V,T) \, {\rm tree}
                   \end{scarray}}
      \; \prod\limits_{e \in T} z_e
      \sum\limits_{T \subseteq E' \subseteq {\sf R}(T)}
      \; \prod\limits_{e \in E' \setminus T} z_e
         \nonumber \\[3mm]
      & = &
\sum\limits_{\begin{scarray}
                      T \subseteq E  \\
                      (V,T) \, {\rm tree}
                   \end{scarray}}
      \; \prod\limits_{e \in T} z_e
      \; \prod\limits_{e \in {\sf R}(T) \setminus T} (1 + z_e)
      \;.
    \label{penrose_identity}
\end{eqnarray}
\end{proposition}

\noindent
This identity (for one specific choice of ${\sf R}$)
is due originally to Penrose \cite{Penrose_67}.

One immediate consequence of the identity \reff{penrose_identity}
is the following inequality valid in the
``complex repulsive'' regime $|1+z_e| \le 1$:

\begin{proposition}[Penrose \protect\cite{Penrose_67}]
     \label{prop_penrose}
Let $H=(V,E)$ be a finite undirected graph
equipped with complex edge weights $\{ z_e \}_{e \in E}$
satisfying $|1 + z_e| \le 1$ for all $e$.
Then
\be
      |C_H(\z)|
      \;\,\le\;\,
      \sum\limits_{\begin{scarray}
                      T \subseteq E  \\
                      (V,T) \, {\rm tree}
                   \end{scarray}}
      \; \prod\limits_{e \in T} |z_e|
      \;.
    \label{penrose_bound}
\ee
\end{proposition}

\noindent
This bound plays a major role in several traditional (pre-Dobrushin)
proofs of convergence of the Mayer expansion
\cite{Penrose_67,Cammarota_82,Seiler_82,Brydges_86,%
Simon_93,Brydges_99,Sokal_Mayer_in_prep},
as well as in a recent bound on the zeros of chromatic polynomials
\cite{Sokal_chromatic_bounds}.

Let us now define the ``generalized connected sum''
\begin{subeqnarray}
      C_H(\lambda;\z)   & = &
\!\!\sum\limits_{\begin{scarray}
                          E' \subseteq E  \\
                          (V,E') \, {\rm connected}
                       \end{scarray}}
      \!\lambda^{c(E')} \,
      \prod\limits_{e \in E'} z_e
           \\[2mm]
      & = & \lambda^{-(|V|-1)} C_H(\lambda \z)
         \slabel{gen_conn_sum_b}
\end{subeqnarray}
where $c(E') = |E'| - |V| + 1$
is the cyclomatic number of the connected subgraph $(V,E')$
[more generally, for a subgraph $(V,E')$ with $k(E')$ components,
    the cyclomatic number is $c(E') = |E'| - |V| + k(E')$].
Of course, \reff{gen_conn_sum_b} shows that $C_H(\lambda;\z)$
contains no more information than $C_H(\z)$;
it is just a convenient way of
scaling all the variables $z_e$ simultaneously.
The function $C_H(\lambda;\z)$ interpolates between the tree sum ($\lambda=0$)
and the connected sum ($\lambda=1$);
and Proposition~\ref{prop_penrose} can be rephrased as saying that
$|C_H(1;\z)| \le C_H(0;|\z|)$.
One of us has conjectured that
the absolute-value signs can in fact be put outside the sum,
i.e.\ $|C_H(1;\z)| \le |C_H(0;\z)| \;$
\cite[Remark 2 in Section 4.1]{Sokal_chromatic_bounds}.
In general this is still an open problem;
but if the $z_e$ are {\em real}\/ and lie in the interval $[-1,0]$
--- which corresponds to the physical repulsive regime $0 \le W \le 1$ ---
then a vastly {\em stronger}\/ result is true:

\begin{proposition}[Sokal \protect\cite{Sokal_chromatic_bounds}]
     \label{prop_alternating}
Let $H=(V,E)$ be a finite undirected graph
equipped with real edge weights $\{ z_e \}_{e \in E}$
satisfying $-1 \le z_e \le 0$ for all $e$.
Then
\be
      (-1)^{k+|V|-1} \, {d^k \over d\lambda^k} \, C_H(\lambda; \z)
      \;\ge\; 0
    \label{eq.alternating}
\ee
on $0 \le \lambda \le 1$, for all integers $k \ge 0$.
In particular, setting $k=0$ and $\lambda=1$ we have
\be
      (-1)^{|V|-1}  C_H(\z)  \;\ge\; 0
      \;.
    \label{eq.alternating_k=0}
\ee
\end{proposition}

\proof
By \reff{penrose_identity} and \reff{gen_conn_sum_b}, we have
\be
      C_H(\lambda; \z)   \;=\;
      \sum\limits_{\begin{scarray}
                      T \subseteq E  \\
                      (V,T) \, {\rm tree}
                   \end{scarray}}
      \; \prod\limits_{e \in T} z_e
      \; \prod\limits_{e \in {\sf R}(T) \setminus T} (1 + \lambda z_e)
  \label{eq.proof.prop_alternating.1}
\ee
and hence
\be
      {d^k \over d\lambda^k} \, C_H(\lambda; \z)   \;=\;
k! \!\sum\limits_{\begin{scarray}
                      T \subseteq E  \\
                      (V,T) \, {\rm tree}
                   \end{scarray}}
      \sum\limits_{\begin{scarray}
                      \widetilde{T} \subseteq
                           \hbox{\sf\scriptsize R}(T) \setminus T \\
                      |\widetilde{T}| = k
                   \end{scarray}}
      \; \prod\limits_{e \in T \cup \widetilde{T}} z_e
      \; \prod\limits_{e \in {\sf R}(T) \setminus (T \cup \widetilde{T})}
                                                           (1 + \lambda z_e)
      \;,
    \label{eq.alternating.final}
\ee
which has the claimed sign whenever $0 \le \lambda \le 1$
and $-1 \le z_e \le 0$ for all $e$.\footnote{
      Equation (4.7) of \cite{Sokal_chromatic_bounds}
      inadvertently omitted the prefactor $k!$ in \reff{eq.alternating.final}.
}
\qed

We shall also need a variant of Proposition~\ref{prop_alternating}
with strict inequality for the case $k=0$:

\begin{proposition}
     \label{prop_alternating.strict}
Let $H=(V,E)$ be a finite undirected graph
equipped with real edge weights $\{ z_e \}_{e \in E}$
satisfying $-1 \le z_e \le 0$ for all $e$.
Assume furthermore that the subgraph $(V, \supp \z)$ is connected
[here $\supp \z = \{e \in E \colon\, z_e \neq 0\}$].
Then
\be
      (-1)^{|V|-1} \, C_H(\z) \;>\; 0   \;.
    \label{eq.alternating.strict}
\ee
\end{proposition}

\proof
Choose $T_0 \subseteq \supp \z$ such that $(V,T_0)$ is a spanning tree,
and use Lemma~\ref{lemma.partitionability} to choose ${\sf R}$
so that ${\sf R}(T_0) = T_0$.
Then, in the sum \reff{eq.proof.prop_alternating.1},
the term $T=T_0$ is nonzero with sign $(-1)^{|V|-1}$,
and all the other terms either have the sign $(-1)^{|V|-1}$ or else are zero.
\qed

We can also prove some inequalities on the partial derivatives of $C_H(\z)$
with respect to individual weights $z_e$.
Note first that the deletion-contraction identity \reff{eq.delcon}
implies that
\be
      {\partial C_H(\z) \over \partial z_e}  \;=\;
      C_{H/e}(\z_{\neq e})
      \;.
\ee
The graph $H/e$ has $|V|$ vertices if $e$ is a loop,
and $|V|-1$ vertices if $e$ is not a loop.
Repeated application of these facts together with \reff{eq.alternating_k=0}
yields the following result:

\begin{proposition}
     \label{prop_alternating_2}
Let $H=(V,E)$ be a finite undirected graph
equipped with real edge weights $\{ z_e \}_{e \in E}$
satisfying $-1 \le z_e \le 0$ for all $e$.
Let $k\ge 1$ and $e_1,\ldots,e_k \in E$.
\begin{itemize}
      \item[(a)]  If $e_1,\ldots,e_k$ are not all distinct,
         we have
\be\label{2.26}
       {\partial C_H(\z) \over \partial z_{e_1} \cdots \partial z_{e_k}}
       \;=\;  0   \;.
\ee
      \item[(b)]  If $e_1,\ldots,e_k$ are all distinct
and form
a subgraph with cyclomatic number $c$, we have
\be\label{2.27}
       (-1)^{|V|-k+c-1} \,
       {\partial C_H(\z) \over \partial z_{e_1} \cdots \partial z_{e_k}}
       \;\ge\;  0   \;.
\ee
\end{itemize}
\end{proposition}

Let us now specialize these results
to the Mayer expansion \reff{eq.mayer_coeff.a}
by taking $H=K_n$
(where $n = |\n|$ and $K_n$ denotes the complete graph on $n$ vertices),
$z_{ij} = F(x_i,x_j) = W(x_i,x_j) - 1$ and
then summing over $x_1,\ldots,x_n \in X$
with the specified cardinalities:  we get
\be
    c_{\n}(W)  \;=\;
    {1 \over n!}
    \sum\limits_{\begin{scarray}
                     x_1,\ldots,x_n \in X \\
                     \#\{i \colon\, x_i=x\} = n_x \, \forall x
                 \end{scarray}}
    C_{K_n}(\z({\sf x}))
  \label{eq.mayer.CKn}
\ee
where ${\sf x} = (x_1,\ldots,x_n)$ and $\z({\sf x})_{ij} = F(x_i,x_j)$.
Since any subgraph of $K_n$ with 0, 1 or 2 edges
has cyclomatic number 0, we can use
Propositions~\ref{prop_alternating} and \ref{prop_alternating_2}
to deduce the following:

\begin{proposition}[signs of Mayer coefficients]
      \label{prop_derivs_of_cn}
Suppose that the lattice gas is repulsive,
i.e.\ $0 \le W(x,y) \le 1$ for all $x,y \in X$.
Then, for all $x,y,x',y'\in X$, the Mayer coefficients $c_{\n}(W)$ satisfy
\begin{eqnarray}
      (-1)^{|\n| - 1} \, c_{\n}(W)  & \ge & 0
         \label{eq.sign_of_cn}     \\[2mm]
      (-1)^{|\n| - 1} \,
      {\partial c_{\n}(W) \over \partial W(x,y)} & \le & 0
         \label{eq.first_deriv_of_cn}     \\[2mm]
      (-1)^{|\n| - 1} \,
      {\partial^2 c_{\n}(W) \over \partial W(x,y) \, \partial W(x',y')}
				 & \ge & 0
         \label{eq.second_deriv_of_cn}
\end{eqnarray}
\end{proposition}

\proof
\reff{eq.sign_of_cn} is an immediate consequence of \reff{eq.mayer.CKn}
and \reff{eq.alternating_k=0}.  For \reff{eq.first_deriv_of_cn}, note first
that for any fixed $x_1,\ldots,x_n\in X$, we have
\be
\frac{\partial C_{K_n}(\z({\sf x})) }{\partial W(x,y)}
=\sum_{\{i,j\}\in E(x,y;x_1,\ldots,x_n)}
\frac{\partial C_{K_n}(\z) }{\partial z_{ij}}{\Bigg|}_{\z=\z({\sf x})}  \;,
\ee
where $E(x,y;x_1,\ldots,x_n)$ is the set of unordered pairs $\{i,j\}$
($i \neq j$) such that $\{x_i,x_j\}=\{x,y\}$.
Since any subgraph of $K_n$ with one edge has cyclomatic number 0,
the inequality \reff{eq.first_deriv_of_cn} now follows from \reff{2.27}.
Similarly, the left-hand side of \reff{eq.second_deriv_of_cn}
gives rise to a double sum over pairs $\{i,j\} \in E(x,y;x_1,\ldots,x_n)$
and $\{i',j'\} \in E(x',y';x_1,\ldots,x_n)$.
The diagonal terms (if any) in this sum vanish by \reff{2.26},
and the other terms have the claimed sign by \reff{2.27}.
\qed

Please note that \reff{eq.second_deriv_of_cn} can be nonzero even when
$x=x'$ and $y=y'$, because distinct edges of $K_n$ could correspond
to the same pair $x,y$.
Note also that the sign of the third and higher derivatives
cannot be controlled by this method,
because a subgraph of $K_n$ ($n \ge 4$) containing three or more edges could
have cyclomatic number of either parity.
Indeed, for the simple case $X = \{x\}$ and $W(x,x)=W$ we have
\be
      Z_W(w)   \;=\;  \sum_{n=0}^\infty {w^n W^{n(n-1)/2}  \over n!}
\ee
and hence
\begin{subeqnarray}
    \log Z_W(w)   & = &
       w \,+\, {F \over 2} w^2 \,+\, {3F^2 + F^3 \over 6} w^3   \nonumber \\
       &  & \quad +\,
               {16F^3 + 15F^4 + 6F^5 + F^6 \over 24} w^4 \,+\, O(w^5)  \\[2mm]
       & = &
       w \,+\, {W-1 \over 2} w^2 \,+\, {(W+2)(W-1)^2 \over 6} 
w^3   \nonumber \\
       &  & \quad +\,
               {(W^3+3W^2+6W+6)(W-1)^3 \over 24} w^4 \,+\, O(w^5)  \;, \quad
\end{subeqnarray}
so that
\be
      {\partial^3 c_4(W) \over \partial W^3}  \;=\;  5W^3 - 1
\ee
has no fixed sign on $0 \le W \le 1$.

We can also determine the precise conditions under which
the inequality \reff{eq.sign_of_cn} is strict:

\begin{proposition}[condition for Mayer coefficients to be nonzero]
      \label{prop_sign_of_cn.strict}
Suppose that the lattice gas is repulsive,
i.e.\ $0 \le W(x,y) \le 1$ for all $x,y \in X$.
Let $\n = (n_x)_{x \in X}$ be a multi-index with support
$\supp\n = \{x \in X \colon\, n_x \neq 0\}$.
Then $(-1)^{|\n| - 1} \, c_{\n}(W)  > 0$
if and only if one of the following is true:
\begin{itemize}
    \item[(a)]  $|\n| = 1$;
    \item[(b)]  $|\n| \ge 2$, $\supp\n = \{x\}$ and $W(x,x) \neq 1$; or
    \item[(c)]  $|\supp\n| \ge 2$ and the induced subgraph $G_W[\supp\n]$
       is connected.
\end{itemize}
\end{proposition}

\proof
In case (a), $c_{\n}(W) = 1 > 0$;
in case (b) or (c), for each term in \reff{eq.mayer.CKn}
[actually they are all the same!]
the subgraph of $K_n$ with edge set $\supp \z({\sf x})$ is connected,
so that $(-1)^{|\n| - 1} \, C_{K_n}(\z({\sf x})) > 0$
by Proposition~\ref{prop_alternating.strict}.

There are only two other possibilities:
\begin{itemize}
    \item[(d)]  $\n = 0$;
    \item[(e)]  $|\supp\n| \ge 2$ and $G_W[\supp\n]$ is disconnected.
\end{itemize}
In case (d), clearly $c_{\n}(W) = 0$;
and in case (e), for each term in \reff{eq.mayer.CKn}
the subgraph of $K_n$ with edge set $\supp \z({\sf x})$ is disconnected,
so that $(-1)^{|\n| - 1} \, C_{K_n}(\z({\sf x})) = 0$.
\qed

\medskip
\par\noindent
{\bf Historical remark.}
The alternating-sign property \reff{eq.sign_of_cn}
for the Mayer coefficients of a repulsive gas
has been known in the physics literature for over 40 years:
see Groeneveld \cite{Groeneveld_62} for a brief sketch of the proof,
which uses methods quite different from ours.\footnote{
   Groeneveld \cite{Groeneveld_62} writes that
   ``a more detailed account of this work, containing also the
   corresponding results for the pressure and the distribution functions
   will be published in Physica'',
   but to our knowledge that more detailed paper never appeared.
   See also Penrose \cite[eq.~(8.1)]{Penrose_63} for a related result.
}
Nevertheless, this result does not seem to be as well known as it should be.
To the best of our knowledge, the inequalities \reff{eq.first_deriv_of_cn}
and \reff{eq.second_deriv_of_cn} are new.
We think that the Mayer coefficients $c_{\n}(W)$,
and more generally the ``connected sums'' $C_H(\z)$,
merit further study from a combinatorial point of view;
we would not be surprised if new identities or inequalities
were waiting to be discovered.

\subsection{The fundamental theorem}
   \label{sec2.fundamental}

Let us now state the principal result of this section:

\begin{theorem}[The fundamental theorem]
      \label{thm2.fund}
Consider any repulsive lattice gas,
and let $\R = \{R_x\} _{x \in X} \ge 0$.
Then the following are equivalent:
\begin{itemize}
      \item[(a)] There exists a connected set $C \subseteq (-\infty,0]^X$
          that contains both $\mathbf 0$ and $-\R$, such that
          $Z_W(\w) > 0$ for all $\w \in C$.
          [Equivalently, $-\R$ belongs to the connected component of
           $Z_W^{-1}(0,\infty) \cap (-\infty,0]^X$ containing $\mathbf 0$.]
      \item[(b)]  $Z_W(\w) > 0$ for all $\w$ satisfying
		$-\R \le \w \le \mathbf 0$.
      \item[(c)]  $Z_W(\w) \neq 0$ for all $\w$ satisfying
          $|\w| \le \R$.\footnote{
                Here we use the notation $|\w|=\{|w_x|\}_{x\in X}$.
	        This conflicts slightly with our notation
                $|\n|=\sum_{x\in X}n_x$ for multi-indices,
                but we trust that it will not lead to any confusion. }
      \item[(d)]  The Taylor series for $\log Z_W(\w)$
	  around $\mathbf 0$ is convergent
         at $\w = -\R$.
      \item[(e)]  The Taylor series for $\log Z_W(\w)$
	  around $\mathbf 0$ is absolutely
         convergent for $|\w| \le \R$.
\end{itemize}
Moreover, when these conditions hold, we have
$|Z_W(\w)| \ge Z_W(-\R) > 0$ for all $\w$ satisfying $|\w| \le \R$.

In the case of hard-core self-repulsion, (a)--(e) are also equivalent to
\begin{itemize}
      \item[(b\textprime)]  $Z_W(-\R \, \, {\bf 1}_S) > 0$ for all $S 
\subseteq X$,
where
\be
      (\R \, {\bf 1}_S)_x  \;=\;  \cases{ R_x   & if $x \in S$  \cr
                                       0     & otherwise     \cr
                                     }
\label{eqrestrictR}
\ee
      \item[(f)]  $Z_W(-\R) > 0$, and $(-1)^{|S|}Z_W(-\R;S) \ge 0$
for all $S \subseteq X$, where
\be
      Z_W(\w;S) \;=\;  \sum_{S \subseteq X' \subseteq X}
                       \left( \prod\limits_{x \in X'} w_x \right)
                       \left( \prod\limits_{\{x,y\} \subseteq X'}  W(x,y)
                       \right)
      \;.
\label{zsemicol}
\ee
      \item[(g)]  There exists a probability measure $P$ on $2^X$
satisfying $P(\emptyset) > 0$ and
\be
      \sum\limits_{T \supseteq S} P(T)   \;=\;
                       \left( \prod\limits_{x \in S} R_x \right)
                       \left( \prod\limits_{\{x,y\} \subseteq S}  W(x,y)
                       \right)
     \label{def.PT}
\ee
for all $S \subseteq X$.
[This probability measure is unique
    and is given by $P(S) = (-1)^{|S|} Z_W(-\R;S)$.
    In particular, $P(\emptyset) = Z_W(-\R) > 0$.]
\end{itemize}
\end{theorem}

\medskip
\par\noindent
{\bf Remarks.}
1.  The conditions (b\textprime), (f) and (g)
are inspired in part by Shearer \cite[Theorem 1]{Shearer_85}.

2.  Suppose that the univariate entire function $Z_W(w)$,
defined by setting $w_x = w$ for all $x$,
is strictly positive whenever $-R \le w \le 0$.
Then in fact $Z_W(\w) > 0$ whenever $-R \le w_x \le 0$ for all $x$:
this follows from (a)$\implies$(b)
by taking $C$ to be the segment $[-R,0]$ of the diagonal.

\bigskip

The proof of Theorem~\ref{thm2.fund} will hinge on the
alternating-sign property \reff{eq.sign_of_cn}
for the Taylor coefficients of $\log Z_W$.
In preparation for this proof,
let us recall the Vivanti--Pringsheim theorem
in the theory of analytic functions
of a single complex variable \cite[Theorem 5.7.1]{Hille_73}:
if a power series $f(z) = \sum_{n=0}^\infty a_n z^n$
with {\em nonnegative}\/ coefficients
has a finite nonzero radius of convergence,
then the point of the circle of convergence
lying on the positive real axis is a singular point of the function $f$.
Otherwise put, if $f$ is a function whose Taylor series at 0
has all nonnegative coefficients
and $f$ is analytic on some complex neighborhood
of the real interval $[0,R)$,
then $f$ is in fact analytic on the open disc of radius $R$
centered at the origin
and its Taylor series is absolutely convergent there.
Here we will need the following multidimensional generalization
of the Vivanti--Pringsheim theorem:

\begin{proposition}[multidimensional Vivanti--Pringsheim theorem]
      \label{prop.pringsheim}
Let $C$ be a connected subset of $[0,\infty)^n$ containing $\mathbf 0$,
let $U$ be an open neighborhood of $C$ in $\C^n$,
and let $f$ be a function analytic on $U$ whose Taylor series around
$\mathbf 0$ has all nonnegative coefficients.
Then the Taylor series of $f$ around $\mathbf 0$
converges absolutely on the set
${\rm hull}(C) \equiv \bigcup\limits_{\R \in C} \bar{D}_{\R}$,
where $\bar{D}_{\R}$ denotes the closed polydisc
    $\{ \w \in \C^n \colon\; |w_i| \le R_i \hbox{ for all } i \}$,
and it defines a function that is continuous on ${\rm hull}(C)$
and analytic on its interior.
\end{proposition}

\medskip
\par\noindent
{\bf Remark.}
We could equally well start from an open neighborhood $U$ of $\mathbf 0$
in $\C^n$,
and {\em define}\/ $C$ to be the connected component of
$U \cap [0,\infty)^n$ containing $\mathbf 0$.
This is the maximal set $C$ compatible with the given $U$;
note that it is open in $[0,\infty)^n$,
and that ${\rm hull}(C)$ is then open in $\C^n$.
\medskip

\proof
Let $f(\z) = \sum_{\n} a_{\n} \z^{\n}$ be the Taylor series of $f$
around $\mathbf 0$.
And let us define
\begin{subeqnarray}
      S  & = & \{ \R \in [0,\infty)^n \colon\;
                   \hbox{$\sum_{\n} a_{\n} \z^{\n}$ converges
                         absolutely at $\z = \R$}  \}
         \\
         & = & \{ \R \in [0,\infty)^n \colon\;
                   \hbox{$\sum_{\n} a_{\n} \z^{\n}$ converges
                         absolutely for $\z \in \bar{D}_{\R}$}  \}
\end{subeqnarray}
Note that $S$ is a down-set
(that is, $\R \in S$ and ${\mathbf 0} \le \R' \le \R$ imply $\R' \in S$)
and that ${\mathbf 0} \in S$; we shall show that $C\subset S$.

Consider any point $\z_0 \in \bar{S} \cap C$
(here $\bar{S}$ denotes the closure of $S$).
Choose $\epsilon > 0$ such that the closed polydisc
of radius $\epsilon$ in each direction around $\z_0$
--- call it $\bar{D}(\z_0,\epsilon)$ --- is contained in $U$.
Then choose $\z_1 \in S \cap \bar{D}(\z_0,\epsilon/5)$,
and choose $\z_2\in\bar{D}(z_1,\epsilon/5)$ such that
${\mathbf 0}\le\z_2\le\z_1$ and $(\z_2)_i<(\z_1)_i$ for
all coordinates $i$ with $(\z_1)_i>0$.
It follows that $\bar{D}(\z_2,3\epsilon/5) \subset U$.
Now, since $\z_1 \in S$ and $\z_2$ is strictly below
$\z_1$ in all nonzero coordinates,
the Taylor series around $\mathbf 0$
for $f$ and all its derivatives
converge absolutely at $\z_2$.
And the Taylor series for $f$ around $\z_2$ converges absolutely in
$\bar{D}(\z_2,3\epsilon/5)$.
If $\z_3 \in \bar{D}(\z_2,3\epsilon/5)$ with $\z_3 - \z_2 \ge {\mathbf 0}$,
we can write
\begin{subeqnarray}
      f(\z_3)  & = & \sum_{\k} {f^{(\k)}(\z_2) \over \k!} (\z_3 - \z_2)^{\k}
         \slabel{eq.prings.a} \\[2mm]
               & = & \sum_{\k} (\z_3 - \z_2)^{\k}
                      \sum_{\n \ge \k} {\n \choose \k} a_{\n} \z_2^{\n-\k}
         \slabel{eq.prings.b} \\[2mm]
               & = & \sum_{\n} a_\n
                      \sum_{{\mathbf 0} \le \k \le \n} {\n \choose \k} 
(\z_3 - \z_2)^{\k}
                             \, \z_2^{\n-\k}
         \\[2mm]
               & = & \sum_{\n} a_\n \z_3^{\n}  \;,
\end{subeqnarray}
where the rearrangements are justified
because all the terms are nonnegative,
so that the convergence of the iterated sum \reff{eq.prings.b}
implies the absolute convergence of the corresponding double sum.
It follows that $\z_3 \in S$.
In particular we can choose
$\z_3 = \z_2 + (3\epsilon/5,\ldots,3\epsilon/5)$.
Since $S$ is a down-set,
and since the the real points of the polydisc $\bar{D}(\z_0,\epsilon/5)$
all lie below $\z_3$,
we conclude that $\bar{D}(\z_0,\epsilon/5) \cap [0,\infty)^n \subset S$.
Hence $\bar{D}(\z_0,\epsilon/5) \cap C \subset S \cap C$.
Since $\z_0$ is an arbitrary point of $\bar{S} \cap C$,
it follows that $S \cap C$ is both closed and open in $C$
(and nonempty since ${\mathbf 0} \in S \cap C$).
Since $C$ is connected, we conclude that $S \cap C = C$.
\qed

For the proof of Theorem~\ref{thm2.fund}, we need an elementary topological 
lemma:

\begin{lemma}
      \label{lemma.topological}
Let $K$ be a convex set in $\RR^n$,
let $C$ be an open connected subset of $K$ (in the relative topology),
let $U$ be an open neighborhood of $C$ in $\RR^n$,
and let $x_1,x_2 \in C$.
Then there exists a finite polygonal path $P \subset C$
running from $x_1$ to $x_2$
and a simply connected open set $U'$ in $\RR^n$
satisfying $P \subset U' \subset U$.
\end{lemma}

\proof
It is a well-known fact
that any two points in $C$ can be connected by a finite polygonal path
$P$ lying in $C$.
[Sketch of proof:  Define an equivalence relation $\sim$ on $C$
  by setting $x \sim y$ iff there exists a finite polygonal path in $C$
  connecting $x$ to $y$.
  Because $K$ is convex and $C$ is open in $K$,
  the equivalence classes of $\sim$ are (relatively) open subsets of $C$.
  Since $C$ is connected, there is just one equivalence class.]
By removing loops, we can assume that $P$ is non-self-intersecting.

Then we can take $U'$ to be a sufficiently small tube centered on $P$.
[Let $\delta$ be the minimal distance between non-adjacent segments of $P$,
 and let $\delta'$ be the distance from $P$ to the complement of $U$;
 we have $\delta' > 0$ by compactness of $P$.
 Then let $U'$ be the set of all points whose distance from $P$
 is less than $\min(\delta/3, \delta'/2)$.
 It is not hard to see that $U'$ is simply connected.\footnote{
     It suffices to use repeatedly the following fact:
     If $U_1$ and $U_2$ are simply connected open subsets of $\RR^n$,
     and $U_1 \cap U_2$ is nonempty and connected,
     then $U_1 \cup U_2$ is simply connected.
     For a proof of this lemma, see e.g.\ \cite[pp.~46--47]{Berenstein_91}.
     This is a special case of the Seifert--van Kampen theorem,
     which gives a recipe for computing the fundamental group
     (first homotopy group) of $U_1 \cup U_2$ in terms of
     the fundamental groups of $U_1$, $U_2$ and $U_1 \cap U_2$
     \cite[Chapter 4]{Massey_90}.
}]
\qed

We shall also make use of the following elementary result:

\begin{lemma}
      \label{lemma.inclexcl}
Let $F$ be a function on $2^X$, and define
\begin{eqnarray}
      F_-(S)  & = & \sum\limits_{X' \subseteq S}  F(X')   \\[2mm]
      F_+(S)  & = & \sum\limits_{X' \supseteq S}  F(X')
\end{eqnarray}
Then
\be
      F_-(S)  \;=\;  \sum_{Y \subseteq S^c}  (-1)^{|Y|} F_+(Y)
\ee
where $S^c \equiv X \setminus S$.
\end{lemma}

\proof
This is a straightforward application of inclusion-exclusion
(see e.g.\ \cite[Section 2.1]{Stanley_86} or \cite[Chapter IV]{Aigner_79}).
We have
\be
      F(X')  \;=\;  \sum_{Y \supseteq X'}  (-1)^{|Y \setminus X'|}  F_+(Y)
      \;.
\ee
Hence
\begin{subeqnarray}
      F_-(S)  \;=\;  \sum\limits_{X' \subseteq S}  F(X')
         & = & \sum\limits_{X' \subseteq S}
                 \sum_{Y \supseteq X'}  (-1)^{|Y \setminus 
X'|}  F_+(Y)  \\[2mm]
         & = & \sum\limits_{Y}  F_+(Y)
                 \sum\limits_{X' \subseteq Y \cap S}
                 (-1)^{|Y \setminus 
X'|}                                 \\[2mm]
         & = & \sum\limits_{Y}  F_+(Y) (-1)^{|Y \cap S^c|}
                 \sum\limits_{X' \subseteq Y \cap S}
                 (-1)^{|(Y \cap S) \setminus 
X'|}                        \\[2mm]
         & = & \sum\limits_{Y}  F_+(Y) (-1)^{|Y \cap S^c|}
                 \delta_{Y \cap 
S,\emptyset}                             \\[2mm]
         & = & \sum_{Y \subseteq S^c}  (-1)^{|Y|} F_+(Y)   \;.
\end{subeqnarray}
\qed

\proofof{Theorem~\ref{thm2.fund}}
(c)$\implies$(b)$\implies$(a) is trivial.

(e)$\implies$(d) is trivial, while (d)$\implies$(e)
follows from the alternating-sign property \reff{eq.sign_of_cn}.

(e) implies that the sum of the Taylor series for $\log Z_W(\w)$
defines an analytic function on the open polydisc $D_{\R}$
and a continuous function on the closed polydisc $\bar{D}_{\R}$.
Its exponential equals $Z_W(\w)$ on $D_{\R}$
and hence by continuity also on $\bar{D}_{\R}$.
Therefore (e)$\implies$(c).

Finally, assume (a).  Since $Z_W$ is continuous on ${\mathbb C}^X$ (and
has real coefficients),
we can find an open connected neighborhood $C'$ of $C$ in $(-\infty,0]^X$
on which $Z_W > 0$,
and an open neighborhood $U$ of $C'$ in $\C^X \simeq \RR^{2|X|}$
on which $Z_W \neq 0$.
Applying Lemma~\ref{lemma.topological} (with $n = 2|X|$), we can find a
finite polygonal path $P \subset C'$ from 0 to $-\R$
and a simply connected open set $U'$ in $\C^X$
satisfying $P \subset U' \subset U$.
Then $\log Z_W$ is a well-defined single-valued analytic function on $U'$,
once we specify $\log Z_W({\mathbf 0}) = 0$.
Applying Proposition~\ref{prop.pringsheim} to $\log Z_W$ on $P$ and $U'$
[using the alternating-sign property \reff{eq.sign_of_cn}],
we conclude that the Taylor series for $\log Z_W$ around $\mathbf 0$
is absolutely convergent on $\bar{D}_{\R}$.
Therefore (a)$\implies$(e).

The bound $|Z_W(\w)| \ge Z_W(-\R)$ for $|\w| \le \R$,
which is equivalent to $\Re \log Z_W(\w) \ge \log Z_W(-\R)$,
is an immediate consequence of
the alternating-sign property \reff{eq.sign_of_cn}.

Now consider the special case of a hard-core self-repulsion.
(b)$\implies$(b\textprime) is trivial,
and (b\textprime)$\implies$(b) follows from the fact that
$Z_W$ is multiaffine (i.e.\ of degree $\le 1$ in each $w_x$ separately)
because the value of $Z_W$ at any point $\w$ in the rectangle
$-\R \le \w \le {\mathbf 0}$
is a convex combination of the values at the vertices.

To show that (b)$\implies$(f), note that
\be
      Z_W(\w;S)  \;=\;  \left( \prod\limits_{x \in S} w_x \right)
                        \left( \prod\limits_{\{x,y\} \subseteq S} W(x,y) 
\right)
                        Z_W(W(S,\cdot) \w)
\ee
where we have defined
\be
      [W(S,\cdot) \w]_y  \;=\;
      \left( \prod\limits_{x \in S} W(x,y) \right)  w_y
\ee
(note in particular that this vanishes whenever $y \in S$).
Hence
\be
      (-1)^{|S|} Z_W(-\R;S)  \;=\;
      \left( \prod\limits_{x \in S} R_x \right)
      \left( \prod\limits_{\{x,y\} \subseteq S} W(x,y) \right)
      Z_W(-W(S,\cdot) \R)
      \;\ge\; 0
\ee
since $-\R \le -W(S,\cdot) \R \le 0$,
with strict inequality when $|S| = 0$ or 1
[since the product over $W(x,y)$ is in that case empty].

To show that (f)$\implies$(b\textprime),
use Lemma~\ref{lemma.inclexcl} applied to the set function
\be
      F(S)  \;=\;   \left( \prod\limits_{x \in S} -R_x \right)
                    \left( \prod\limits_{\{x,y\} \subseteq S}  W(x,y) \right)
      \;.
\ee
We have
\begin{eqnarray}
      F_-(S)  & = & Z_W(-\R \, {\bf 1}_S)   \\[2mm]
      F_+(S)  & = & Z_W(-\R;S)
\end{eqnarray}
so that Lemma~\ref{lemma.inclexcl} asserts the identity
\be
      Z_W(-\R \, {\bf 1}_S)  \;=\;
      \sum_{Y \subseteq S^c}  (-1)^{|Y|} Z_W(-\R;Y)
      \;.
\ee
By (f), the $Y=\emptyset$ term is $>0$ and the other terms are $\ge 0$,
so $Z_W(-\R \, {\bf 1}_S) > 0$ for all $S$.

Finally, let us show that (f)$\iff$(g).
By inclusion-exclusion, there are unique numbers $P(T)$
satisfying \reff{def.PT}, namely $P(T) = (-1)^{|T|} Z_W(-\R;T)$.
Moreover, taking $S=\emptyset$ in \reff{def.PT} we see that
$\sum_T P(T) = 1$.
Therefore, $P$ is a probability measure if and only if
$(-1)^{|T|} Z_W(-\R;T) \ge 0$ for all $T$;
and $P(\emptyset) > 0$ if and only if $Z_W(-\R;\emptyset) = Z_W(-\R) > 0$.
\qed

\subsection{Properties of the set $\scrr(W)$}  \label{sec2.scrr}

The following definition plays a central role in the
remainder of this paper:

\begin{definition}[definition of $\scrr(W)$]
  \label{def.scrr}
We define $\scrr(W)$ to be the set of all vectors $\R \ge 0$
satisfying the equivalent conditions (a)--(e) of Theorem~\ref{thm2.fund}.
When $W$ is the hard-core pair interaction for a graph $G$,
\be
      W(x,y)  \;=\;  \cases{ 0   & if $x=y$ or $xy \in E(G)$  \cr
                             1   & if $x \neq y$ and $xy \notin E(G)$ \cr
                           }
\ee
we also write $\scrr(G)$.
\end{definition}

\begin{proposition}[elementary properties of $\scrr(W)$]
     \label{prop.scrrW.properties}
For any repulsive lattice gas, the set $\scrr(W)$ is
\begin{itemize}
      \item[(a)] open in $[0,\infty)^X$
      \item[(b)] a down-set
         [i.e.\ $\R \in \scrr(W)$ and $0 \le \R' \le \R$
          imply $\R' \in \scrr(W)$]
      \item[(c)] logarithmically convex
         [i.e.\ $\R,\R' \in \scrr(W)$ and $0 \le \lambda \le 1$
           imply $\R^\lambda (\R')^{1-\lambda} \in \scrr(W)$].
\end{itemize}
Moreover, the set $\scrr(W)$ is an increasing function of each $W(x,y)$
on $0 \le W \le 1$.
\end{proposition}

\proof
It is obvious from Theorem~\ref{thm2.fund}(b) or (c)
that $\scrr(W)$ is open in $[0,\infty)^X$  and is a down-set.

Given any formal power series $f(\z) = \sum_{\n} a_{\n} \z^{\n}$
with complex coefficients,
let us define its set of radii of absolute convergence by
\be
      {\rm cvg}(f)  \;=\;
      \{\R \ge 0 \colon\;  \sum_{\n} |a_{\n}| \R^{\n} < \infty \}
      \;.
\ee
It then follows immediately from H\"older's inequality for infinite series
that ${\rm cvg}(f)$ is logarithmically convex.\footnote{
      See \cite[Chapters B and G]{Gunning_90},
      \cite[Sections 2.3 and 3.4.3]{Krantz_92},
      \cite[Section II.3.8]{Range_86} and \cite[pp.~116--122]{Vladimirov_66}
      for further discussion of related questions
      in the theory of analytic functions
      of several complex variables.
}
By Theorem~\ref{thm2.fund}(e) we have $\scrr(W) = {\rm cvg}(\log Z_W)$,
so $\scrr(W)$ is logarithmically convex.

Finally, from \reff{eq.sign_of_cn}/\reff{eq.first_deriv_of_cn}
we see that $|c_{\n}(W)|$ is a decreasing function of
each $W(x,y)$ on $0 \le W \le 1$,
so $\scrr(W) = {\rm cvg}(\log Z_W)$
is an increasing function of each $W(x,y)$ on $0 \le W \le 1$.
\qed

See Corollary~\ref{cor.strict_monotonicity} below
for a strict monotonicity that strengthens
Proposition~\ref{prop.scrrW.properties}(b);
and see Corollary~\ref{cor.2Dconvexity} for an additional convexity property
in the special case of a hard-core self-repulsion.

\medskip
\par\noindent
{\bf Remark.}
If $\Lambda$ is any nonempty subset of $X$,
we can obviously define a lattice gas on $\Lambda$
with interaction $W \restrict \Lambda$,
and there will be a corresponding set
$\scrr(W \restrict \Lambda) \subseteq [0,\infty)^\Lambda$.
Now, for any $\w \in \C^\Lambda$, we trivially have
$Z_{W \restrict \Lambda}(\w) = Z_W(\w,{\bf 0})$
where ${\bf 0} \in \C^{X \setminus \Lambda}$;
hence $\R \in \scrr(W \restrict \Lambda)$
if and only if $(\R,{\bf 0}) \in \scrr(W)$.
So the sets $\scrr(W \restrict \Lambda)$
are the sections through ${\bf 0}$ of $\scrr(W)$.

\bigskip

Next let us show that the sets $\scrr(W)$ are bounded
in a suitable sense.
In the case of hard-core self-repulsion this is immediate,
because $\scrr(W) \subseteq [0,1)^X$.
[This containment is trivial if $|X|=1$, and otherwise follows from
  Proposition~\ref{prop.scrrW.properties}(b).
  Alternatively, it is trivial if $W(x,y) = 1$ for all $x \neq y$,
  and otherwise follows from the monotonicity statement
  in the last sentence of Proposition~\ref{prop.scrrW.properties}.]
However, in the general case the set $\scrr(W)$ can be unbounded,
as is shown by the following two examples:
\begin{itemize}
    \item[(a)]  Let $X = \{x\}$ and $W(x,x) = 1$;
       then $Z_W(w) = e^w$ and $\scrr(W) = [0,\infty)$.
    \item[(b)]  Let $X = \{x,y\}$, $W(x,x) = W(y,y) = 1$ and $W(x,y) = 0$;
       then $Z_W(w_x,w_y) = e^{w_x} + e^{w_y} - 1$
       and $\scrr(W) = \{(R_x,R_y) \colon\, R_x,R_y \ge 0 \hbox{ and }
            e^{-R_x} + e^{-R_y} > 1\}$.
       This set is unbounded, since $R_x$ can go to $\infty$
       as $R_y \to 0$ (and vice versa).
       [When $R_y = 0$ this reduces to example (a).]
\end{itemize}
We now claim that these are essentially the only ways
in which $\scrr(W)$ can be unbounded:

\begin{proposition}[boundedness of $\scrr(W)$]
   \label{prop.scrr.bounded}
Consider any repulsive lattice gas, and let $x \in X$.
\begin{itemize}
    \item[(a)]  If $W(x,x) < 1$, then there exists $C_x < \infty$ such that
       $\scrr(W) \subseteq [0,C_x) \times [0,\infty)^{X \setminus x}$.
       [In particular, if $W(x,x) < 1$ for all $x \in X$,
       then $\scrr(W)$ is bounded.]
    \item[(b)]  Suppose that $W(x,x) = 1$,
       and let $\a \in [0,\infty)^{X \setminus x}$.
       If there exists at least one $y \in X$
       such that $W(x,y) < 1$ and $a_y > 0$,
       then the section $\{R_x \colon\, (R_x,\a) \in \scrr(W)\}$
       is bounded.  Moreover, the converse is true provided that
       $\a \in \scrr(W \restrict (X \setminus x))$.
\end{itemize}
\end{proposition}

\medskip
\par\noindent
{\bf Remark.}
Since $\scrr(W)$ is a down-set,
this proposition can alternatively be formulated as follows:
Let $\a,\b \in [0,\infty)^X$;
then the set $\{ \lambda\ge 0 \colon\, \a + \lambda\b \in \scrr(W)\}$
is bounded if
there exists $x \in \supp\b$ and $y \in \supp\a \cup \supp\b$
such that $W(x,y) < 1$;
and the converse is true provided that $\a \in \scrr(W)$.

\medskip

\proof
(a) Using the fact that $\scrr(W)$ is a down-set,
it suffices to prove the claim for the single-site partition function
\be
    Z_W(w)  \;=\;  \sum_{n=0}^\infty  {W^{n(n-1)/2} \over n!} \, w^n
\ee
where $0 \le W < 1$.
If $W=0$, we have $Z_W(w) = 1 +w$, so that $\scrr(W) = [0,1)$.
If $0 < W < 1$, then $Z_W$ is a nonpolynomial entire function
of order 0 \cite[Theorem I.2]{Levin_64};
and by the Hadamard factorization theorem \cite[Theorem I.13]{Levin_64},
any such function must have infinitely many zeros.
In particular, $\scrr(W) = [0,\alpha)$
where $\alpha$ is the smallest absolute value of a zero of $Z_W$.\footnote{
   It turns out (though we do not need this fact here)
   that all the zeros of $Z_W$ are negative real numbers
   \cite{Morris_72,Iserles_93}.
}

(b)  If there does not exist $y \in X$ such that $W(x,y) < 1$ and $a_y > 0$,
then $Z_W(\w) = e^{w_x} Z_{W,X \setminus x}(\w_{\neq x})$
whenever $\supp\w \subseteq \supp\a \cup \{x\}$.
It follows that the section $\{R_x \colon\, (R_x,\a) \in \scrr(W)\}$
is either empty [in case $\a \notin \scrr(W \restrict (X \setminus x))$]
or all of $[0,\infty)\,$
[in case $\a \in \scrr(W \restrict (X \setminus x))$].

Now suppose that there does exist such a $y$.
Using the fact that $\scrr(W)$ is a down-set,
it suffices to prove the claim for the two-site partition function
with $\Lambda = \{x,y\}$.
Moreover, by the monotonicity statement in the last sentence of
Proposition~\ref{prop.scrrW.properties},
we may assume that $W(y,y) = 1$.
Writing $W \equiv W(x,y) \in [0,1)$, we need to treat
\begin{subeqnarray}
    Z_W(w_x,w_y)  & = &
    \sum_{n_x,n_y=0}^\infty  {W^{n_x n_y} \over n_x! n_y!} \,
            w_x^{n_x} w_y^{n_y}
      \\[2mm]
    & = &
    \sum_{n_x=0}^\infty  {w_x^{n_x} \, \exp(w_y W^{n_x})  \over n_x!}
\end{subeqnarray}
If $w_y \neq 0$, this is an entire function of order 1 in $w_x$
that is not of the form $e^{\alpha w_x} P(w_x)$
for any $\alpha \in \C$ and polynomial $P$.\footnote{
    {\sc Proof.}  If $P(z) = \sum_{k=0}^K a_k z^k$, then
    \begin{eqnarray*}
       e^{\alpha z} P(z)  & = &
       \sum\limits_{n=0}^\infty {(\alpha z)^n \over n!} \,
         \left[ a_0 + n {a_1 \over \alpha} + n(n-1) {a_2 \over \alpha^2}
                +\, \ldots\, + n(n-1)\cdots(n-K+1) {a_K \over \alpha^K}
         \right]
       \\[1mm]
     & \equiv & \sum\limits_{n=0}^\infty {\alpha^n Q(n) \over n!} \, z^n
    \end{eqnarray*}
    where $Q$ is a polynomial.
    But $F(n) = \exp(w_y W^{n})$ is not of the form $\alpha^n Q(n)$
    for any $\alpha \in \C$ and polynomial $Q$.

    {\sc Alternate proof.}
    We have
    $$  Z_W(w_x,w_y) \,-\, e^{w_x} \;=\;
        \sum_{n_x=0}^\infty  {w_x^{n_x} \over n_x!} \, [\exp(w_y W^{n_x}) - 1]
    $$
    Since $w_y$ is fixed and $0 \le W \le 1$,
    we have $|\exp(w_y W^{n_x}) - 1| \le C W^{n_x}$
    for some constant $C < \infty$, from which it follows that
    $$  |Z_W(w_x,w_y) \,-\, e^{w_x}|  \;\le\; C e^{W |w_x|} \;. $$
    If $Z_W(w_x,w_y)$ were equal to  $e^{\alpha w_x} P(w_x)$,
    then by taking $w_x \to +\infty$ we conclude that
    we would have to have $\alpha = 1$ and $P \equiv 1$
    (since $W < 1$);  but $Z_W$ is not in fact of this form
    when $w_y \neq 0$.
}
It again follows from the Hadamard factorization theorem
that $Z_W(\,\cdot\,, w_y)$ has infinitely many zeros.
Choosing any $w_y \in \C$ with $|w_y| = a_y$
(which by hypothesis is nonzero),
we conclude from Theorem~\ref{thm2.fund}(c) that
the section $\{R_x \colon\, (R_x,a_y) \in \scrr(W)\}$ is bounded.
\qed

Let $\overline{\scrr(W)}$ be the closure of $\scrr(W)$,
and let $\partial\scrr(W) = \overline{\scrr(W)} \setminus \scrr(W)$.
[Note that this is the boundary of $\scrr(W)$
  in $[0,\infty)^X$, not in $\RR^X$.]
Since $\scrr(W)$ is a down-set,
it is obvious that $\overline{\scrr(W)}$ is a down-set as well.

\begin{proposition}[properties of closure and boundary of $\scrr(W)$]
  \label{prop.closure_scrr}
For any repulsive lattice gas:
\begin{itemize}
    \item[(a)] If $-\w \in \overline{\scrr(W)}$, then
\be
      {\partial^n Z_W(\w)
       \over
       \partial w_{x_1} \cdots \partial w_{x_n}
      }
      \;\ge\;  0
  \label{eq.prop.closure_scrr}
\ee
for all $n \ge 0$ and all $x_1,\ldots,x_n \in X$.
    \item[(b)] If $-\w \in \partial\scrr(W)$, then $Z_W(\w) = 0$.
\end{itemize}
\end{proposition}

\proof
(a)  We have $Z_W(\w) > 0$ for $-\w \in \scrr(W)$,
so by continuity $Z_W(\w) \ge 0$ for $-\w \in  \overline{\scrr(W)}$.
The general claim \reff{eq.prop.closure_scrr}
then follows from the multiple differentiation identity
\reff{eq.differentiation.multiple},
using the facts that $0 \le W \le 1$
and that $\overline{\scrr(W)}$ is a down-set.

(b) If $Z_W(\w)$ were strictly positive,
then $-\scrr(W) \cup \{\w\}$ would be a connected subset of $(-\infty,0]^X$
on which $Z_W$ is strictly positive,
so by Theorem~\ref{thm2.fund}(a) we would have $-\w \in \scrr(W)$.
\qed

Under an appropriate connectivity condition,
we can prove that the inequality \reff{eq.prop.closure_scrr}
is strict for $n=1$:

\begin{proposition}
  \label{prop.strict_monotonicity}
Consider any repulsive lattice gas.
Let $\R \in \overline{\scrr(W)}$,
and suppose that the induced subgraph $G_W[\supp \R]$ is connected.
Then
\be
    \left. {\partial Z_W(\w) \over \partial w_x} \right| _{\w = -\R}
    \;>\; 0
    \qquad \hbox{for all } x \in \supp \R \;.
  \label{eq.prop.strict_monotonicity}
\ee
\end{proposition}

\noindent
Here the connectedness hypothesis is essential,
as is shown by taking $G$ to be the edgeless graph $\bar{K}_n$,
for which $Z_G(\w) = \prod_{i=1}^n (1+w_i)$,
and taking $\R = (1,\ldots,1)$.
It is also essential to consider the induced subgraph $G_W[\supp \R]$,
since a site can be effectively eliminated by setting its fugacity to zero:
for example, when $G$ is the path 123,
we have $Z_G(\w) = (1+w_1)(1+w_3) + w_2$,
and a counterexample to \reff{eq.prop.strict_monotonicity}
is obtained by taking $R_2 = 0$ and $R_1 = R_3 = 1$.

Before proving Proposition~\ref{prop.strict_monotonicity},
let us state and prove some corollaries.

\begin{corollary}[strict monotonicity]
    \label{cor.strict_monotonicity}
Consider any repulsive lattice gas.
Let $\R \in \overline{\scrr(W)}$,
and suppose that the induced subgraph $G_W[\supp \R]$ is connected.
If ${\bf 0} \le \R' \le \R$ with $R'_x < R_x$ for at least one $x$,
then $\R' \in \scrr(W)$.
\end{corollary}

\proof
Since $Z_W(-\R) \ge 0$ and
$\left. {\partial Z_W(\w) / \partial w_x} \right| _{\w = -\R} > 0$
by Propositions~\ref{prop.closure_scrr}(a) and \ref{prop.strict_monotonicity},
we have $Z_W(-(\R - \epsilon \bdelta_x)) > 0$
for all sufficiently small $\epsilon > 0$
[here $\bdelta_x$ is the vector with $x$th coordinate 1
  and all other coordinates 0].
Since $\overline{\scrr(W)}$ is a down-set,
we have $\R - \epsilon \bdelta_x \in \overline{\scrr(W)}$;
but by Proposition~\ref{prop.closure_scrr}(b),
$\R - \epsilon \bdelta_x$ cannot lie in $\partial \scrr(W)$,
so we must have $\R - \epsilon \bdelta_x \in \scrr(W)$.
Now pick $\epsilon$ small enough so that
$\R' \le \R - \epsilon \bdelta_x$
and use the fact that $\scrr(W)$ is a down-set.
\qed

\begin{corollary}
    \label{cor.closure_of_scrr}
Consider any repulsive lattice gas.
\begin{itemize}
    \item[(a)]  If $\R \in \partial\scrr(W)$,
       then every neighborhood of $-\R$ in $(-\infty,0]^X$
       contains points $\w \neq -\R$ with $\supp \w = \supp \R$
       where $Z_W > 0$, where $Z_W = 0$ and where $Z_W < 0$.
    \item[(b)]  $\R \in \overline{\scrr(W)}$ if and only if
       $Z_W(\w) \ge 0$ for all $\w$ satisfying $-\R \le \w \le {\bf 0}$.
\end{itemize}
\end{corollary}

\proof
(a)  Let $X_1,\ldots,X_k$ be the vertex sets of the components
of $G_W[\supp\R]$.  For any $\w$ with $\supp \w \subseteq \supp \R$,
we have $Z_W(\w) = \prod_{i=1}^k Z_W(\w \, {\bf 1}_{X_i})$.
Pick a vertex $x_i$ in each $X_i$,
and let $\w_\pm = -\R \pm \epsilon \bdelta_{x_1} +
                           \epsilon \sum_{i=2}^k \bdelta_{x_i}$.
By Proposition~\ref{prop.strict_monotonicity},
for all sufficiently small $\epsilon > 0$ we have
$\pm Z_W(\w \, {\bf 1}_{X_1}) > 0$ and
$Z_W(\w \, {\bf 1}_{X_i}) > 0$ for $2 \le i \le k$.
This proves the existence of points near $-\R$
with $Z_W > 0$ and $Z_W < 0$.
The existence of points $\w \neq -\R$ with $Z_W=0$
then follows by the intermediate value theorem.

(b) ``Only if'' is obvious from Proposition~\ref{prop.closure_scrr}(a)
together with the fact that $\overline{\scrr(W)}$ is a down-set.
To prove ``if'', suppose that $Z_W(\w) \ge 0$
for all $\w$ satisfying $-\R \le \w \le {\bf 0}$.
Consider the line segment $\{\lambda\R\}_{0 \le \lambda \le 1}$,
and let
$\lambda_{\rm max} =
  \sup \{\lambda \in [0,1] \colon\, \lambda\R \in \scrr(W)\}$.
If $\lambda_{\rm max} = 1$, then $\R \in \overline{\scrr(W)}$.
If $\lambda_{\rm max} < 1$, then
$\lambda_{\rm max}\R \in \partial \scrr(W)$,
so by part (a) we can choose $\w$ arbitrarily close to $-\lambda_{\rm max}\R$
with $\supp \w = \supp \R$
(and in particular satisfying $-\R \le \w \le {\bf 0}$)
such that $Z_W(\w) < 0$ --- a contradiction.
\qed

\medskip
\par\noindent
{\bf Remark.}
Corollary~\ref{cor.closure_of_scrr}(b) is an analogue of
Theorem~\ref{thm2.fund}(b) for the closure of $\scrr(W)$.
It is worth noting that the analogue of Theorem~\ref{thm2.fund}(a)
is {\em false}\/:  for example, for $G=K_2$ we have
$Z_G(\w) = (1+w_1)(1+w_2)$, so that the set
$\{ \R \ge {\bf 0} \colon\, Z_G(-\R) \ge 0\}$
is connected and equals $[0,1]^2 \cup [1,\infty)^2$,
while $\overline{\scrr(W)} = [0,1]^2$ only.

\bigskip

\proofof{Proposition~\ref{prop.strict_monotonicity}}
If $\R \in \scrr(W)$, then $W(x,\cdot) \R \in \scrr(W)$ as well
[since $0 \le W \le 1$ and $\scrr(W)$ is a down-set],
so \reff{eq.prop.strict_monotonicity} follows from the
differentiation identity \reff{eq.differentiation}.
So we can assume henceforth that $\R \in \partial\scrr(W)$.
By deleting sites where $R_x = 0$, we can also assume
without loss of generality that $\supp\R = X$.
Let us define
\begin{eqnarray}\label{2.58}
    S  & = & \{ x \in X \colon\;
                \left. {\partial Z_W(\w) \over \partial w_x} \right| _{\w = 
-\R}
                \,=\, 0 \}   \\[2mm]
	\label{2.59}
    S' & = & \{ x \in X \colon\;
                Z_W(w_x,-\R_{\neq x}) = 0 \hbox{ for all } w_x \in \C \}
\end{eqnarray}
Clearly $S' \subseteq S \subseteq X$.  The Proposition will then be
an immediate consequence of Lemmas \ref{lemma.strict_monotonicity.1}
and \ref{lemma.strict_monotonicity.2} below.
\qed

\begin{lemma}
    \label{lemma.strict_monotonicity.1}
Let $\R \in \partial\scrr(W)$, and let $S$, $S'$
be defined as in \reff{2.58}/\reff{2.59}.
If $\supp\R = X$ and $G_W$ is connected,
then either $S' = S = \emptyset$ or else $S' = S = X$.
\end{lemma}

\begin{lemma}
    \label{lemma.strict_monotonicity.2}
Suppose $\R \in \overline{\scrr(W)}$ and
for all $x \in X$ and all $w_x \in \C$, we have
$Z_W(w_x,-\R_{\neq x}) = 0$.
Then $G_W[\supp\R]$ is disconnected.
\end{lemma}

\proofof{Lemma~\ref{lemma.strict_monotonicity.1}}
Suppose $S \neq \emptyset$, and consider any $x \in S$.
Then $Z_W(-\R) = 0$ and
$Z_W(-W(x,\cdot)\R) =
  \left. {\partial Z_W(\w) / \partial w_x} \right| _{\w = -\R} = 0$
[using the differentiation identity \reff{eq.differentiation}].
By monotonicity of $Z_W$ in $\overline{\scrr(W)}$,
we have $Z_W(-\R') = 0$ whenever $-\R \le -\R' \le -W(x,\cdot)\R$.
In particular, if $y$ is adjacent to $x$ in $G_W$ [i.e.\ $W(x,y) < 1$],
then $Z_W(w_y, -\R_{\neq y}) = 0$ for all $w_y \in [-R_y, -W(x,y)R_y]$.
By analyticity it follows that
$Z_W(w_y, -\R_{\neq y}) = 0$ for all $w_y \in \C$,
i.e.\ $y \in S' \subseteq S$.
Since $G_W$ is connected, it easily follows that $S' = S = X$.
\qed

\proofof{Lemma~\ref{lemma.strict_monotonicity.2}}
As before, we may assume without loss of generality that $\supp\R = X$.
The proof is by induction on $|X|$.
If $|X|=1$, the hypothesis is impossible, since $Z_W(0) = 1 \neq 0$.
So let $|X| > 1$ and suppose that $G_W$ is connected.
Choose a vertex $x \in X$ such that $G_W \setminus x$ is connected
[e.g.\ let $x$ be any endvertex of a spanning tree in $G_W$].
By the fundamental identity \reff{eq.fund_general}, we have
\begin{eqnarray}
    Z_W(w_x,-\R_{\neq x})  & = &
      Z_W(0,-\R_{\neq x})  \,+\, w_x Z_W(0, -W(x,\cdot) \R_{\neq x})
         \nonumber \\
    & & \qquad +\; \hbox{terms of order $w_x^2$ and higher}
  \label{eq.lemma.strict_monotonicity.2}
\end{eqnarray}
Since $Z_W(\cdot,-\R_{\neq x})$ is identically zero,
all coefficients in the Taylor expansion \reff{eq.lemma.strict_monotonicity.2}
must vanish;  in particular we must have
\begin{subeqnarray}
    Z_W(0,-\R_{\neq x})  & = &   0   \\
    Z_W(0,-W(x,\cdot) \R_{\neq x})  & = &   0
\end{subeqnarray}
Pick $y \neq x$ with $W(x,y) < 1$
[which is possible since $x$ is not an isolated vertex of $G_W$].
Since $R_y>0$ and $Z_W(0,w_y,-\R_{\neq x,y})$ is monotonic in $w_y$ for
$w_y\in[-R_y,0]$, it must be vanishing on the nontrivial
interval $[-R_y,-W(x,y)R_y]$.
By analyticity of $Z_W$, it follows that
$Z_W(0,w_y,-\R_{\neq x,y}) = 0$ for all $w_y \in \C$,
so in particular
\begin{subeqnarray}
    Z_W(0, -R_y, -\R_{\neq x,y})   & = &  0
      \slabel{eq.xxx.a} \\[2mm]
    \left. {\partial Z_W(0,w_y, -\R_{\neq x,y}) \over \partial w_y}
       \right| _{w_y = -R_y}       & = &  0
\end{subeqnarray}
Notice that $\R \equiv (R_x, \R_{\neq x}) \in \overline{\scrr(W)}$
implies $(0,\R_{\neq x}) \in \overline{\scrr(W)}$
[since $\overline{\scrr(W)}$ is a down-set],
or in other words
$\R_{\neq x} \in \overline{\scrr(W \restrict (X \setminus x))}$.
Moreover, by \reff{eq.xxx.a} $\R_{\neq x}$ must lie in
$\partial \scrr(W \restrict (X \setminus x))$.
So we can apply Lemma~\ref{lemma.strict_monotonicity.1}
to $X \setminus x$ to conclude that $S' = X \setminus x$;
therefore, by the inductive hypothesis,
$G_W \setminus x$ is disconnected,
which contradicts the choice of $x$.
\qed

\medskip
\par\noindent
{\bf Remark.}
One might think that Lemma~\ref{lemma.strict_monotonicity.2}
would hold not only for $\R \in \overline{\scrr(W)}$
but for arbitrary $\w^{(0)} \in \C^X$.
However, this is false:
Consider the star $K_{1,4}$ with center $x$ and endvertices $y_1,\ldots,y_4$
with interactions $W \in (0,1)$ along the edges
[and $W(x,x) = W(y_i,y_i) = 0$].
Then
\be
    Z_W(w_x,w_{y_1},\ldots,w_{y_4})  \;=\;
       \prod\limits_{i=1}^4 (1 + w_{y_i})  \,+\,
       w_x \prod\limits_{i=1}^4 (1 + W w_{y_i})
    \;.
\ee
Setting $w^{(0)}_{y_1} = w^{(0)}_{y_2} = -1$ and
$w^{(0)}_{y_3} = w^{(0)}_{y_4} = -1/W$
[here $w^{(0)}_x$ is arbitrary],
we see that $Z_W(w_z,\w^{(0)}_{\neq z}) = 0$
for every vertex $z$ and every $w_z \in \C$.

\subsection{Further consequences of the alternating-sign property}
    \label{sec2.alternating}

Let us now exploit systematically the
alternating-sign property \reff{eq.sign_of_cn}
for the Taylor coefficients of $\log Z_W$.
The general context is the following:

\begin{definition}[absolute monotonicity]
      \label{def.absolutely_monotone}
Let $U \subset \C^n$ be a union of open polydiscs centered at $\mathbf 0$
(a ``complete Reinhardt domain'').
We say that $f$ is {\em absolutely monotone in $U$}\/
if it is analytic in $U$ and
all the Taylor coefficients of $f$ at $\mathbf 0$ are nonnegative.
\end{definition}

\noindent
We will use the following essentially trivial result:

\begin{lemma}[elementary consequences of absolute monotonicity]
      \label{lemma.absolutely_monotone}
Let $f$ be absolutely monotone in a union $U$ of open polydiscs
centered at ${\mathbf 0}\in\C^n$.
Then:
\begin{itemize}
      \item[(a)]  All the derivatives $D^{\smallbalpha} f$
(where $\balpha$ is a multi-index) are absolutely monotone in $U$.
      \item[(b)]  For every $\blambda \in [0,1]^n$, the function
\be
      f_{\smallblambda}(\z)  \;=\;  f(\z) - f(\blambda\z)
\ee
is absolutely monotone in $U$.
[Here $\blambda\z$ is the pointwise product,
  so that $(\blambda \z)_x = \lambda_x z_x$.]
      \item[(c)]  For all multi-indices $\balpha$, we have
\be
      |(D^{\smallbalpha} f)(\z)|  \;\le\;  (D^{\smallbalpha} f)(\R)
\ee
whenever $|\z| \le \R \in U$.  In particular,
$(D^{\smallbalpha} f)(\R) \ge 0$ whenever $0 \le \R \in U$.
\end{itemize}
\end{lemma}

Now, the alternating-sign property \reff{eq.sign_of_cn}
is precisely the statement that
\be
      f(\z)  \;=\;  -\log Z_W(-\z)
\ee
is absolutely monotone in the domain
$D_{\scrr(W)} = \!\bigcup\limits_{\R \in \scrr(W)} \! D_{\R}
                 = \!\bigcup\limits_{\R \in \scrr(W)} \! \bar{D}_{\R}$.
Applying Lemma~\ref{lemma.absolutely_monotone}(c) to $f$, we obtain:

\begin{proposition}
     \label{prop.derivs_of_Z}
For any repulsive lattice gas:
\begin{itemize}
      \item[(a)] When $-\w \in \scrr(W)$, we have
\be
      (-1)^{n-1} \, {\partial^n \log Z_W(\w)
                     \over
                     \partial w_{x_1} \cdots \partial w_{x_n}
                    }
      \;\ge\;  0
    \label{eq.derivs_of_Z}
\ee
for all $n \ge 0$ and all $x_1,\ldots,x_n \in X$.
      \item[(b)]  When $|\w| \le \R \in \scrr(W)$, we have
\be
      \left|
                    {\partial^n \log Z_W(\w)
                     \over
                     \partial w_{x_1} \cdots \partial w_{x_n}
                    }
      \right|
      \;\le\;
      \left.
      (-1)^{n-1} \, {\partial^n \log Z_W(\w')
                     \over
                     \partial w'_{x_1} \cdots \partial w'_{x_n}
                    }
      \right| _{\w' = -\R}
    \label{eq.derivs_of_Z.bound}
\ee
for all $n \ge 0$ and all $x_1,\ldots,x_n \in X$.
\end{itemize}
\end{proposition}

\proof
This is an immediate consequence of the
alternating-sign property \reff{eq.sign_of_cn}
together with
Lemma~\ref{lemma.absolutely_monotone}(c) and Theorem~\ref{thm2.fund}(d).
\qed

Let us now define
\be
      Y_W(\w;\blambda)  \;=\;  {Z_W(\blambda\w) \over Z_W(\w)}
      \;.
    \label{def.YW}
\ee
By Lemma~\ref{lemma.absolutely_monotone}(b), the function
\be
      f_{\smallblambda}(\z)  \;=\;  \log Y_W(-\z;\blambda)  \;=\;
           \log {Z_W(-\blambda\z) \over Z_W(-\z)}
\ee
is absolutely monotone on $D_{\scrr(W)}$ whenever $\blambda \in [0,1]^X$.
Applying Lemma~\ref{lemma.absolutely_monotone}(c) to $f_{\smallblambda}$,
we obtain:

\begin{proposition}
     \label{prop.derivs_of_Y}
For any repulsive lattice gas and any $\blambda \in [0,1]^X$:
\begin{itemize}
      \item[(a)] When $-\w \in \scrr(W)$, we have
\be
      (-1)^{n} \,   {\partial^n \log Y_W(\w;\blambda)
                     \over
                     \partial w_{x_1} \cdots \partial w_{x_n}
                    }
      \;\ge\;  0
    \label{eq.derivs_of_Y}
\ee
for all $n \ge 0$ and all $x_1,\ldots,x_n \in X$.
      \item[(b)]  When $|\w| \le \R \in \scrr(W)$, we have
\be
      \left|
                    {\partial^n \log Y_W(\w;\blambda)
                     \over
                     \partial w_{x_1} \cdots \partial w_{x_n}
                    }
      \right|
      \;\le\;
      \left.
      (-1)^{n} \,   {\partial^n \log Y_W(\w';\blambda)
                     \over
                     \partial w'_{x_1} \cdots \partial w'_{x_n}
                    }
      \right| _{\w' = -\R}
    \label{eq.derivs_of_Y.bound}
\ee
for all $n \ge 0$ and all $x_1,\ldots,x_n \in X$.
In particular, setting $n=0$ we have
\be
      \left| {Z_W(\blambda\w) \over Z_W(\w)} \right|   \;\le\;
      {Z_W(-\blambda\R) \over Z_W(-\R)}  \;<\;  \infty
\ee
for all $\w \in \bar{D}_{\R}$.
That is, the quantity $|Z_W(\blambda\w) / Z_W(\w)|$
takes its maximum on the polydisc $\bar{D}_{\R}$
at the point $\w = -\R$,
and this maximum value $Z_W(-\blambda\R) / Z_W(-\R)$
is therefore an increasing function of $\R \in \scrr(W)$.
\end{itemize}
\end{proposition}

Let us define the function
$\scrz_W \colon\, [0,\infty)^X \to [0,\infty)$ by
\begin{subeqnarray}
      \scrz_W(\R)
      & = & \inf\limits_{-\R \le \w \le 0} \max[Z_W(\w), 0]
         \slabel{def.scrzW.a}  \\[3mm]
      & = & \inf\limits_{\w \in \bar{D}_{\R}}  |Z_W(\w)|
         \slabel{def.scrzW.b}  \\[3mm]
      & = & \cases{Z_W(-\R)   & for $\R \in \scrr(W)$  \cr
                     0          & for $\R \notin \scrr(W)$ \cr
                    }
         \slabel{def.scrzW.c}
         \label{def.scrzW}
\end{subeqnarray}
where the equivalence of these three expressions
is an immediate consequence of Theorem~\ref{thm2.fund}.
We then have:

\begin{corollary}
      \label{cor.log_submodular}
For any repulsive lattice gas, the function $\scrz_W$ is
\begin{itemize}
      \item[(a)] continuous
      \item[(b)] decreasing
      \item[(c)] satisfies $\scrz_W(\blambda_1 \blambda_2 \R) \, \scrz_W(\R)
                   \,\le\, \scrz_W(\blambda_1 \R) \, \scrz_W(\blambda_2 \R)$
                   for all $\blambda_1,\blambda_2 \in [0,1]^X$ and
                   $\R \in [0,\infty)^X$
      \item[(d)] log submodular
         [i.e.\ $\scrz_W(\R_1 \wedge \R_2) \, \scrz_W(\R_1 \vee \R_2) \,\le\,
                 \scrz_W(\R_1) \, \scrz_W(\R_2)$,
          where $\wedge$ (resp.\ $\vee$) denotes the elementwise
          min (resp.\ max) of vectors]
\end{itemize}
\end{corollary}

\proof
(a)  The continuity of $\scrz_W$ follows easily from
\reff{def.scrzW.a} or \reff{def.scrzW.b}
and the continuity of $Z_W$.
[For any $r < \infty$, the function $Z_W$ is uniformly continuous
  on the compact ball $B = \{\w \in \C \colon\, |\w| \le r\}$.
  If $\R,\R' \in B \cap [0,\infty)^X$ with $|\R-\R'| \le \delta$,
  then $|\scrz_W(\R) - \scrz_W(\R')| \le
        \sup\limits_{\begin{scarray}
                        \w,\w' \in B \\
                        |\w - \w'| \le \delta
                     \end{scarray}}
        |Z_W(\w) - Z_W(\w')|$.
  Uniform continuity of $Z_W$ on $B$ therefore implies
  uniform continuity of $\scrz_W$ on $B \cap [0,\infty)^X$.]

(b)  The decreasing property is an immediate consequence of
\reff{def.scrzW.a} or \reff{def.scrzW.b}.

(c) This is trivial if $\R \notin \scrr(W)$, so assume $\R \in \scrr(W)$.
Then $Y_W(-\R;\blambda_2) = \scrz_W(\blambda_2 \R)/\scrz_W(\R)$,
while $Y_W(-\blambda_1\R;\blambda_2) =
        \scrz_W(\blambda_1 \blambda_2 \R)/\scrz_W(\blambda_1 \R)$,
so that the claim follows from Proposition~\ref{prop.derivs_of_Y}
with $n=1$.

(d) The log submodularity follows immediately from (c)
by setting $\R = \R_1 \vee \R_2$ and $\blambda_i = \R_i/\R$ for $i=1,2$
(setting $0/0=1$ where needed).
[Alternatively, for $\R_1 \vee \R_2 \in \scrr(W)$ it follows
by integrating \reff{eq.derivs_of_Z}
with $n=2$ and $x_1 \neq x_2$.]
\qed


We can also prove a strict version of the cases $n=0$ of
Propositions~\ref{prop.derivs_of_Z}(a) and \ref{prop.derivs_of_Y}(a)
as well as Corollary~\ref{cor.log_submodular}(c).
Note first that
\begin{eqnarray}
    -\log Z_W(-\R)   & = &
        \sum\limits_{\n} (-1)^{|\n| - 1} \, c_{\n}(W) \, \R^{\n}
    \label{eq.logZmR.1}   \\[2mm]
    -\log Z_W(-\R) \,+\, \log Z_W(-\blambda\R)   & = &
        \sum\limits_{\n} (-1)^{|\n| - 1} \, c_{\n}(W) \,
                               (1 - \blambda^{\n}) \, \R^{\n}
    \label{eq.logZmR.2}   \\[2mm]
    -\log Z_W(-\R) \,+\, \log Z_W(-\blambda_1\R)   \!\! & + & \!\!
       \log Z_W(-\blambda_2\R) \,-\, \log Z_W(-\blambda_1\blambda_2\R)
        \nonumber \\
    &  & \hspace*{-4cm} =\;
        \sum\limits_{\n} (-1)^{|\n| - 1} \, c_{\n}(W) \,
             (1 - \blambda_1^{\n}) (1 - \blambda_2^{\n}) \, \R^{\n}
    \label{eq.logZmR.3}
\end{eqnarray}
By Proposition~\ref{prop_derivs_of_cn},
all three quantities are nonnegative
whenever $\R \in \scrr(W)$ and
$\blambda,\blambda_1,\blambda_2 \in [0,1]^X$.
We can determine when they are strictly positive:

\begin{proposition}
Consider any repulsive lattice gas,
and let $\R \in \scrr(W)$ and $\blambda,\blambda_1,\blambda_2 \in [0,1]^X$.
Then:
\begin{itemize}
    \item[(a)]  $-\log Z_W(-\R) > 0$ if and only if $\R \neq {\bf 0}$.
    \item[(b)]  $-\log Z_W(-\R) + \log Z_W(-\blambda\R) > 0$ if and only if
        $\supp \R \cap \supp({\bf 1} - \blambda) \neq \emptyset$.
    \item[(c)]  $-\log Z_W(-\R) + \log Z_W(-\blambda_1\R) +
                  \log Z_W(-\blambda_2\R) - \log Z_W(-\blambda_1\blambda_2\R)
                 > 0$ if and only there exists a component of $G_W[\supp\R]$
                that meets both $\supp({\bf 1} - \blambda_1)$ and
                $\supp({\bf 1} - \blambda_2)$.
\end{itemize}
\end{proposition}

\proof
We examine the terms in \reff{eq.logZmR.1}--\reff{eq.logZmR.3}
and determine (for some of them) when they are nonzero.

(a) ``Only if'' is trivial;
for ``if'', just consider the term $\n = \bdelta_x$
for some $x \in \supp\R$, and use Proposition~\ref{prop_sign_of_cn.strict}(a).

(b) ``Only if'' is again trivial;
for ``if'', just consider $\n = \bdelta_x$
for some $x \in \supp\R \cap \supp({\bf 1} - \blambda)$,
and again use Proposition~\ref{prop_sign_of_cn.strict}(a).

(c)  For ``if'', let $S$ be the vertex set of a component of $G_W[\supp\R]$
that meets both $\supp({\bf 1} - \blambda_1)$
and $\supp({\bf 1} - \blambda_2)$,
take $\n = {\bf 1}_S$,
and use Proposition~\ref{prop_sign_of_cn.strict}(a)
or \ref{prop_sign_of_cn.strict}(c).
For ``only if'', suppose that $c_{\n}(W) \neq 0$.
Then, by  Proposition~\ref{prop_sign_of_cn.strict},
either $\n = \bdelta_x$ for some $x \in X$,
or $\n = k\bdelta_x$ for some $k \ge 2$ and
  some $x \in X$ having $W(x,x) \neq 1$,
or $|\supp\n| \ge 2$ and $G_W[\supp\n]$ is connected.
In all three cases, $G_W[\supp\n]$ is connected.
Furthermore, to make a nonzero contribution to \reff{eq.logZmR.3},
$\supp\n$ has to meet both
$\supp({\bf 1} - \blambda_1)$ and $\supp({\bf 1} - \blambda_2)$
and be contained in $\supp\R$.
Therefore, there is a component of $G_W[\supp\R]$
(namely, the one containing $\supp\n$)
that meets both
$\supp({\bf 1} - \blambda_1)$ and $\supp({\bf 1} - \blambda_2)$.
\qed

For a repulsive lattice gas with hard-core self-repulsion
[i.e.\ $W(x,x)=0$ for all $x \in X$],
the set $\scrr(W)$ has an additional convexity property:

\begin{corollary}[convexity of two-dimensional sections]
  \label{cor.2Dconvexity}
For any repulsive lattice gas with hard-core self-repulsion,
the two-dimensional sections of $\scrr(W)$
parallel to the coordinate axes are
convex and (when nonempty)
are bounded either by a hyperbola or by a straight line.
\end{corollary}

\proof
For $x,y\in X$ ($x \neq y$)
and a nonnegative vector $\balpha=(\alpha_z)_{z \in X \setminus\{x,y\}}$,
consider the section
\be
    P_{x,y,\smallbalpha} \scrr(W)   \;=\;
    \{(R_x,R_y) \colon\;  (R_x,R_y,\balpha) \in \scrr(W) \}
    \;.
\ee
If $(0,0,\balpha) \notin \scrr(W)$,
then clearly the section is empty;
so let us assume henceforth that $(0,0,\balpha) \in \scrr(W)$.
Let $Z_W(w_x,w_y;-\balpha)$ be the polynomial obtained from
$Z_W(\w)$ by setting $w_z=-\alpha_z$ for $z\ne x,y$.
Then, as $Z_W$ is multiaffine, we have
\be
    Z_W(w_x,w_y;-\balpha)  \;=\;
    A + B w_x + C w_y + D w_x w_y
\ee
where
\begin{subeqnarray}
    A  & = &  Z_W(0,0;-\balpha)   \\
    B  & = &  Z_W(0,0;-W(x,\cdot) \balpha)   \\
    C  & = &  Z_W(0,0;-W(y,\cdot) \balpha)   \\
    D  & = &  W(x,y) \, Z_W(0,0;-W(x,\cdot) W(y,\cdot) \balpha)
\end{subeqnarray}
as a consequence of the multiple differentiation identity
\reff{eq.differentiation.multiple}.
Since $(0,0,\balpha) \in \scrr(W)$,
$0 \le W(x,y) \le 1$ for all $x,y$, and $\scrr(W)$ is a down-set,
we have $A,B,C > 0$ and $D \ge 0$
[with $D = 0$ if and only if $W(x,y) = 0$].
Moreover, by Corollary~\ref{cor.log_submodular}(c)
we have $AD \le BC$.\footnote{
    In the case of a hard-core pair interaction,
    this follows alternatively from Corollary~\ref{cor.log_submodular}(d).
}
It follows from Theorem~\ref{thm2.fund}(a)
that $P_{x,y,\smallbalpha}\scrr(W)$ is the component of
$[0,\infty)^2 \cap \{(R_x,R_y) \colon\; Z_W(-R_x,-R_y;-\balpha)>0\}$
containing $(0,0)$
[here we have used the fact that $(0,0,\balpha) \in \scrr(W)$].
Thus,
\be
    P_{x,y,\smallbalpha}\scrr(W)  \;=\;
    \{(R_x,R_y) \colon\;  R_x,R_y \ge 0 \hbox{ and }
                          BR_x + CR_y < A \}
\ee
in case $D=0$, and
\begin{eqnarray}
    P_{x,y,\smallbalpha}\scrr(W)
    & = &
    \{(R_x,R_y) \colon\; 0 \le R_x \le C/D \hbox{ and }
                         0 \le R_y \le B/D \hbox{ and }  \nonumber \\
    &  & \hspace{0.3in}
                         (C/D - R_x)(B/D - R_y) > (BC-AD)/D^2 \} \qquad
\end{eqnarray}
in case $D > 0$.
\qed

\medskip
\par\noindent
{\bf Remarks.}
1.  The proof shows that the boundary of $P_{x,y,\smallbalpha}\scrr(W)$
is a straight line when $W(x,y) = 0$
and is a hyperbola when $0 < W(x,y) \le 1$.

2.  Consider $\w = (w_x,w_y)$ and $\w' = (w'_x,w'_y)$,
and let $0 \le \lambda \le 1$;  then
\begin{eqnarray}
   & &
    Z_W(\lambda\w + (1-\lambda)\w' ; -\balpha)  \,-\,
    [\lambda Z_W(\w;-\balpha) + (1-\lambda) Z_W(\w';-\balpha)]
   \nonumber \\
   & & \qquad\qquad
    \;=\;
    -D \lambda(1-\lambda)(w_x - w'_x)(w_y - w'_y)
    \;,
\end{eqnarray}
so that $Z_W(\,\cdot\,;-\balpha)$ is concave along line segments
of negative slope in the $(w_x,w_y)$-plane.
This provides an alternate proof that the projected section
$P_{x,y,\smallbalpha}\scrr(W)$ is convex
[because $\scrr(W)$ is a down-set,
  only line segments of negative slope need be considered].

3.  The convexity of two-dimensional sections
is false without the hypothesis of hard-core self-repulsion:
consider $X=\{1,2\}$, $W(1,1)=W(2,2)=1$ and $W(1,2)=0$.
Then $Z_W(\w) = e^{w_1} + e^{w_2} - 1$,
so that $\scrr(W) = \{(R_1,R_2) \colon\, R_1,R_2 \ge 0 \hbox{ and }
            e^{-R_1} + e^{-R_2} > 1\}$,
which is not convex.

4.  Even in the case of hard-core self-repulsion,
$\scrr(W)$ itself is not in general convex.
To see this, let $X = \{1,2,3\}$ and let $G$ be the path 123;
its independent-set polynomial is $Z_G(w)=w_2+(1+w_1)(1+w_3)$.
The diagonal section of $\scrr(G)$ given by $R_1=R_3=x$ and $R_2=y$
is the region of $[0,1]^2$ where $y<(1-x)^2$,
which is not convex.

\subsection{Algebraic irreducibility of $Z_W(\w)$}  \label{sec2.irred}

In this subsection ---
which is a digression from the main thread
of the paper and can be omitted on a first reading ---
we discuss the algebraic irreducibility of the multivariate
partition function $Z_W(\w)$.
We restrict attention to the case of hard-core self-repulsion,
in which $Z_W$ is a multiaffine polynomial.

Let $R$ be a commutative ring with identity,
and let $W \colon\, X \times X \to R$ be symmetric
and satisfy $W(x,x) = 0$ for all $x \in X$.
Define the support graph $G_W$
by setting $xy \in E(G)$
if and only if $W(x,y) \neq 1$ and $x \neq y$.
Define the polynomial $Z_W \in R[\w]$ by
\be
      Z_W   \;=\;  \sum_{X' \subseteq X}
                   \left( \prod\limits_{\{x,y\} \subseteq X'}  W(x,y) \right)
                   \left( \prod\limits_{x \in X'} w_x \right)
      \;.
\ee

\begin{proposition}
  \label{prop.irred}
Suppose that $R$ is an integral domain.
Then $Z_W$ is irreducible over $R$
if and only if $G_W$ is connected.
\end{proposition}

The proof is based on an easy but crucial lemma \cite[Lemma~4.7]{Choe_02}:

\begin{lemma}
   \label{lemma.irreducible}
Let $P_1$ and $P_2$ be nonzero polynomials in the indeterminates
$\{w_x\}_{x \in X}$ with coefficients in an integral domain $R$.
Suppose that $P_1 P_2$ is multiaffine
(i.e.\ of degree 0 or 1 in each variable separately).  Then:
\begin{itemize}
    \item[(a)]  There exist {\em disjoint}\/ subsets $X_1,X_2 \subseteq X$
       such that $P_i$ uses only the variables $\{w_x\}_{x \in X_i}$
       ($i=1,2$).
    \item[(b)]  $P_1$ and $P_2$ are both multiaffine.
\end{itemize}
\end{lemma}

\proof
Suppose there exists $x \in X$ such that both $P_1$ and $P_2$ use
the variable $w_x$.  For $i=1,2$, let $d_i \ge 1$ be the degree of $P_i$
in the variable $w_x$, and let $Q_i \neq 0$ be the coefficient of
$w_x^{d_i}$ in $P_i$, considered as an element of the polynomial ring
$R[\w_{\neq x}]$.  Then $Q_1 Q_2 \neq 0$ because $R[\w_{\neq x}]$ is an
integral domain \cite[Theorem III.5.1 and Corollary III.5.7]{Hungerford_74}.
But this shows that the coefficient of $w_x^{d_1 + d_2}$ in $P_1 P_2$
is nonzero, contradicting the hypothesis that $P_1 P_2$ is multiaffine
(since $d_1 + d_2 \ge 2$).  This proves (a);  and (b) is an easy consequence.
\qed

\proofof{Proposition~\ref{prop.irred}}
If $G_W$ is disconnected, it is obvious that $Z_W$ is reducible.
To prove the converse,
suppose that $Z_W$ is reducible, i.e.\ $Z_W = P_1 P_2$
where $P_1$ and $P_2$ are nonconstant polynomials over $R$.
It follows from Lemma~\ref{lemma.irreducible} that there exist
disjoint subsets $X_1,X_2 \subseteq X$
such that $P_i$ uses only the variables $\{w_x\}_{x \in X_i}$
(we can assume without loss of generality that $X_1 \cup X_2 = X$);
moreover, neither $X_1$ nor $X_2$ can be empty
(since $P_1$ and $P_2$ are nonconstant).
Let $a_i \in R$ be the constant term of $P_i$ ($i=1,2$).
Since $Z_W$ has constant term 1, we have $a_1 a_2 = 1$.
Replacing $P_1$ by $a_2 P_1$ and $P_2$ by $a_1 P_2$,
we can assume without loss of generality that $a_1 = a_2 = 1$.
For $x \in X_1$ (resp.\ $x \in X_2$),
let $b_x$ be the coefficient of the linear term $w_x$ in $P_1$
(resp.\ in $P_2$).
Since the linear term $w_x$ has coefficient 1 in $Z_W$
and $a_1 = a_2 = 1$, we have $b_x = 1$ for all $x \in X$.
It follows that, for $x \in X_1$ and $y \in X_2$,
the term $w_x w_y$ in $P_1 P_2$ has coefficient 1;
since this term has coefficient $W(x,y)$ in $Z_W = P_1 P_2$,
we have $W(x,y) = 1$ whenever $x \in X_1$ and $y \in X_2$.
Hence $G_W$ is disconnected.
\qed

\medskip
\par\noindent
{\bf Remarks.}
1. The {\em univariate}\/ polynomial $Z_W(w)$ is obviously reducible over $\C$
whenever $|X| \ge 2$;  and it is in many cases reducible over $\RR$ as well.
Indeed, it can be reducible over the integers:
e.g.\ for $G = P_4$ (the 4-vertex path) we have
$Z_G(w) = 1 + 4w + 3w^2 = (3w+1)(w+1)$.

2.  Results analogous to Proposition~\ref{prop.irred}
hold for some other combinatorial polynomials.
For example, the irreducibility over $\C$ for the (bivariate)
Tutte polynomial $Z_M(x,y)$ of a connected matroid $M$
was proven recently by Merino, de Mier and Noy \cite{Merino_01}.
Likewise, the irreducibility over any integral domain $R$
for the {\em multivariate}\/ basis generating polynomial
of a connected matroid $M$ ---
or more generally, for any multiaffine polynomial whose
support is exactly the collection of bases of $M$ ---
was proven recently by Choe, Oxley, Sokal and Wagner \cite{Choe_02}.

3.  It would be interesting to know whether analogous results
can be obtained without the hypothesis of hard-core self-repulsion.
The difficulty is to find the right ring in which to work.
Irreducibility in the ring $R[[\w]]$ of formal power series
is too trivial, because {\em every}\/ formal power series with
constant term 1 is invertible.
On the other hand, it might conceivably be possible to prove irreducibility
in the ring of entire functions (or entire functions of order 1)
on $\C^X$ --- but only for $|X| \ge 2$, since it is manifestly false
if $|X|=1$ and $0 < W(x,x) < 1$ (a nonpolynomial entire function
of order 0 has infinitely many zeros \cite[Theorem I.13]{Levin_64}).

\subsection{Convexity of $\log Z$ at nonnegative fugacity}
   \label{sec2.convexity}

In this paper we are primarily concerned with the behavior of $Z_W(\w)$
for {\em complex}\/ fugacities $\w$.
However, the regime of {\em nonnegative}\/ fugacities $\w$
is of particular interest to probabilists and statistical mechanicians,
since the Boltzmann weights [cf.\ \reff{defZ}]
are there nonnegative and so can be interpreted,
after normalization, as a probability measure on lattice-gas configurations.
In this ``probabilistic regime'',
the logarithm of the partition function is, in very great generality,
a convex function of all the interaction energies \cite[p.~12]{Israel_79}.
Let us prove the specialization of this statement to our model:

Let $0 \le W,W' \le 1$ and $\w, \w' \ge 0$.
Then, for $0 \le \lambda \le 1$, define
$W_\lambda(x,y) = W(x,y)^\lambda W'(x,y)^{1-\lambda}$
and $(w_\lambda)_x = w_x^\lambda (w'_x)^{1-\lambda}$.
By H\"older's inequality applied to \reff{defZ_1} or \reff{defZ_2},
we obtain:

\begin{lemma}
   \label{lemma.sec2.convexity}
Suppose that $0 \le W,W' \le 1$ and $\w, \w' \ge 0$.
Then, for $0 \le \lambda \le 1$,
\be
    Z_{W_\lambda}(\w_\lambda)  \;\le\;
    Z_W(\w)^\lambda Z_{W'}(\w')^{1-\lambda}   \;.
\ee
\end{lemma}

\noindent
That is, $\log Z_{W_\lambda}(\w_\lambda)$ is a convex function of $\lambda$;
or in other words, $\log Z_W(\w)$ is a convex function of the vector
$\left\< \{\log W(x,y)\}_{x,y \in X}, \{\log w_x\}_{x \in X} \right\>$.

\clearpage 
\section{The lattice gas with hard-core self-repulsion}   \label{sec3}

We now restrict attention to the case of a repulsive lattice gas
with hard-core self-repulsion,
i.e.\ $0 \le W(x,y) \le 1$ for all $x,y$
and $W(x,x) = 0$ for all $x$.
This means, as noted in the Introduction,
that $Z_W(\w)$ can be written as a sum over subsets:
\be
      Z_W(\w)   \;=\;  \sum_{X' \subseteq X}
                       \prod\limits_{x \in X'} w_x
                       \prod\limits_{\{x,y\} \subseteq X'}  W(x,y)
      \;.
\ee
Here we shall exploit this special structure,
which implies that $Z_W$ is a {\em multiaffine polynomial}\/,
i.e.\ a polynomial of degree 1 in each $w_x$ separately.
Since $W$ will be fixed throughout, we shall henceforth often omit it
from the notation.

\subsection{The fundamental identity}  \label{sec3.1}

Let us define, for each subset $\Lambda \subseteq X$,
the restricted partition function
\be
      Z_\Lambda(\w)  \;=\;  \sum_{X' \subseteq \Lambda}  \;
              \prod_{x \in X'} w_x
              \prod_{\{x,y\} \subseteq X'}   W(x,y)
      \;.
      \label{def_Z_Lambda}
\ee
Of course this notation is redundant,
since the same effect can be obtained by setting $w_x = 0$
for $x \in X \setminus \Lambda$,
but it is useful for the purpose of inductive computations and proofs.
We have, for any $x \in \Lambda$, the {\em fundamental identity}\/
\be
      Z_\Lambda(\w) \;=\;
         Z_{\Lambda \setminus x}(\w)  \,+\,
         w_x Z_{\Lambda \setminus x}(W(x,\cdot) \w)
     \label{eq.dob_basic}
\ee
where
\be
      [W(x,\cdot) \w]_y  \;=\;  W(x,y) \, w_y  \;;
\ee
here the first term on the right-hand side of \reff{eq.dob_basic}
covers the summands in \reff{def_Z_Lambda} with $X' \not\ni x$,
while the second covers $X' \ni x$.
[Note that \reff{eq.dob_basic} is a special case of \reff{eq.fund_general}.]
In the special case of a hard-core interaction
(= independent-set polynomial) for a graph $G$,
\reff{eq.dob_basic} reduces to
\be
      Z_\Lambda(\w) \;=\;
         Z_{\Lambda \setminus x}(\w)  \,+\,
         w_x Z_{\Lambda \setminus \Gamma^*(x)}(\w)   \;,
     \label{eq.dob_basic_G}
\ee
where we have used the notation $\Gamma^*(x) = \Gamma(x) \cup \{x\}$
[here $\Gamma(x)$ denotes the set of vertices of $G$ adjacent to $x$].
The fundamental identity \reff{eq.dob_basic}/\reff{eq.dob_basic_G}
will play an important role both in the inductive proof of
the Lov\'asz local lemma [cf.\ \reff{eq.lov.fund} and \reff{eq.lov.fund2}]
and in the Dobrushin--Shearer inductive argument
for the nonvanishing of $Z_W$ in a polydisc (Section~\ref{sec.dob}).

Let us also remark that because $Z_{\Lambda \setminus x}$ is multiaffine,
the last term in \reff{eq.dob_basic} can be expanded to rewrite
$Z_{\Lambda \setminus x}(W(x,\cdot) \w)$ as a linear combination
of values at the vertices of a rectangle:\footnote{
   If $F \colon\, \C^X \to \C$ is multiaffine,
   then $F$ can be reconstructed from its values at the corners
   of any rectangle with positive volume.
   For example, if the rectangle is $[0,1]^X$, we have
   $$ F(\z)  \;=\;  \sum\limits_{Y \subseteq X}
                    \left( \prod\limits_{i \in Y} z_i \right)
                    \left( \prod\limits_{i \in X \setminus Y} (1-z_i) \right)
                    F({\bf 1}_Y)  \;.
   $$
   This is clear when $\z \in \{0,1\}^X$,
   and equality for all $\z$ then follows because both sides
   are multiaffine.
}
\be
      Z_\Lambda(\w) \;=\;
         Z_{\Lambda \setminus x}(\w)  \,+\,
         w_x \sum\limits_{Y \subseteq \Lambda \setminus x}
         \left( \prod\limits_{y \in Y} W(x,y) \right)
         \left( \prod\limits_{y \in (\Lambda \setminus x) \setminus Y}
            [1 - W(x,y)] \right)
        Z_{Y}(\w)
      \;.
     \label{eq.dob_basic_expanded}
\ee
In the hard-core case, the only term with a nonzero coefficient
is $Y = \Lambda \setminus \Gamma^*(x)$,
so that \reff{eq.dob_basic_expanded} reduces to \reff{eq.dob_basic_G}.

\medskip
\par\noindent
{\bf Remark.}
Repeated use of \reff{eq.dob_basic} obviously gives an algorithm
to compute $Z_W(\w)$.   But this algorithm takes in general exponential time.
In fact, calculating $Z_G(\w)$ for general graphs $G$
(or even for cubic planar graphs)
is NP-hard (as noted by Shearer \cite{Shearer_85}),
since even calculating the {\em degree}\/ of $Z_G(\w)$
--- that is, the maximum size of an independent set ---
is NP-hard \cite[pp.~194--195]{Garey_79}.
Therefore, if ${\rm P} \neq {\rm NP}$
it is impossible to calculate $Z_G(\w)$ for general graphs
in polynomial time.

\bigskip

A key role in \reff{eq.dob_basic} is manifestly played by
the rational function
\be
      \scrk_{x,\Lambda}(\w)  \;\equiv\;
         {Z_{\Lambda \setminus x}(W(x,\cdot) \w)
          \over
          Z_{\Lambda \setminus x}(\w)
         }
      \;.
    \label{def.scrk}
\ee
Note that $\scrk_{x,\Lambda}(\w)$ depends only on
$\{w_y\}_{y \in \Lambda \setminus x}$;
in particular, it does not depend on $w_x$.
We shall sometimes write it as $\scrk_{x,\Lambda}(\w_{\neq x})$
to emphasize this fact.
Please note also that $Z_\Lambda(\w)$ is nonvanishing in the region
$|w_x| < 1/|\scrk_{x,\Lambda}(\w_{\neq x})|$
and vanishes at $w_x = - 1/\scrk_{x,\Lambda}(\w_{\neq x})$.
We have
\be
      {\partial \log Z_\Lambda(\w)  \over  \partial w_x}
      \;=\;
      {\scrk_{x,\Lambda}(\w_{\neq x})
       \over
       1 \,+\, \scrk_{x,\Lambda}(\w_{\neq x}) w_x
      }
      \;.
    \label{eq.logZLambda.KxLambda}
\ee
%
%
%
Since $\scrk_{x,\Lambda}(\w)$ is a special case of the function
$Y_W(\w;\blambda)$ defined in \reff{def.YW},
we obtain as an immediate corollary of Proposition~\ref{prop.derivs_of_Y}(b):

\begin{proposition}
      \label{prop.scrk}
Consider any repulsive lattice gas with hard-core self-repulsion.
If $\R \in \scrr(W)$, then
\be
      |\scrk_{x,\Lambda}(\w)|   \;\le\;
      \scrk_{x,\Lambda}(-\R)    \;<\;  \infty
    \label{eq.prop.scrk}
\ee
for all $\w \in \bar{D}_{\R}$.
That is, the maximum of $|\scrk_{x,\Lambda}(\w)|$ over $\bar{D}_{\R}$
is attained at $\w = -\R$, and this maximum value $\scrk_{x,\Lambda}(-\R)$
is an increasing function of $\R \in \scrr(W)$.
\end{proposition}

\noindent
The function $\scrk_{x,\Lambda}(-\R)$ will play
an important role in Section~\ref{sec.dob}.

We can translate the fundamental identity \reff{eq.dob_basic}
into a recursion for the rational functions $\scrk_{x,\Lambda}(\w)$.
Let us order arbitrarily the sites
$\Lambda \setminus x = \{y_1,\ldots,y_k\}$.
Then we write \reff{def.scrk} as a telescoping product:
\be
      \scrk_{x,\Lambda}(\w)  \;=\;
         \prod\limits_{i=1}^k
         {Z_{\Lambda \setminus x}(\wtilde^{(i)})
          \over
          Z_{\Lambda \setminus x}(\wtilde^{(i-1)})
         }
    \label{eq.scrk.telescoping1}
\ee
where the vectors $\wtilde^{(i)}$ are defined by
\be
      (\wtilde^{(i)})_y  \;=\;
      \cases{ W(x,y) w_y  & if $y=y_j$ for some $j \le i$  \cr
              w_y         & otherwise \cr
            }
    \label{def.wtildei_sec3}
\ee
[Note that it is irrelevant how we define $(\wtilde^{(i)})_y$
    for $y=x$, since we set this weight to zero anyway
    by considering $Z_{\Lambda \setminus x}$.]
Applying the fundamental identity \reff{eq.dob_basic} to $y_i$, we obtain
\be
      {Z_{\Lambda \setminus x}(\wtilde^{(i)})
       \over
       Z_{\Lambda \setminus x}(\wtilde^{(i-1)})
      }
      \;=\;
      {1 \,+\,  W(x,y_i) w_{y_i} \scrk_{y_i,\Lambda \setminus 
x}(\wtilde^{(i-1)})
       \over
       1 \,+\,  w_{y_i}  \scrk_{y_i,\Lambda \setminus x}(\wtilde^{(i-1)})
      }
\ee
and hence
\be
\scrk_{x,\Lambda}(\w)  \;=\;
      \prod\limits_{i=1}^k
      {1 \,+\,  W(x,y_i) w_{y_i} \scrk_{y_i,\Lambda \setminus 
x}(\wtilde^{(i-1)})
       \over
       1 \,+\,  w_{y_i} \scrk_{y_i,\Lambda \setminus x}(\wtilde^{(i-1)})
      }
      \;.
    \label{eq.scrk.telescoping2}
\ee
This identity will play a central role at the end of Section~\ref{sec.dob}.
In the special case of a hard-core pair interaction,
\reff{eq.scrk.telescoping2} becomes
\be
      \scrk_{x,\Lambda}(\w)  \;=\;
      \prod\limits_{i=1}^l
      {1 \over
       1 \,+\, w_{y_i}
               \scrk_{y_i,\Lambda \setminus x \setminus 
\{y_1,\ldots,y_{i-1}\}}(\w)
      }
    \label{eq.scrk.telescoping2_G}
\ee
where $\{y_1,\ldots,y_l\}$ is an ordering of $\Lambda \cap \Gamma(x)$.

Note, finally, that the partition functions $Z_\Lambda(\w)$
can be reconstructed from the rational functions $\scrk_{x,\Lambda}(\w)$.
For, by the fundamental identity \reff{eq.dob_basic}
and the definition \reff{def.scrk}, we have
\be
      {Z_\Lambda(\w)  \over Z_{\Lambda \setminus x}(\w)}
      \;=\;
      1  \,+\, w_x \scrk_{x,\Lambda}(\w_{\neq x})
      \;;
\ee
and hence, setting $\Lambda = \{x_1,\ldots,x_n\}$ (in arbitrary order)
we can write
\be
      Z_\Lambda(\w)  \;=\;
      \prod\limits_{i=1}^n
      {Z_{\{x_1,\ldots,x_i\}}(\w) \over Z_{\{x_1,\ldots,x_{i-1}\}}(\w)}
      \;=\;
      \prod\limits_{i=1}^n
      \left[ 1 \,+\, w_x \scrk_{x_i,\{x_1,\ldots,x_i\}}(\w) \right]
      \;.
    \label{eq.reconstructZ}
\ee

\subsection{Examples}  \label{sec3.2}

Let us take a moment to compute a few examples
of partition functions $Z_W(\w)$.
We shall assume hard-core self-repulsion
[i.e.\ $W(x,x)=0$ for all $x$] throughout,
and shall mostly restrict attention to the case of a
hard-core pair interaction,
i.e.\ the independent-set polynomial $Z_G(\w)$ for a graph $G$.
In some cases we shall, for simplicity,
compute only the univariate polynomial $Z_G(w)$
obtained by setting $w_x = w$ for all vertices $x$.

\bigskip

{\bf Example \thesection.1.}
{\em The complete graph $K_n$.}
Clearly $Z_{K_n}(\w) = 1 + w_1 + \ldots + w_n$.
In particular, $\scrr(K_n) = \{\R \colon\; R_1 + \ldots + R_n < 1 \}$.

\bigskip

{\bf Example \thesection.2.}
{\em The $n$-vertex path $P_n$.}
We limit attention to the univariate independent-set polynomial.
Applying the fundamental identity \reff{eq.dob_basic_G}
to an endvertex, we obtain the recursion relation
\be
      Z_{P_n}(w)  \;=\;  Z_{P_{n-1}}(w) \,+\, w Z_{P_{n-2}}(w)
      \;,
    \label{eq.Pn.recursion}
\ee
which is valid for all $n \ge 0$ if we define
$Z_{P_0} \equiv Z_{P_{-1}} \equiv 1$ and $Z_{P_{-2}} \equiv 0$.
The solution of \reff{eq.Pn.recursion} is
\be
      Z_{P_n}(w)  \;=\;
      {1 \over \sqrt{1+4w}} \, (\lambda_+^{n+2} - \lambda_-^{n+2})
    \label{eq.Pn.solution}
\ee
where
\be
      \lambda_\pm  \;=\;  {1 \pm \sqrt{1+4w}  \over 2}
      \;.
    \label{eq.Pn.solution_def_lambda}
\ee
(An alternative method of obtaining this result is given in
  Example~\thesection.2${}'$ below.)
The zeros of $Z_{P_n}$ are located at
\be
      w  \;=\;  - \, {1  \over  4 \cos^2 {\pi k \over n+2}}
\ee
for $k=1,2,\ldots, \lfloor {n+1 \over 2} \rfloor$.
The zero nearest the origin converges from below to $w= -1/4$
as $n \to\infty$.
Elementary counting arguments give also the explicit formula
\be
      Z_{P_n}(w)  \;=\;  \sum\limits_{k=0}^{\lfloor {n+1 \over 2} \rfloor}
        {n+1-k \choose k} \, w^k
      \;.
    \label{eq.Pn.explicit}
\ee
\bigskip

{\bf Example \thesection.3.}
{\em The $n$-vertex cycle $C_n$.}
We again limit attention to the univariate independent-set polynomial.
Applying the fundamental identity \reff{eq.dob_basic_G},
we find
\be
      Z_{C_n}(w)  \;=\;  Z_{P_{n-1}}(w) \,+\, w Z_{P_{n-3}}(w)
      \;,
    \label{eq.Cn.recursion}
\ee
valid for $n \ge 2$.
Inserting \reff{eq.Pn.solution}/\reff{eq.Pn.solution_def_lambda},
we obtain
\be
      Z_{C_n}(w)  \;=\;  \lambda_+^n \,+\, \lambda_-^n  \;.
\ee
(An alternative method of obtaining this result is given in
  Example~\thesection.3${}'$ below.)
The zeros of $Z_{C_n}$ are located at
\be
      w  \;=\;  - \, {1  \over  4 \cos^2 {\pi (k + \half) \over n}}
\ee
for $k=0,1,\ldots, \lfloor {n \over 2} \rfloor - 1$.
The zero nearest the origin converges from below to $w= -1/4$
as $n \to\infty$.
Elementary counting arguments
[or \reff{eq.Pn.explicit}/\reff{eq.Cn.recursion}]
give also the explicit formula
\be
      Z_{C_n}(w)  \;=\;  \sum\limits_{k=0}^{\lfloor {n \over 2} \rfloor}
        \frac{n}{n-k}{n-k \choose k} \, w^k
      \;.
\ee

\medskip
\par\noindent
{\bf Remark.}
There is a one-to-one correspondence between matchings on a graph $G$
and independent sets on the line graph $L(G)$.
Since $L(P_n) = P_{n-1}$ and $L(C_n) = C_n$,
these formulae for $Z_{P_n}$ and $Z_{C_n}$
can also be found in the work of Heilmann and Lieb
on matching polynomials \cite[pp.~196--197]{Heilmann_72}.
Heilmann and Lieb also noted that $Z_{P_n}$ and $Z_{C_n}$
can be written as Chebyshev polynomials.

\bigskip

{\bf Example \thesection.3\textprime.}
{\em The ``soft'' $n$-vertex cycle $C_n$.}
Let us now generalize Example \thesection.3 by considering the cycle $C_n$
with edge weights
\be
      W(x,y)  \;=\;  \cases{W  & if $xy$ is an edge of $C_n$  \cr
                            1  & otherwise  \cr
                           }
\ee
where $W$ is a constant in $[0,1]$
(the case $W=0$ corresponds to the independent-set polynomial of $C_n$,
    while the case $W=1$ corresponds to the independent-set polynomial of
    the edgeless graph on $n$ vertices).
Then the univariate polynomial $Z_W(w)$ is given by
a ``transfer matrix''
\cite[Section 2.1]{Baxter_82}
\cite[Sections 2.2--2.4]{Biggs_77}
\cite[Section 4.7]{Stanley_86}:
\be
      Z_W(w)  \;=\;  \tr \left(\!\! \begin{array}{cc}
                                        1 & 1 \\
                                        w & wW
                                    \end{array}
                         \!\!\right) ^{\! n}
\ee
where the first (resp.\ second) row or column corresponds to
an empty (resp.\ occupied) site, and the fugacity $w$ is attributed
to the row only.  It follows that
\be
      Z_W(w)  \;=\;  \lambda_+^n \,+\, \lambda_-^n   \;,
\ee
where
\be
      \lambda_\pm  \;=\;
      {1 + wW \pm \sqrt{(1-wW)^2 + 4w}
       \over
       2}
     \label{def.lambdapm.W}
\ee
are the eigenvalues of the transfer matrix.
Simple algebra then shows that all of the zeros of $Z_W$
are real and negative (when $0 \le W \le 1$).

\bigskip

{\bf Example \thesection.2\textprime.}
{\em The ``soft'' $n$-vertex path $P_n$.}
Let us now generalize Example \thesection.2 analogously.
The univariate polynomial $Z_W(w)$ is again given by a transfer matrix:
\be
      Z_W(w)  \;=\;  (1 \quad 1) \; \left(\!\! \begin{array}{cc}
                                                   1 & 1 \\
                                                   w & wW
                                               \end{array}
                                    \!\!\right) ^{\! n-1}
                                 \left(\!\! \begin{array}{c}
                                                1  \\
                                                w
                                            \end{array}
                                 \!\!\right)
      \;\;.
\ee
After some algebra we find
\be
      Z_W(w)  \;=\;  A_+ \lambda_+^{n+2} \,+\, A_- \lambda_-^{n+2}   \;,
\ee
with $\lambda_\pm$ given by \reff{def.lambdapm.W} and
\be
    A_\pm  \;=\;
    {2 \over
     \pm (2-W+wW^2) \sqrt{(1-wW)^2 + 4w} \;+\; W [(1-wW)^2 + 4w]
    }
      \;.
\ee

\bigskip

{\bf Example \thesection.4.}
{\em The star $K_{1,r}$.}
Let $x$ be the center vertex of the star
and let $y_1,\ldots,y_r$ be the leaves.
The independent-set polynomial is
\be
      Z_{K_{1,r}}(\w)  \;=\;
      \prod\limits_{i=1}^r (1+w_{y_i})   \;+\;  w_x
      \;.
\ee
More generally, suppose we have a pair interaction matrix $W(x,y_i)$
[with $W(y_i,y_j) = 1$ for all $i \neq j$
    and $W(x,x) = W(y_i,y_i) = 0$].
Then
\be
      Z_{W}(\w)  \;=\;
      \prod\limits_{i=1}^r (1+w_{y_i})
      \;+\;
      w_x  \prod\limits_{i=1}^r [1+ W(x,y_i) w_{y_i}]
      \;.
    \label{eq.ZW.star}
\ee

\bigskip

{\bf Example \thesection.5.}
{\em The complete bipartite graph $K_{m,n}$.}\/
Let $X = \{x_1,\ldots,x_m\}$ and $Y = \{y_1,\ldots,y_n\}$
be the bipartition.  Then the independent-set polynomial is
\be
      Z_{K_{m,n}}(\w)  \;=\;
      \prod\limits_{i=1}^m (1+w_{x_i})   \;+\;
      \prod\limits_{j=1}^n (1+w_{y_j})   \;-\;  1
      \;.
\ee

\bigskip

{\bf Example \thesection.6.}
{\em The complete $r$-ary rooted tree}\/ \cite{Runnels_67,Shearer_85}.
Let $T_n^{(r)}$ be the complete rooted tree
with branching factor $r$ and depth $n$.
We limit attention to the univariate independent-set polynomial.
Fix $r \ge 1$; and to lighten the notation,
let us write $Z_n$ as a shorthand for $Z_{T_n^{(r)}}$.
Applying the fundamental identity \reff{eq.dob_basic_G}
to the root vertex, we obtain the nonlinear recursion
\be
      Z_n(w)  \;=\;  Z_{n-1}(w)^r  \,+\,  w Z_{n-2}(w)^{r^2}
      \;,
    \label{eq.Zn.tree}
\ee
which is valid for all $n \ge 0$
if we set $Z_{-1} \equiv Z_{-2} \equiv 1$.
By defining
\be
      Y_n(w)  \;=\;  {Z_n(w) \over Z_{n-1}(w)^r}
      \;,
    \label{def.Yn.tree}
\ee
we can convert the second-order recursion \reff{eq.Zn.tree}
to a first-order recursion
\be
      Y_n(w)  \;=\;  1  \,+\,  {w \over Y_{n-1}(w)^r}
    \label{eq.Yn.tree}
\ee
with initial condition $Y_{-1} \equiv 1$.
The polynomials $Z_n(w)$ can be reconstructed from
the rational functions $Y_n(w)$ by
\be
      Z_n(w)  \;=\; \prod\limits_{k=0}^n Y_k(w)^{r^{n-k}}
      \;.
    \label{reconstruct.Zn.tree}
\ee

Let $w_n < 0$ be the negative real root of $Z_n$ of smallest magnitude
(set $w_n = -\infty$ if $Z_n$ has no negative real root).
Note that $w_{-1} = -\infty$ and $w_0 = -1$.
Let us prove by induction that $w_{n-1} < w_n$ for $n \ge 0$.
It is true for $n=0$.  For $n \ge 1$ we have
\be
      Z_n(w_{n-1})  \;=\;  Z_{n-1}(w_{n-1})^r  \,+\,
                           w_{n-1} Z_{n-2}(w_{n-1})^{r^2}
                    \;<\;  0
\ee
since $Z_{n-1}(w_{n-1}) = 0$, $w_{n-1} < 0$
and $Z_{n-2}(w_{n-1}) > 0$ by the inductive hypothesis.
Therefore $Z_n$ vanishes somewhere between $w_{n-1}$ and 0.

It follows that the $w_n$ increase to a limit $w_\infty \le 0$ as $n\to\infty$.
Let us show, following Shearer \cite{Shearer_85}, that
\be
      w_\infty  \;=\;  - {r^r \over (r+1)^{r+1}}
\ee
by proving the two inequalities:

{\sc Proof of $\ge$:}
If $w \in [w_\infty,0)$,
we have $Z_n(w) > 0$ for all $n$ and hence also
$Y_n(w) > 0$ for all $n$.
Since $Y_{-1}>Y_0$, it follows from the monotonicity of \reff{eq.Yn.tree} that
$\{Y_n(w)\}_{n \ge 0}$ is a strictly decreasing sequence
of positive numbers, hence converges to a limit $y_* \ge 0$
satisfying the fixed-point equation $y_* = 1 + w/y_*^r$,
or equivalently $w = y_*^{r+1} - y_*^r$.
Elementary calculus then shows that $w \ge -r^r / (r+1)^{r+1}$;
taking $w=w_\infty$ we obtain $w_\infty \ge -r^r / (r+1)^{r+1}$.

{\sc Proof of $\le$:}
If $-r^r / (r+1)^{r+1} \le w < 0$,
the equation $w = y_*^{r+1} - y_*^r$
has a unique solution $y_* \in [r/(r+1), 1)$.
It then follows by induction
[using \reff{eq.Yn.tree} and the initial condition $Y_{-1} = 1$] that
$1 = Y_{-1}(w) > Y_0(w) > \ldots > Y_{n-1}(w) > Y_n(w) > \ldots > y_*$
for all $n \ge 0$.
In particular, $Y_n(w) > 0$ for all $n$,
so that $w_n < w$ for all $n$.
This shows that $w_\infty \le -r^r / (r+1)^{r+1}$.

Let us conclude by observing that \reff{eq.Yn.tree}
defines a degree-$r$ rational map
$R_w \colon\, y \mapsto 1 + w/y^r$
parametrized by $w \in \C \setminus 0$.
Moreover, the zeros of $Z_W(w)$ correspond to those values $w$
for which $R_w$ has a (superattractive) orbit
$0 \mapsto \infty \mapsto 1 \mapsto 1+w \mapsto \ldots \mapsto 0$
of period $n+3$ (or some submultiple of $n+3$).
As $n\to\infty$, these points accumulate on a ``Mandelbrot-like'' set
in the complex $w$-plane.
For further information on the maps $y \mapsto 1 + w/y^r$,
see \cite{Milnor_bicritical,Bamon_99,Bobenrieth_tesis,Bobenrieth_1,%
Bobenrieth_2,Bobenrieth_3}.

\subsection{Reduction formulae}   \label{sec3.3}

Recall that, given $W$, we have defined a simple loopless graph $G = G_W$
(``the support graph of $W$'') by setting $xy \in E(G)$
if and only if $W(x,y) \neq 1$ and $x \neq y$.
When the support graph $G$ has a very simple structure,
the polynomial $Z_W(\w)$ can be simplified:

1) {\em Disconnected graph.}\/
Suppose that $G$ can be written as the disjoint union
of $G_1$ and $G_2$.  Set $\Lambda_i = V(G_i)$.
Then clearly $Z_W$ factorizes:
\be
      Z_{\Lambda_1 \cup \Lambda_2}(\w)  \;=\;
      Z_{\Lambda_1}(\w) \, Z_{\Lambda_2}(\w)
      \;.
    \label{eq.factorization}
\ee
Note that $Z_{\Lambda_i}(\w)$ depends only on $\{w_x\}_{x \in \Lambda_i}$.

2) {\em Cut vertex.}\/
Suppose that $G = G_1 \cup G_2$ where $V(G_1) \cap V(G_2) = \{x\}$.
[If $G_1$ and $G_2$ have at least two vertices, this means that
    $x$ is a cut vertex of $G$.  But the formulae below hold also
    in the trivial cases $V(G_1) = \{x\}$ or $V(G_2) = \{x\}$ or both.]
Again set $\Lambda_i = V(G_i)$.
The fundamental identity \reff{eq.dob_basic} asserts that
\be
      Z_{\Lambda_1 \cup \Lambda_2}(\w)  \;=\;
      Z_{(\Lambda_1 \cup \Lambda_2) \setminus x}(\w)  \,+\,
      w_x Z_{(\Lambda_1 \cup \Lambda_2) \setminus x}(W(x,\cdot) \w)
      \;.
    \label{eq.cut_vertex}
\ee
But since the graphs $G_1 \setminus x$ and $G_2 \setminus x$
are disjoint\footnote{
   We recall that $G \setminus x$ denotes the graph obtained from $G$
   by deleting the vertex $x$ and all edges incident with it.
},
we can apply the factorization \reff{eq.factorization}
to the right-hand side of \reff{eq.cut_vertex}, yielding
\begin{subeqnarray}
      Z_{\Lambda_1 \cup \Lambda_2}(\w)  & = &
Z_{\Lambda_1 \setminus x}(\w) \, Z_{\Lambda_2 \setminus x}(\w)
         \;+\;
         w_x \, Z_{\Lambda_1 \setminus x}(W(x,\cdot) \w) \,
             Z_{\Lambda_2 \setminus x}(W(x,\cdot) \w)
    \nonumber \\ \slabel{eq.cut_vertex2.a} \\[2mm]
    & = &
Z_{\Lambda_1 \setminus x}(\w)
      \left[ Z_{\Lambda_2 \setminus x}(\w)  \,+\,
             w_x^{{\rm eff},G_1} Z_{\Lambda_2 \setminus x}(W(x,\cdot) \w)
      \right]
    \slabel{eq.cut_vertex2.b} \\[2mm]
    & = &
Z_{\Lambda_1 \setminus x}(\w)  \,
      Z_{\Lambda_2}(\w_{\neq x}, w_x^{{\rm eff},G_1})
    \slabel{eq.cut_vertex2.c}
    \label{eq.cut_vertex2}
\end{subeqnarray}
where we have defined
\be
      w_x^{{\rm eff},G_1}  \;=\;
      w_x {Z_{\Lambda_1 \setminus x}(W(x,\cdot) \w)
           \over
           Z_{\Lambda_1 \setminus x}(\w)
          }
      \;=\;
      w_x \scrk_{x,\Lambda_1}(\w)
      \;.
     \label{def.wxeff}
\ee
This can be interpreted as ``integrating out'' the variables
in $V(G_1) \setminus x$, leaving an ``effective fugacity''
$w_x^{{\rm eff},G_1}$ for the vertex $x \in V(G_2)$.

\bigskip

{\bf Example \thesection.4 revisited}.
Integrating out the leaves $y_1,\ldots,y_r$ of the star $K_{1,r}$,
we find from \reff{eq.ZW.star}:
\be
      w_x^{{\rm eff},K_{1,r}}  \;=\;
      w_x  \prod\limits_{i=1}^r {1+ W(x,y_i) w_{y_i}  \over  1+w_{y_i}}
      \;.
    \label{eq.wxeff.star}
\ee

\subsection{When $G$ is a tree \ldots}  \label{sec.tree_algorithm}

When the support graph $G = G_W$ is a tree,
every vertex of $G$ is either a cut vertex or a leaf,
so the partition function $Z_W(\w)$ can be calculated
by repeated use of the reduction formula
\reff{eq.cut_vertex2}/\reff{def.wxeff}.
This can be done in many ways, but the simplest is probably
to ``roll up'' the tree ``from the leaves up'', as follows:

We say that a vertex $x \in V(G)$ is a {\em near-leaf}\/ of $G$
if it is adjacent to at least one leaf of $G$.
If $x$ is a near-leaf of $G$,
then we can write $G = G_1 \cup G_2$
where $V(G_1) \cap V(G_2) = \{x\}$,
$G_1$ is a star with center $x$
consisting of $x$ and some or all of the leaves of $G$ adjacent to $x$,
and $G_2$ is a tree.
Integrating out $G_1 \setminus x$
using \reff{eq.wxeff.star}, we obtain
\be
      w_x^{{\rm eff},G_1}  \;=\;
      w_x  \prod\limits_{y \in V(G_1 \setminus x)}
           {1+ W(x,y) w_y  \over  1+w_y}
    \label{eq.wxeff.star_bis}
\ee
and of course
\be
      Z_{\Lambda_1 \setminus x}(\w)  \;=\;
      \prod\limits_{y \in V(G_1 \setminus x)}  (1+w_y)
      \;.
    \label{eq.star2}
\ee
Applying this process repeatedly,
we can obtain a linear-time algorithm for evaluating $Z_W(\w)$
at any point $\w \in \C^X$ whenever the support graph $G_W$ is a tree.
(It is also an algorithm for computing $Z_W(\w)$ as a polynomial in $\w$,
    but in this context it may no longer be linear-time,
    because of the need to multiply rational functions of $\w$.)
The simplest approach is to orient the tree by choosing arbitrarily
one vertex of the tree as the root, and then to work upwards
from the leaves.
We obtain the following algorithm:
\begin{quote}
      {\bf Algorithm T.}  Given numbers $\{w_x\}_{x \in X}$
      and $\{W(x,y)\}_{x,y \in X}$ for which the support graph
      $G = G_W$ is a tree, we compute $Z_W(\w)$ as follows:
\begin{itemize}
      \item[1)]
      Pick (arbitrarily) a root $x_0 \in X$.
      Define
\begin{eqnarray*}
       \hbox{\em depth}(x)   & = & {\rm dist}(x,x_0)              \\
       D    & = & \max\limits_{x \in X} \hbox{\em depth}(x)    \\
       \hbox{\em children}(x)   & = &
\{y \colon\, xy \in E(G)  \hbox{ and }
                \hbox{\em depth}(y) = \hbox{\em depth}(x) + 1 \}
\end{eqnarray*}
      \item[2)]
      For $d = D,D-1,\ldots,0$, do:
\begin{quote}
      For all vertices $x$ of depth $d$, set
\be
         w_x^{\rm eff}  \;\longleftarrow\;
         w_x  \! \prod\limits_{y \in \hbox{\scriptsize\em children}(x)}
              \! {1+ W(x,y) w_y^{\rm eff}  \over  1+w_y^{\rm eff}}
         \;.
         \hfill 
    \label{eq.alg.wxeff}
\ee
      [Note that all the needed $w_y^{\rm eff}$ have been set
       at the previous iteration.
       If $w_y^{\rm eff} = -1$ for any child $y$ of $x$,
       the algorithm is declared to fail.]
\end{quote}
      \item[3)]
      Output
\be
      Z_W(\w)  \;=\;  \prod\limits_{x \in V(G)} (1 + w_x^{\rm eff})
      \;.
         \hfill 
    \label{eq.alg.ZW}
\ee
\end{itemize}
\end{quote}
The correctness of this algorithm (when it succeeds) follows from
\reff{eq.cut_vertex2.b}, \reff{eq.cut_vertex2.c} and \reff{eq.star2}.
Indeed, for any vertex $x$, the $w_x^{\rm eff}$ produced by this algorithm
is $w_x^{{\rm eff},G_x}$ where $G_x$ is the subtree consisting of $x$
and all its descendants.

This also provides an algorithm for testing whether a given vector $\R$
lies in $\scrr(W)$:

\begin{theorem}
      \label{thm.algorithm.tree}
Consider any repulsive lattice gas with hard-core self-repulsion
for which the support graph $G_W$ is a tree.
Then a vector $\R \ge 0$ lies in $\scrr(W)$
if and only if Algorithm T with $\w = -\R$
produces $-1 < w_x^{\rm eff} \le 0$ for all vertices $x$.
\end{theorem}

\proof
If Algorithm T with $\w = -\R$
produces $-1 < w_x^{\rm eff} \le 0$ for all $x$,
then by the monotonicity of \reff{eq.alg.wxeff} in each $w_y^{\rm eff}$
it also does so when $\w = -\R'$ for any vector $0 \le \R' \le \R$.
By \reff{eq.alg.ZW} this means that $Z_W(-\R') > 0$
for all such $\R'$, hence that $\R \in \scrr(W)$.

Conversely, let $x$ be a vertex such that $w_x^{\rm eff} \notin (-1,0]$
but $w_y^{\rm eff} \in (-1,0]$ for all descendants $y$ of $x$.
Since $w_y^{\rm eff} \in (-1,0]$ for all children $y$ of $x$,
and $w_x = -R_x \le 0$, it follows from \reff{eq.alg.wxeff}
that $w_x^{\rm eff} \le 0$, so we must have $w_x^{\rm eff} \le -1$.
Now set $R_z'=R_z$ for $z$ in the subtree consisting of $x$ and
its descendants, and $R_z'=0$ otherwise.  Applying Algorithm T
to $\w'=-\R'$, it follows from \reff{eq.alg.ZW} that
$Z_W(\w')\leq 0$.  Thus $\R'\not\in\scrr(W)$; and since
$\scrr(W)$ is a down-set, we have $\R\not\in\scrr(W)$.
\qed

\bigskip

{\bf Example \thesection.6 revisited}.
Consider again the tree $T_n^{(r)}$.
Comparison of \reff{eq.Yn.tree} with \reff{eq.alg.wxeff}
shows that in the multivariate case $w_x^{\rm eff} = Y_n(w) - 1$
for all vertices $x$ of height $n$ above the leaves.
Then \reff{reconstruct.Zn.tree} is equivalent to \reff{eq.alg.ZW}.

\subsection{Upper bounds on $\scrr(W)$ when $G_W$ is a tree}  \label{sec3.homo}

We can use Theorem~\ref{thm.algorithm.tree}
to prove upper bounds on the set $\scrr(W)$
whenever the support graph $G_W$ is a tree.
Let us begin by considering the special case of a hard-core pair interaction,
i.e.\ the independent-set polynomial for a tree $G$ on the vertex set $X$.
As before, pick (arbitrarily) a root $x_0 \in X$,
and define $X_i = \{x \in X \colon\, \hbox{\em depth}(x) = i\}$
for $i=0,\ldots,D$.
Given any vector $\R \ge 0$,
define $\widetilde{R}_i$ to be the geometric mean of $R_x$
over all vertices $x$ of depth $i$:
\be\label{Rtildedef}
    \widetilde{R}_i  \;=\;
    \left( \prod\limits_{x \in X_i} R_x \right) ^{\! 1/|X_i|}
    \;.
\ee
Now apply Algorithm T to $\w = -\R$,
let $p_x = -w_x^{\rm eff}$, and define
\be\label{ptildedef}
    \widetilde{p}_i  \;=\;
    \left( \prod\limits_{x \in X_i} p_x \right) ^{\! 1/|X_i|}
\ee
(we assume here that $w_x^{\rm eff} \le 0$ for all $x \in X_i$).
By \reff{eq.alg.wxeff} we have
\be
    p_x  \;=\;  {R_x \over
                 \prod\limits_{y \in \hbox{\scriptsize\em children}(x)} (1-p_y)
                }
    \;,
\ee
so that
\be
    \widetilde{p}_i  \;=\;
    {\widetilde{R}_i  \over
     \Biggl( \prod\limits_{y \in X_{i+1}} (1-p_y) \Biggr)^{1/|X_i|}
    }
    \;.
\ee
Now
\begin{subeqnarray}
    \left( \prod\limits_{y \in X_{i+1}} (1-p_y) \right)^{1/|X_{i+1}|}
    & \le &
    1 \,-\, {1 \over |X_{i+1}|} \sum\limits_{y \in X_{i+1}} p_y
       \\[2mm]
    & \le &
    1 \,-\, \left( \prod\limits_{y \in X_{i+1}} p_y \right) ^{\! 1/|X_{i+1}|}
  \label{eq.arith-geom}
\end{subeqnarray}
by two applications of the arithmetic-geometric mean inequality.
It follows that
\be
     \widetilde{p}_i  \;\ge\;
    {\widetilde{R}_i  \over
     (1- \widetilde{p}_{i+1})^{|X_{i+1}|/|X_i|}
    }
    \;.
  \label{ineq.ptilde}
\ee
So, suppose we fix the numbers $(\widetilde{R}_i)_{i=0}^D$
and define $(\widehat{p}_i)_{i=0}^D$ by the recursion
\be\label{precursiondef}
    \widehat{p}_i  \;=\;
    {\widetilde{R}_i  \over
     (1- \widehat{p}_{i+1})^{|X_{i+1}|/|X_i|}
    }
\ee
with initial condition $\widehat{p}_{D+1} = 0$
(which is well-defined as long as $\widehat{p}_{i+1}$ remains $< 1$).
Then it follows immediately from the monotonicity of the
right-hand side of \reff{ineq.ptilde}
that $\widetilde{p}_i \ge \widehat{p}_i$
as long as the former is well-defined.
In particular, if for some level $i$ we have $\widehat{p}_i \ge 1$,
then $\widetilde{p}_i$ is either $\ge 1$ or else ill-defined,
so that there exists $x \in X_i$ with $w_x^{\rm eff} \notin (-1,0]$.
It then follows from Theorem~\ref{thm.algorithm.tree} that
$\R \notin \scrr(W)$.
Moreover, since this calculation depends only on
the numbers $(\widetilde{R}_i)_{i=0}^D$,
the same conclusion holds for {\em any}\/ $\R$ with the given
geometric means on levels.

Let us now generalize this argument to the case of a soft-core
pair interaction $0 \le W(x,y) \le 1$.
It is easily verified that $1-Wp \ge (1-p)^W$
whenever $0 \le p \le 1$ and $0 \le W \le 1$.
Therefore, defining $p_x = -w_x^{\rm eff}$ as before
and applying Algorithm T to $\w = -\R$, we have
\be
    p_x  \;\ge\;
    {R_x \over
     \prod\limits_{y \in \hbox{\scriptsize\em children}(x)} (1-p_y)^{1-W(x,y)}
    }
    \;.
\ee
Now let us assign to each vertex $x$ a weight
\be
     \alpha_x  \;=\; \prod_{j=1}^d [1-W(x_{j-1},x_j)]
  \label{eq.def.alpha}
\ee
where $x_0,x_1,\ldots,x_d \equiv x$ is the unique path in $G_W$
connecting $x$ to the root $x_0$.
[In particular, $\alpha_{x_0}=1$ and if $y$ is a child of $x$,
  then $\alpha_y = [1-W(x,y)] \alpha_x$.]
Setting
\be
    \widetilde{\alpha}_i  \;=\; \sum_{x \in X_i} \alpha_x
    \;,
  \label{eq.def.alphatilde}
\ee
let us define $\widetilde{R}_i$ and $\widetilde{p}_i$
to be the weighted geometric means
\begin{eqnarray}
    \widetilde{R}_i  & = &
        \left( \prod\limits_{x \in X_i} R_x^{\alpha_x} \right)
        ^{\! 1/\widetilde{\alpha}_i}
      \\[2mm]
    \widetilde{p}_i  & = &
        \left( \prod\limits_{x \in X_i} p_x^{\alpha_x} \right)
        ^{\! 1/\widetilde{\alpha}_i}
\end{eqnarray}
The preceding argument can then be repeated verbatim
(using the weighted arithmetic-geometric mean inequality),
yielding
\be
     \widetilde{p}_i  \;\ge\;
    {\widetilde{R}_i  \over
     (1- \widetilde{p}_{i+1})
       ^{\widetilde{\alpha}_{i+1}/\widetilde{\alpha}_i}
    }
    \;,
\ee
which can be analyzed analogously to \reff{ineq.ptilde}.
We have therefore proven:

\begin{proposition}
  \label{prop.tree.homo1}
Consider any repulsive lattice gas with hard-core self-repulsion
for which the support graph $G_W$ is a tree.
Define weights $(\alpha_x)_{x \in X}$
and $(\widetilde{\alpha}_i)_{i=0}^D$
by \reff{eq.def.alpha}/\reff{eq.def.alphatilde},
and define numbers $(\widehat{p}_i)_{i=0}^D$ by the recursion
\be
    \widehat{p}_i  \;=\;
    {\widetilde{R}_i  \over
     (1- \widehat{p}_{i+1})
       ^{\widetilde{\alpha}_{i+1}/\widetilde{\alpha}_i}
    }
\ee
(as long as $\widehat{p}_{i+1}$ remains $< 1$),
with initial condition $\widehat{p}_{D+1} = 0$.
Suppose that there exists an $i$ for which $\widehat{p}_i \ge 1$.
Then $\R \notin \scrr(W)$ for all vectors $\R \ge 0$ satisfying
$\prod\limits_{x \in X_i}  R_x^{\alpha_x/\widetilde{\alpha}_i}
  \ge \widetilde{R}_i$
for all $i$.
\end{proposition}


Just as Proposition~\ref{prop.tree.homo1}
``homogenizes'' each level of a tree,
we can also ``homogenize'' {\em between}\/ levels.
Let us again begin by considering the special case of a
hard-core pair interaction.
We first define branching factors $b_i = |X_{i+1}|/|X_i|$.
Choose any number $\bar{b} > 0$ and define
\be
 \label{def.gammai.sec3}
    \gamma_i  \;=\;  |X_i| / \bar{b}^i
    \;.
\ee
Fix an integer $k \ge 1$ (the number of levels to be averaged together)
and define
\be
    \gamma_{i,k}  \;=\;  \sum\limits_{j=i}^{i+k-1} \gamma_j
    \;.
  \label{def.gammabar}
\ee
Then define the weighted inter-level geometric means
\begin{eqnarray}
    \widetilde{p}_{i,k}  & = &  \left( \prod\limits_{j=i}^{i+k-1}
                                  \widetilde{p}_j^{\gamma_j}
                           \right) ^{\! 1/\gamma_{i,k}}
      \label{def.twotilde.p}     \\[2mm]
    \widetilde{R}_{i,k}  & = &  \left( \prod\limits_{j=i}^{i+k-1}
                                  \widetilde{R}_j^{\gamma_j}
                           \right) ^{\! 1/\gamma_{i,k}}
      \label{def.twotilde.R}
\end{eqnarray}
for $i=0,\ldots,D+1-k$,
and $\widetilde{p}_{i,k} = \widetilde{R}_{i,k} = 0$ for $i \ge D+2-k$.
It follows from \reff{ineq.ptilde} that
\begin{eqnarray}
     \widetilde{p}_{i,k}  & \ge &
    {\widetilde{R}_{i,k}  \over
     \prod\limits_{j=i}^{i+k-1}
        (1- \widetilde{p}_{j+1})^{b_j \gamma_j/\gamma_{i,k}}
    }
       \nonumber \\[2mm]
    & = &
    {\widetilde{R}_{i,k}  \over
     \prod\limits_{j=i}^{i+k-1}
        (1- \widetilde{p}_{j+1})^{\bar{b} \gamma_{j+1}/\gamma_{i,k}}
    }
    \;.
  \label{ineq.twotilde}
\end{eqnarray}
Now
\begin{eqnarray}
    \prod\limits_{j=i}^{i+k-1}
    (1- \widetilde{p}_{j+1})^{\gamma_{j+1}/\gamma_{i+1,k}}
    & \le &
    1 \,-\, \sum\limits_{j=i}^{i+k-1}  {\gamma_{j+1} \over \gamma_{i+1,k}}
                \, \widetilde{p}_{j+1}
       \nonumber \\[2mm]
    & \le &
    1 \,-\, \prod\limits_{j=i}^{i+k-1}
             \widetilde{p}_{j+1}^{\gamma_{j+1} / \gamma_{i+1,k}}
       \nonumber \\[2mm]
    & \equiv &  1 \,-\, \twotilde{p}_{i+1}
\end{eqnarray}
by two applications of the arithmetic-geometric mean inequality,
so that
\be
    \widetilde{p}_{i,k}  \;\ge\;
    {\widetilde{R}_{i,k}  \over
     (1 - \widetilde{p}_{i+1,k})^{\bar{b} \gamma_{i+1,k} /\gamma_{i,k}}
    },
  \label{ineq.twotilde.2}
\ee
for $i=0,\ldots,D+1-k$.
This holds true for an arbitrary choice of the numbers $\bar{b}$ and $k$.

In the special case in which the sequence $\{b_0,b_1,\ldots,b_{D-1}\}$
is periodic of period $k$ (where $D \ge k$),
we can choose $\bar{b}$ to be the geometric mean of the $b_i$
over one period:
\be\label{bchoice}
    \bar{b}  \;=\;  \left( \prod_{i=0}^{k-1} b_i \right) ^{\! 1/k}
    \;.
\ee
Then the sequence $\{\gamma_0,\gamma_1,\ldots,\gamma_D\}$
is also periodic of period $k$, so that
$\gamma_{0,k} = \gamma_{1,k} = \ldots = \gamma_{D+1-k,k}$,
and \reff{ineq.twotilde.2} simplifies to
\be
    \widetilde{p}_{i,k}  \;\ge\;
    {\widetilde{R}_{i,k}  \over
     (1 - \widetilde{p}_{i+1,k})^{\bar{b}}
    }
  \label{ineq.twotilde.3}
\ee
for $i=0,\ldots,D+1-k$
[note that for $i=D+1-k$ we have $\widetilde{p}_{i+1,k} = 0$,
  so the value of the exponent $\bar{b} \gamma_{i+1,k} /\gamma_{i,k}$
  is in this case irrelevant].

So we can argue as before:
suppose we fix the numbers $(\widetilde{R}_{i,k})_{i=0}^{D+1-k}$
and define $(\widehat{p}_{i,k})_{i=0}^{D+1-k}$ by the recursion
\be
    \widehat{p}_{i,k}  \;=\;
    {\widetilde{R}_{i,k}  \over
     (1- \widehat{p}_{i+1,k})^{\bar{b} \gamma_{i+1,k} /\gamma_{i,k}}
    }
\ee
(as long as $\widehat{p}_{i+1,k}$ remains $< 1$),
with initial condition $\widehat{p}_{D+2-k,k} = 0$.
Then it follows immediately from the monotonicity of the
right-hand side of \reff{ineq.twotilde.2}
that $\widetilde{p}_{i,k} \ge \widehat{p}_{i,k}$
as long as the former is well-defined.
In particular, if for some level $i$ we have $\widehat{p}_{i,k} \ge 1$,
then $\widetilde{p}_{i,k}$ is either $\ge 1$ or else ill-defined,
so that there exists $x \in \bigcup_{j=i}^{i+k-1} X_j$
with $w_x^{\rm eff} \notin (-1,0]$.
It then follows from Theorem~\ref{thm.algorithm.tree} that
$\R \notin \scrr(W)$.
Moreover, since this calculation depends only on
the numbers $(\widetilde{R}_{i,k})_{i=0}^{D+1-k}$,
the same conclusion holds for {\em any}\/ $\R$ with the given geometric means
on-and-between levels.

The same argument works with soft-core pair interaction
if we replace $|X_i|$ by $\widetilde{\alpha}_i$.
We have therefore proven:

\begin{proposition}
  \label{prop.tree.homo2}
Consider any repulsive lattice gas with hard-core self-repulsion
for which the support graph $G_W$ is a tree.
Define weights $(\alpha_x)_{x \in X}$
and $(\widetilde{\alpha}_i)_{i=0}^D$
by \reff{eq.def.alpha}/\reff{eq.def.alphatilde}.
Fix an integer $k \ge 1$ and a real number $\bar{b} > 0$,
and define $\gamma_i = \widetilde{\alpha}_i/\bar{b}^i$.
Then define $(\gamma_{i,k})_{i=0}^{D+1-k}$,
$(\widetilde{p}_{i,k})_{i=0}^{D+1-k}$ and
$(\widetilde{R}_{i,k})_{i=0}^{D+1-k}$
by \reff{def.gammabar}--\reff{def.twotilde.R}.
Finally, define numbers $(\widehat{p}_{i,k})_{i=0}^D$ by the recursion
\be
    \widehat{p}_{i,k}  \;=\;
    {\widetilde{R}_{i,k}  \over
     (1- \widehat{p}_{i+1,k})
       ^{\bar{b} \gamma_{i+1,k}/\gamma_{i,k}}
    }
\ee
(as long as $\widehat{p}_{i+1,k}$ remains $< 1$),
with initial condition $\widehat{p}_{D+2-k,k} = 0$.
Suppose that there exists an $i$ for which $\widehat{p}_{i,k} \ge 1$.
Then $\R \notin \scrr(W)$ for all vectors $\R \ge 0$ satisfying
$\prod\limits_{j=i}^{i+k-1}
  \left( \prod\limits_{x \in X_j}  R_x^{\alpha_x/\widetilde{\alpha}_j}
  \right) ^{\! \gamma_j/\gamma_{i,k}}
  \ge \widetilde{R}_{i,k}$
for all $i=0,\ldots,D+1-k$.
\end{proposition}

We will return to these ideas in Section~\ref{sec.infinite.trees}.

\section{The Lov\'asz local lemma}  \label{sec.LLL}

\subsection{Hard-core version}

Let $(A_x)_{x\in X}$ be a finite family of events
on some probability space,
and let $G$ be a graph with vertex set $X$.
We say that $G$ is a {\em dependency graph}\/ for the family
$(A_x)_{x\in X}$
if, for each $x\in X$, the event $A_x$ is independent from the
$\sigma$-algebra $\sigma(A_y \colon\; y \in X \setminus \Gamma^*(x))$.
[Here we have used the notation $\Gamma^*(x) = \Gamma(x) \cup \{x\}$,
where $\Gamma(x)$ is the set of vertices of $G$ adjacent to $x$.]
Note that this is much stronger than requiring merely that
$A_x$ be independent of each such $A_y$ separately.

A family of events typically has many possible
dependency graphs:
for instance, if $G$ is a dependency graph for events
$(A_x)_{x\in X}$,
then any graph obtained by adding edges to $G$
is also a dependency graph.
In particular, if the events $A_x$ are independent,
then any graph on $X$ is a dependency graph.
Nor must there be a unique minimal dependency graph.
Consider, for instance, the set of binary strings of
length $n$ with odd digit sum
(giving each such string equal probability),
and let $A_i$ be the event that the $i$th digit is 1.
Any graph without isolated vertices is a dependency
graph for this collection of events.

%
%
%

There is also a stronger notion of a dependency graph $G$
for a collection of events $(A_x)_{x\in X}$,
where we demand that if $Y$ and $Z$ are disjoint subsets of $X$
such that $G$ contains no edges between $Y$ and $Z$,
then the $\sigma$-algebras $\sigma(A_y \colon\, y\in Y)$ and
$\sigma(A_z \colon\, z\in Z)$ are independent.
In this case we shall refer to $G$ as a
{\em strong dependency graph}\/
for the events $(A_x)_{x\in X}$.\footnote{
    For instance, this situation arises in any statistical-mechanical model
    with variables living on the set $X$
    and pair interactions only on the edges of $G$,
    where each $A_x$ depends only on the variable at $x$.
}
Alternatively, the dependency-graph hypothesis
can be replaced by a {\em weaker}\/ hypothesis concerning
conditional probabilities,
as in the lopsided Lov\'asz local lemma
(Theorem~\ref{thm.LLL_ESversion}).
It will follow from Theorem~\ref{thm.LLL.1} below that
all three hypotheses lead to the same lower bounds on
${\mathbb P}(\bigcap_{x\in X}\overline A_x)$.

The various forms of the Lov\'asz local lemma
(e.g.\ Theorems~\ref{thm.LLL} and \ref{thm.LLL_ESversion})
provide a {\em sufficient}\/ (but not necessary) condition
to have ${\mathbb P}(\bigcap_{x\in X}\overline A_x) > 0$,
in terms of the existence of numbers $(r_x)_{x\in X}$
satisfying $p_x \le r_x\prod_{y\in\Gamma(x)}(1-r_y)$
for suitable probabilities or conditional probabilities
${\mathbf p} = (p_x)_{x \in X}$.
As discussed in the Introduction,
we shall approach the problem by dividing our
analysis into two parts.  First, in this section,
we examine a best-possible condition to have
${\mathbb P}(\bigcap_{x\in X}\overline A_x) > 0$,
in terms of the independent-set polynomial $Z_G(-{\mathbf p})$;
then, in the next section, we discuss
a {\em sufficient}\/ condition to have
$Z_G(-{\mathbf p}) > 0$, in terms of the existence of
such numbers $(r_x)_{x\in X}$ or generalizations thereof.

The following result is a development of
Shearer \cite[Theorem 1]{Shearer_85}.

\begin{theorem}
    \label{thm.LLL.1}
Let $(A_x)_{x\in X}$ be a family of events on some probability space,
and let $G$ be a graph with vertex set $X$.
Suppose that $(p_x)_{x\in X}$ are real numbers in $[0,1]$
such that, for each $x$ and each $Y \subseteq X \setminus \Gamma^*(x)$,
we have
\be
      {\mathbb P}(A_x|\bigcap_{y\in Y}\overline A_y)  \;\le\;  p_x
      \;.
    \label{eq.weak_dep_graph}
\ee
\begin{itemize}
      \item[(a)] If ${\mathbf p}\in \scrr(G)$, then
\be
      {\mathbb P}(\bigcap_{x\in X}\overline A_x)  \;\ge\;
       Z_G(-{\mathbf p})   \;>\;  0
\label{eq4.2}
\ee
and more generally
\be
      {\mathbb P}(\bigcap_{x\in Y}\overline A_x |
                  \bigcap_{x\in Z}\overline A_x)
      \;\ge\;
      {Z_G(-{\mathbf p} \, {\bf 1}_{Y \cup Z})
       \over
       Z_G(-{\mathbf p} \, {\bf 1}_Z)
      }
      \;>\;  0
   \label{eq4.2a}
\ee
for any subsets $Y,Z \subseteq X$.
Moreover, this lower bound is best possible in the sense that
there exists a probability space on which there can be constructed
a family of events $(B_x)_{x\in X}$ with probabilities
${\mathbb P}(B_x) = p_x$ and strong dependency graph $G$, such that
${\mathbb P}(\bigcap_{x\in X}\overline B_x) = Z_G(-{\mathbf p})$.
      \item[(b)]
If ${\mathbf p} \notin \scrr(G)$, then there exists a probability space
on which there can be constructed:
\begin{itemize}
      \item[(i)]  A family of events $(B_x)_{x\in X}$ with probabilities
         ${\mathbb P}(B_x) = p_x$ and strong dependency graph $G$,
         satisfying ${\mathbb P}(\bigcap_{x\in X}\overline B_x) = 0$; and
      \item[(ii)]  A family of events $(B'_x)_{x\in X}$ with probabilities
         ${\mathbb P}(B'_x) = p'_x \le p_x$ and strong dependency graph $G$,
         satisfying
         ${\mathbb P}(B'_x \cap B'_y) = 0$ for all $xy \in E(G)$
         and ${\mathbb P}(\bigcap_{x\in X}\overline B'_x) = 0$.
\end{itemize}
\end{itemize}
\end{theorem}

\medskip
\par\noindent
{\bf Remarks.}
1.  Please note that $G$ is here an {\em arbitrary}\/ graph
with vertex set $X$;
it need not be a dependency graph for the events $(A_x)_{x\in X}$.
Rather, given $G$, we can regard $\p$ as {\em defined}\/ by
\be
      p_x  \;=\;  \max\limits_{Y \subseteq X \setminus \Gamma^*(x)}
                  {\mathbb P}(A_x|\bigcap_{y\in Y}\overline A_y)
\ee
(this is clearly the minimal choice).
There is then a tradeoff in the choice of $G$:
adding more edges reduces $p_x$
(since there are fewer conditional probabilities to control)
but also shrinks the set $\scrr(G)$
(by the last sentence of Proposition~\ref{prop.scrrW.properties}).

2.  Though \reff{eq.weak_dep_graph} is the {\em weak}\/ hypothesis of the
lopsided Lov\'asz local lemma (Theorem~\ref{thm.LLL_ESversion}),
we will prove in (a) and (b) that the extremal families
$(B_x)_{x \in X}$ and $(B'_x)_{x \in X}$
have $G$ as a {\em strong}\/ dependency graph.
Therefore, all three dependency hypotheses
lead to the same optimal lower bound on
${\mathbb P}(\bigcap_{x\in X}\overline A_x)$.

3.  The proofs given here of Theorems~\ref{thm.LLL.1}
and \ref{thm.softshearer} are logically independent of
nearly all of Theorem~\ref{thm2.fund}.
More precisely, if we define $\scrr(G)$ and $\scrr(W)$
by condition (b) of Theorem~\ref{thm2.fund},
then the only part of Theorem~\ref{thm2.fund} that is used
in the proofs of Theorems~\ref{thm.LLL.1} and \ref{thm.softshearer}
is the (relatively easy) implication (b)$\implies$(f).

\medskip

\proof
For ${\mathbf p}\in \scrr(G)$, we wish to define a family of events
$(B_x)_{x\in X}$ [on a new probability space] such that the
hypotheses of the theorem are satisfied and
${\mathbb P}(\bigcap_{x\in X}\overline B_x)$
is as small as possible.  An intuitively reasonable way to do this
is to make the events $B_x$ as disjoint as possible,
consistent with the condition \reff{eq.weak_dep_graph}
[or with either of the two stronger notions of dependency graph].
With this in mind, for $\Lambda\subseteq X$ let us define
\be
      {\mathbb P}(\bigcap_{x\in\Lambda} B_x)  \;=\;
      \cases{ \prod\limits_{x\in \Lambda}p_x
                         & if $\Lambda$ is independent in $G$ \cr
              \noalign{\vskip 3mm}
              0          & otherwise \cr
            }
    \label{def.Bx_probs}
\ee
This defines a signed measure on the $\sigma$-algebra generated
by $(B_x)_{x\in X}$; indeed, inclusion-exclusion
gives
\begin{subeqnarray}
{\mathbb P}(\bigcap_{x\in\Lambda}B_x\cap\bigcap_{x\not\in\Lambda}
\overline B_x)
&=&\sum_{I\supseteq\Lambda} (-1)^{|I|-|\Lambda|} \,
       {\mathbb P}(\bigcap_{x\in I}B_x)\\
&=&\sum_{I\supseteq\Lambda,\, I \,{\rm independent}}
      (-1)^{|I|-|\Lambda|} \, \prod_{x\in I}p_x  \\
&=& (-1)^{|\Lambda|} \, Z_G(-{\mathbf p};\Lambda)   \;,
    \label{eq.incex}
    \slabel{eq.incexc}
\end{subeqnarray}
where $Z_G(-{\mathbf p};\Lambda)$ is defined as in \reff{zsemicol}.
In particular, taking $\Lambda = \emptyset$, we have
${\mathbb P}( \bigcap_{x \in X} \overline B_x) = Z_G(-{\mathbf p})$.
Theorem~\ref{thm2.fund}(f) implies that \reff{eq.incexc} is nonnegative for
all $\Lambda$, so that \reff{def.Bx_probs} defines a probability measure
on $\sigma(B_x \colon\, x\in X)$.
[This is the probability measure defined in Theorem~\ref{thm2.fund}(g).]

If $Y$ and $Z$ are disjoint subsets of $X$ such that $G$ contains no edges
between $Y$ and $Z$, it follows from \reff{def.Bx_probs} that for
$Y_0\subseteq Y$ and $Z_0\subseteq Z$ the events $\bigcap_{x\in Y_0}B_x$
and $\bigcap_{x\in Z_0}B_x$ are independent.  This implies
(see, for instance, \cite[Theorem 4.2]{Williams_91} or
  \cite[Theorem 4.2]{Billingsley_86})
that $\sigma(B_x \colon\, x\in Y)$ and $\sigma(B_x \colon\, x\in Z)$
are independent, and so $G$ is a strong dependency graph.

We next show that $(B_x)_{x\in X}$ is a family minimizing
${\mathbb P}(\bigcap_{x\in X}\overline B_x)$.
For $\Lambda\subseteq X$, we define
\begin{eqnarray}
      P_\Lambda={\mathbb P}(\bigcap_{x\in\Lambda}\overline A_x)
\label{eq4.6} \\[2mm]
      Q_\Lambda={\mathbb P}(\bigcap_{x\in\Lambda}\overline B_x).
\label{eq4.7}
\end{eqnarray}
Let us now prove by induction on $|\Lambda|$
that $P_\Lambda/Q_\Lambda$ is monotone increasing in $\Lambda$.
Note first that by inclusion-exclusion,
\begin{subeqnarray}
Q_\Lambda
&=&\sum_{I\subseteq\Lambda}  (-1)^{|I|} \,
      {\mathbb P}(\bigcap_{x\in I}B_x)\\
&=&\sum_{I\subseteq\Lambda,\, I \,{\rm independent}}
       (-1)^{|I|} \, \prod_{x \in I} p_x   \\
&=& Z_G(-{\mathbf p} \, {\bf 1}_\Lambda) \;.
\end{subeqnarray}
Thus $Q_\Lambda>0$ for all $\Lambda$, since $\p\in\scrr(G)$ and
$\scrr(G)$ is a down-set.
Furthermore, for $y\notin\Lambda$,
\begin{subeqnarray}
Q_{\Lambda\cup\{y\}}
&=&\sum_{I\subseteq\Lambda\cup\{y\},\, I \,{\rm independent}}
      (-1)^{|I|} \, \prod_{x\in I}p_x\\
&=&Q_\Lambda \,-\, p_y \sum_{I\subseteq\Lambda\setminus\Gamma(y),\,
     I \,{\rm independent}}  (-1)^{|I|} \, \prod_{x\in I}p_x \\
&=&Q_\Lambda \,-\, p_y Q_{\Lambda\setminus\Gamma(y)}   \;.
     \label{eq.lov.fund}
\end{subeqnarray}
[Note that this is just the fundamental identity \reff{eq.dob_basic_G}
applied to $Z_G(-{\mathbf p} \, {\bf 1}_\Lambda)$.]
On the other hand,
\begin{subeqnarray}
P_{\Lambda\cup\{y\}}
&=&P_\Lambda \,-\,
      {\mathbb P}(A_y\cap\bigcap_{x\in\Lambda}\overline A_x)  \\
&\ge& P_\Lambda \,-\,
      {\mathbb P}(A_y\cap\bigcap_{x\in\Lambda\setminus \Gamma(y)}
         \overline A_x)\\
&\ge& P_\Lambda \,-\, p_yP_{\Lambda\setminus\Gamma(y)}
     \label{eq.lov.Pineq}
\end{subeqnarray}
by the hypothesis \reff{eq.weak_dep_graph}.
Now we want to show that
$P_{\Lambda\cup\{y\}} / Q_{\Lambda\cup\{y\}}  \ge P_\Lambda / Q_\Lambda$,
or equivalently that
$P_{\Lambda\cup\{y\}} Q_\Lambda -  Q_{\Lambda\cup\{y\}} P_\Lambda  \ge 0$.
By \reff{eq.lov.fund} and \reff{eq.lov.Pineq} we have
\begin{subeqnarray}
      P_{\Lambda\cup\{y\}} Q_\Lambda -  Q_{\Lambda\cup\{y\}} P_\Lambda
      & \ge &
[P_\Lambda - p_y P_{\Lambda\setminus\Gamma(y)}] Q_\Lambda
      \,-\,
      [Q_\Lambda - p_y Q_{\Lambda\setminus\Gamma(y)}] P_\Lambda   \qquad 
\\[1mm]
      & = &
p_y \, [P_\Lambda Q_{\Lambda\setminus\Gamma(y)} -
              Q_\Lambda P_{\Lambda\setminus\Gamma(y)} ]           \\[1mm]
      & \ge & 0
\end{subeqnarray}
since
\be
\frac{P_\Lambda}{Q_\Lambda} \;\ge\;
      \frac{P_{\Lambda\setminus\Gamma(y)}}{Q_{\Lambda\setminus\Gamma(y)}}
\ee
by the inductive hypothesis.

Since $P_\Lambda/Q_\Lambda$ is monotone increasing in $\Lambda$,
we have $P_X/Q_X \ge P_\emptyset/Q_\emptyset = 1$,
which proves \reff{eq4.2}.
More generally, for any subsets $Y,Z \subseteq X$,
we have $P_{Y \cup Z}/Q_{Y \cup Z} \ge P_Z/Q_Z$
and hence $P_{Y \cup Z}/P_Z \ge Q_{Y \cup Z}/Q_Z$,
which gives \reff{eq4.2a}.

For ${\mathbf p}\not\in\scrr(G)$, choose a minimal vector
${\mathbf p'}\le\mathbf p$ such that ${\mathbf p'}\ge 0$
and $Z_G(-{\mathbf p'})=0$
[such a ${\mathbf p'}$ is in general nonunique].
Then the family of events $(B'_x)_{x\in X}$ defined by \reff{def.Bx_probs}
with $p_x$ replaced by $p'_x$ satisfies
${\mathbb P}(\bigcap_{x\in X}\overline B'_x)=Z_G(-{\mathbf p'})=0$
[by \reff{eq.incexc} with $\Lambda=\emptyset$].
Since $\mathbf p'$ is in the closure of $\scrr(G)$,
it follows by the minimality of $\p'$ and the continuity of $Z_G$
that this is a well-defined probability measure; note that if $x$ and $y$
are adjacent then ${\mathbb P}(B_x'\cap B_y')=0$ by \reff{def.Bx_probs}.
Thus we have constructed a collection of events
satisfying part (ii) of the Theorem.

To construct a collection of events satisfying part (i),
let $(C_x)_{x\in X}$ be an (independent) collection of independent events
satisfying
\be
     [1-{\mathbb P}(B'_x)] \, [1-{\mathbb P}(C_x)]  \;=\;  1-p_x  \;.
\ee
Then the events $B_x=B'_x\cup C_x$ satisfy ${\mathbb P}(B_x)=p_x$ and
${\mathbb P}(\bigcap\overline B_x) \le
    {\mathbb P}(\bigcap \overline B'_x)=0$.
\qed

\medskip
\noindent
{\bf Remarks.}
1. If $(A_x)_{x\in X}$ is a family of events satisfying
\reff{eq4.2} with equality, then we have $P_X=Q_X$ in the foregoing proof;
and since $P_\emptyset = Q_\emptyset = 1$,
the monotonicity of $P_\Lambda/Q_\Lambda$ implies that we have
$P_\Lambda=Q_\Lambda$ for every $\Lambda\subseteq X$.
Thus, if $(A_x)_{x\in X}$ is an extremal family,
the probabilities of all events in $\sigma(A_x \colon\, x\in X)$
are completely determined and are given by \reff{def.Bx_probs}/\reff{eq.incex}.

2. The Lov\'asz local lemma can be formulated more generally
for families of events with a dependency {\em digraph}\/:
each event $A_x$ is independent from the
$\sigma$-algebra $\sigma(A_y \colon\; y \in X \setminus \Gamma^*_+(x))$,
where $\Gamma^*_+(x) = \Gamma_+(x) \cup \{x\}$
and $\Gamma_+(x)$ is the out-neighborhood of $x$.
See e.g.\ \cite[Lemma 5.1.1]{AS} or \cite[Theorem 1.17]{Bollobas_01}.
It would be interesting to have a digraph analogue of
Theorem \ref{thm.LLL.1}, but we do not know how to do this.

3.  There are other probabilistic inequalities that are expressed in terms of
a dependency graph (see for instance Suen \cite{Suen_90} or
Janson \cite{Janson_98}); it would be interesting to know if any
of these have counterparts in statistical mechanics.  One obstacle
here is the need for a counterpart of Theorem \ref{thm.LLL.1}.
However, even without such a result, there may be scope for proving
further inequalities in the presence of weak dependency conditions
of the form discussed in the next subsection.

\subsection{Soft-core version}

Let us now consider how to extend Theorem \ref{thm.LLL.1} to the more
general case of a soft-core pair interaction,
i.e.\ to allow ``soft edges'' $xy$ of strength $1-W(x,y) \in [0,1]$.
The first step here is to replace
the hard-core dependency condition \reff{eq.weak_dep_graph}
by an appropriate soft-core version.

Let $W \colon\, X \times X \to [0,1]$ be symmetric and satisfy
$W(x,x) = 0$ for all $x \in X$;
and let $(A_x)_{x\in X}$ be a collection of events in some probability space.
For each $x\in X$, let $S_x$ be a random subset of $X$,
independent of the $\sigma$-algebra $\sigma(A_x \colon\, x \in X)$,
defined by the probabilities
\be
{\mathbb P}(y\in S_x) \;=\; W(x,y)
\ee
independently for each $y\in X$.
[Thus in the case of a hard-core pair interaction,
  we have $S_x=X\setminus \Gamma^*(x)$ with probability 1.]
Let $(p_x)_{x\in X}$ be real numbers in $[0,1]$.  We say that
$(A_x)_{x\in X}$ satisfies the {\em weak dependency conditions
with interaction $W$ and probabilities $(p_x)_{x\in X}$} if, for
each $x\in X$ and each $Y\subseteq X\setminus x$ we have
\be
{\mathbb E}
\left(
{\mathbb P} (A_x\cap\bigcap_{y\in Y\cap S_x} {\overline A_y})
\right)
   \;\le\;
p_x
{\mathbb E}
\left(
    {\mathbb P} (\bigcap_{y\in Y\cap S_x} {\overline A_y})
\right)
    \;.
\label{wdc}
\ee
[Note that in the special case of a hard-core pair interaction,
  we have $Y\cap S_x=Y \setminus \Gamma^*(x)$ with probability 1,
  so that \reff{wdc} reduces to \reff{eq.weak_dep_graph}.]
Of course, the reference here to a random subset $S_x$ can be replaced
by an explicit expression for the probabilities
${\mathbb P}(Y\cap S_x = Y')$,
so that \reff{wdc} is equivalent to
\begin{eqnarray}
   & &
   \sum\limits_{Y' \subseteq Y}
   \left( \prod\limits_{y \in Y'} W(x,y) \right)
   \left( \prod\limits_{y \in Y \setminus Y'} [1 - W(x,y)] \right)
   {\mathbb P}(A_x\cap\bigcap_{y\in Y'} {\overline A_y})
   \;\le\;
       \nonumber \\
   & & \qquad
   p_x \sum\limits_{Y' \subseteq Y}
   \left( \prod\limits_{y \in Y'} W(x,y) \right)
   \left( \prod\limits_{y \in Y \setminus Y'} [1 - W(x,y)] \right)
   {\mathbb P}(\bigcap_{y\in Y'} {\overline A_y})
    \;.
\label{wdc2}
\end{eqnarray}

We can now state a soft-core version of Theorem \ref{thm.LLL.1}:

\begin{theorem}\label{thm.softshearer}
Let $(A_x)_{x\in X}$ be a family of events in some probability space,
and let $W \colon\, X \times X \to [0,1]$ be symmetric and satisfy
$W(x,x) = 0$ for all $x \in X$.
Suppose that $(A_x)_{x\in X}$ satisfies the weak
dependency conditions \reff{wdc}/\reff{wdc2} with interaction $W$ and
probabilities $(p_x)_{x\in X}$.
\begin{itemize}
\item[(a)]  If $\p\in\scrr(W)$, then
\be
{\mathbb P}(\bigcap_{x\in X}{\overline A_x}) \;\ge\; Z_W(-\p) \;>\; 0
   \label{softeq4.2}
\ee
and more generally
\be
      {\mathbb P}(\bigcap_{x\in Y}\overline A_x |
                  \bigcap_{x\in Z}\overline A_x)
      \;\ge\;
      {Z_W(-{\mathbf p} \, {\bf 1}_{Y \cup Z})
       \over
       Z_W(-{\mathbf p} \, {\bf 1}_Z)
      }
      \;>\;  0
   \label{softeq4.2a}
\ee
for any subsets $Y,Z \subseteq X$.
Furthermore, this bound is best possible in the sense that
there exists a family $(B_x)_{x\in X}$ with probabilities
${\mathbb P}(B_x)=p_x$ that satisfies the weak dependency
conditions \reff{wdc}/\reff{wdc2} with interaction $W$ and probabilities
$(p_x)_{x\in X}$, has strong dependency graph $G_W$,
and has ${\mathbb P}(\bigcap_{x\in X} {\overline B_x})=Z_W(-\p)$.
\item[(b)]
If ${\mathbf p} \notin \scrr(W)$, then there exists a probability space
on which there can be constructed:
\begin{itemize}
      \item[(i)]  A family of events $(B_x)_{x\in X}$ with probabilities
         ${\mathbb P}(B_x) = p_x$ and satisfying the weak dependency conditions
	  \reff{wdc}/\reff{wdc2} with interaction $W$, such that
	  ${\mathbb P}(\bigcap_{x\in X}\overline B_x) = 0$; and
      \item[(ii)]  A family of events $(B'_x)_{x\in X}$ with probabilities
         ${\mathbb P}(B'_x) = p'_x \le p_x$ and
	  satisfying the weak dependency conditions
	  \reff{wdc}/\reff{wdc2} with interaction $W$,
	  such that
         ${\mathbb P}(B'_x \cap B'_y) =
          W(x,y){\mathbb P}(B'_x){\mathbb P}(B'_y)$
         for all $x,y$
         and ${\mathbb P}(\bigcap_{x\in X}\overline B'_x) = 0$.
\end{itemize}
\end{itemize}
\end{theorem}

\proof
For $\p\in\scrr(W)$, we define a family of events $(B_x)_{x\in X}$
[on a new probability space] by
\be\label{4.20}
{\mathbb P}(\bigcap_{x\in\Lambda}B_x)  \;=\;
    \left(\prod_{x\in \Lambda}p_x \right)
    \left(\prod_{\{x,y\}\subseteq\Lambda}W(x,y) \right)
\ee
[here the second product runs over all two-element subsets
  $\{x,y\} \subseteq X$ ($x \neq y$)].
As before, inclusion-exclusion gives
\begin{subeqnarray}\label{4.21}
{\mathbb P}(\bigcap_{x\in\Lambda}B_x\cap\bigcap_{x\not\in\Lambda}
{\overline B_x})
&=&\sum_{I\supseteq\Lambda}(-1)^{|I|-|\Lambda|}{\mathbb P}
(\bigcap_{x\in I}B_x)\\
&=&\sum_{I\supseteq\Lambda}(-1)^{|\Lambda|}\prod_{x\in I}(-p_x)
\prod_{\{x,y\}\subseteq I}W(x,y)\\
&=&(-1)^{|\Lambda|}Z_W(-\p;\Lambda)  \;.
\end{subeqnarray}
In particular, we have ${\mathbb P}(\bigcap_{x\in X}{\overline B_x})
=Z_W(-\p)>0$, and more generally for $\Lambda\subseteq X$ we have
\begin{eqnarray*}
{\mathbb P}(\bigcap_{x\in \Lambda}{\overline B_x})
&=&\sum_{I\subseteq \Lambda}
(-1)^{|I|}{\mathbb P}(\bigcap_{x\in I}{\overline B_x})\\
&=&\sum_{I\subseteq \Lambda}
(-1)^{|I|}\left(\prod_{x\in I}p_x\right)
\left(\prod_{\{x,y\}\subseteq \Lambda}W(x,y)\right)\\
&=&Z_W(-\p \, {\bf 1}_\Lambda)\\
&>&0.
\end{eqnarray*}

We define $P_{\Lambda}$ and $Q_{\Lambda}$ as in \reff{eq4.6}/\reff{eq4.7}.
Note that $Q_\Lambda=Z_W(-\p{\mathbf 1}_\Lambda)>0$, as
$Q_X=Z_W(-\p)>0$ and $\scrr(W)$ is a down-set.
Let us now prove by induction on $|\Lambda|$
that $P_\Lambda/Q_\Lambda$ is monotone increasing in $\Lambda$.
For $\Lambda\subseteq X$ and $y \in X \setminus \Lambda$,
we have
\begin{subeqnarray}
    Q_{\Lambda\cup\{y\}}  & = &
       Z_W \bigl(-{\mathbf p}  \, {\bf 1}_{\Lambda\cup\{y\}} \bigr)   \\
    & = &
       Z_W(-{\mathbf p} \, {\bf 1}_\Lambda) \,-\,
          p_y Z_W(-W(y,\cdot) {\mathbf p} \, {\bf 1}_\Lambda)   \\
    & = &
       Z_W(-{\mathbf p} \, {\bf 1}_\Lambda) \,-\,
          p_y
   \sum_{Y\subset\Lambda}
   \left( \prod\limits_{y \in Y} W(x,y) \right)
   \left( \prod\limits_{y \in \Lambda \setminus Y} [1 - W(x,y)] \right)
          Z_W(-{\mathbf p} \, {\bf 1}_Y)    \nonumber \\ \\
    & = &
       Q_\Lambda \,-\,
          p_y
   \sum_{Y\subset\Lambda}
   \left( \prod\limits_{y \in Y} W(x,y) \right)
   \left( \prod\limits_{y \in \Lambda \setminus Y} [1 - W(x,y)] \right)
          Q_Y
  \label{eq.lov.fund2}
\end{subeqnarray}
%
%
where we have used the fundamental identity
\reff{eq.dob_basic}/\reff{eq.dob_basic_expanded}.
On the other hand,
\begin{subeqnarray}
P_{\Lambda\cup\{y\}}
&=& P_\Lambda-
{\mathbb P}(A_y\cap\bigcap_\Lambda{\overline A_x}) \\
&=& P_\Lambda-\sum_{Y\subseteq\Lambda} \left(\prod_{x\in Y}W(x,y) \right)
\left(\prod_{x\in\Lambda\setminus Y}[1-W(x,y)] \right)
{\mathbb P}(A_y\cap\bigcap_{x \in \Lambda}{\overline A_x})
    \nonumber \\ \\
&\ge& P_\Lambda-\sum_{Y\subseteq\Lambda} \left(\prod_{x\in Y}W(x,y) \right)
\left( \prod_{x\in\Lambda\setminus Y} [1-W(x,y)] \right)
{\mathbb P}(A_y\cap\bigcap_{x \in Y}{\overline A_x})
    \nonumber \\ \\
&\ge& P_\Lambda-p_y\sum_{Y\subseteq\Lambda} \left( \prod_{x\in Y}W(x,y) \right)
\left( \prod_{x\in\Lambda\setminus Y} [1-W(x,y)] \right)
{\mathbb P}(\bigcap_{x \in Y}{\overline A_x})  \;.
    \nonumber \\
  \label{eq.Plam}
\end{subeqnarray}
where the last line uses the weak dependency condition \reff{wdc}/\reff{wdc2}.
We then have
\begin{subeqnarray}
\frac{P_\Lambda}{Q_\Lambda}Q_{\Lambda\cup\{y\}}
&=& P_\Lambda-p_y\sum_{Y\subseteq\Lambda} \left( \prod_{x\in Y}W(x,y) \right)
\left( \prod_{x\in\Lambda\setminus Y} [1-W(x,y)] \right)
\frac{P_\Lambda}{Q_\Lambda}Q_Y
    \nonumber \\ \\
&\le& P_\Lambda-p_y\sum_{Y\subseteq\Lambda} \left( \prod_{x\in Y}W(x,y) \right)
\left( \prod_{x\in\Lambda\setminus Y} [1-W(x,y)] \right) P_Y \\
&\le& P_{\Lambda\cup\{y\}}  \;,
\end{subeqnarray}
where the first inequality uses the inductive hypothesis
and the second inequality uses \reff{eq.Plam}.
Hence $P_{\Lambda\cup\{y\}}/Q_{\Lambda\cup\{y\}} \ge P_\Lambda/Q_\Lambda$
as claimed.

The bounds \reff{softeq4.2} and \reff{softeq4.2a}
follow as in Theorem \ref{thm.LLL.1}.

Part (b) follows as in Theorem \ref{thm.LLL.1}.
For $\p\notin\scrr(W)$ choose a minimal vector $\p'$ with
${\mathbf 0}\leq\p'\le\p$ and $Z_W(-\p')=0$.  The
events $(B_x')_{x\in X}$ defined by \reff{4.20} with
$p_x$ replaced by $p_x'$ satisfy
${\mathbb P}(\bigcap_{x\in X} \overline B_x')=0$ and
${\mathbb P}(B'_x\cap B'_y)=W(x,y){\mathbb P}(B'_x){\mathbb P}(B'_y)$
for all $x,y$.  A collection of events satisfying (i) is
constructed as before.
\qed

\noindent
{\bf Remark.} The uniqueness of a family of events satisfying
\reff{softeq4.2} with equality follows as before: since
$P_X/Q_X=P_\emptyset/Q_\emptyset=1$ and $P_\Lambda/Q_\Lambda$ is
increasing we have $P_\Lambda=Q_\Lambda$ for all $\Lambda\subseteq X$
and so the probabilities of events are given by \reff{4.21}.

\section{Sufficient conditions for $Z_W \neq 0$ in a polydisc}
      \label{sec.dob}

In this section we shall exhibit sufficient conditions
on a set of radii $\R = \{R_x\}_{x \in X}$
so that the partition function $Z_W(\w)$ is nonvanishing
in the closed polydisc $|\w| \le \R$.
We shall restrict attention to the case of a repulsive lattice gas
with hard-core self-repulsion,
i.e.\ $0 \le W(x,y) \le 1$ for all $x,y$
and $W(x,x) = 0$ for all $x$,
and we shall use the notation introduced in Section~\ref{sec3.1}.
Our main tool will be the fundamental identity \reff{eq.dob_basic},
applied inductively.
%

\subsection{Basic bound}  \label{sec.dob.1}

Our first (and most basic) bound is due to
Dobrushin \cite{Dobrushin_96a,Dobrushin_96b}
in the case of a hard-core interaction;
the generalization to a soft repulsive interaction
was proven recently by one of us \cite{Sokal_chromatic_bounds}.
The method of proof is, however, already implicit
(in more powerful form) in Shearer \cite[Theorem 2]{Shearer_85}.

\begin{theorem}[Dobrushin \protect\cite{Dobrushin_96a,Dobrushin_96b},
                   Sokal \protect\cite{Sokal_chromatic_bounds}]
     \label{thm.dobrushin}
Let $X$ be a finite set, and let $W$ satisfy
\begin{itemize}
      \item[(a)]  $0 \le W(x,y) \le 1$ for all $x,y \in X$
      \item[(b)]  $W(x,x) = 0$ for all $x \in X$
\end{itemize}
Let $\R = \{R_x\}_{x \in X} \ge 0$.
Suppose that there exist constants $\{K_x\}_{x \in X}$
satisfying $0 \le K_x < 1/R_x$ and
\be
      K_x  \;\ge\;
      \prod_{y \neq x}   {1 - W(x,y) K_y R_y \over  1 - K_y R_y}
    \label{eq.dob1}
\ee
for all $x \in X$.
Then, for each subset $\Lambda \subseteq X$,
$Z_\Lambda(\w)$ is nonvanishing in the closed polydisc
$\bar{D}_{\R} =
    \{ \w \in \C^X \colon\; |w_x| \le R_x \hbox{ for all } x \}$
and satisfies there
\be
     \left|  {\partial\log Z_\Lambda (\w) \over \partial w_x} \right|
     \;\le\;
     \cases{ {\displaystyle {K_x \over 1 - K_x |w_x|} }
                                 & for all $x \in \Lambda$  \cr
              \noalign{\vskip 4mm}
             0                   & for all $x \in X \setminus \Lambda$ \cr
           }
    \label{eq.dob2}
\ee
Moreover, if $\w, \w' \in \bar{D}_{\R}$
and $w'_x / w_x \in [0,+\infty]$ for each $x \in \Lambda$,
then
\be
     \left|  \log {Z_\Lambda(\w') \over Z_\Lambda(\w)}  \right|
     \;\le\;  \sum_{x \in \Lambda}
     \left|  \log {1 - K_x |w'_x|  \over  1 - K_x |w_x|}  \right|
    \label{eq.dob3}
\ee
where on the left-hand side we take the standard branch of the log,
i.e.\ $|\imag\log \cdots| \le \pi$.
\end{theorem}

\medskip
\par\noindent
{\bf Remark.}  It follows from \reff{eq.dob1} that $K_x \ge 1$
and hence that $R_x < 1$.

\proof
Note first that \reff{eq.dob2} for any given $\Lambda$
implies \reff{eq.dob3} for the same $\Lambda$, by integration.

The proof is by induction on the cardinality of $\Lambda$.
If $\Lambda = \emptyset$ the claims are trivial.
So let us assume that \reff{eq.dob2}
[and hence also \reff{eq.dob3}]
holds for all sets of cardinality $< n$,
and let a set $\Lambda$ of cardinality $n$ be given.
Let $x$ be any element of $\Lambda$.
Let us apply the fundamental identity \reff{eq.dob_basic},
and observe that $W(x,\cdot)\w \in \bar{D}_{\R}$ since $|W(x,y)| \le 1$.
Therefore, by the inductive hypothesis we have
$Z_{\Lambda \setminus x}(\w) \neq 0$ and
$Z_{\Lambda \setminus x}(W(x,\cdot)\w) \neq 0$;
and from \reff{eq.dob_basic} we have
\be
      {\partial \over \partial w_x} \log Z_\Lambda(\w)
      \;=\;
      {\scrk_{x,\Lambda}(\w)  \over  1 + \scrk_{x,\Lambda}(\w) w_x}
\ee
where
\be
      \scrk_{x,\Lambda}(\w)  \;=\;
         {Z_{\Lambda \setminus x}(W(x,\cdot)\w)  \over
          Z_{\Lambda \setminus x}(\w)}  \;.
\ee
Now by the inductive hypothesis \reff{eq.dob3} for $\Lambda \setminus x$,
and using the fact that $\w' = W(x,\cdot)\w$ satisfies
$w'_y/w_y = W(x,y) \ge 0$, we have
\be
      |\scrk_{x,\Lambda}(\w)|  \;\le\;
      \prod_{y \in \Lambda \setminus x}
          {1 - W(x,y) K_y |w_y| \over  1 - K_y |w_y|}
      \;\le\;
      \prod_{y \in X \setminus x}
          {1 - W(x,y) K_y |w_y| \over  1 - K_y |w_y|}
      \;,
\ee
which is $\le K_x$ by the hypothesis \reff{eq.dob1}.
This proves \reff{eq.dob2} for $\Lambda$, and hence completes the induction.
\qed

It is convenient to rewrite Theorem~\ref{thm.dobrushin}
in terms of the new variables $r_x = K_x R_x$:

\begin{corollary}
    \label{cor.dobrushin_1}
Let $X$ be a finite set, and let $W$ satisfy
$0 \le W(x,y) \le 1$ for all $x,y \in X$
and $W(x,x) = 0$ for all $x \in X$.
Suppose that there exist constants $0 \le r_x < 1$ satisfying
\be
      R_x  \;\le\;  r_x \prod_{y \neq x}  {1-r_y  \over  1 - W(x,y) r_y}
    \label{eq.dob1a}
\ee
for all $x \in X$.
Then, for all $\w$ satisfying $|\w| \le \R$,
the partition function $Z_W$ satisfies
\be
      |Z_W(\w)| \;\ge\; Z_W(-\R) \;\ge\; \prod_{x \in X} (1-r_x)  \;>\; 0
    \label{eq.dob3a}
\ee
and more generally
\be
      \left| {Z_W(\w \, {\bf 1}_{Y \cup Z})
              \over
              Z_W(\w \, {\bf 1}_Z)
             }
      \right|
      \;\ge\; \prod_{x \in Y} (1-r_x)  \;>\; 0  \;.
    \label{eq.dob3a_bis}
\ee

In particular, if we define the ``maximum weighted degree''
\be\label{weighted.degree}
    \Delta_W  \;=\;
    \max\limits_{x \in X} \sum\limits_{y \neq x} [1-W(x,y)]
\ee
and write
\begin{eqnarray}
    F(\Delta_W)  & = & {2 + \Delta_W - \sqrt{\Delta_W^2 + 4\Delta_W}
                        \over 2}
        \\[2mm]
    R(\Delta_W)  & = & F(\Delta_W) \, e^{-[1-F(\Delta_W)]}
  \label{def.Rdelta}
\end{eqnarray}
we have
\be
    |Z_W(\w)| \;\ge\; [1-F(\Delta_W)]^{|X|} > 0
\ee
whenever $|w_x| \le R(\Delta_W)$ for all $x \in X$.
\end{corollary}

\proof
Setting $r_x = K_x R_x$,
we find that \reff{eq.dob1} becomes \reff{eq.dob1a},
and \reff{eq.dob3} with $\Lambda=X$ and $\w' = 0$ becomes \reff{eq.dob3a}.

To obtain the last claim, note first that
\be
    {1-r \over 1-Wr}  \;=\;
    {1-r \over 1-r+(1-W)r}  \;=\;
    {1 \over 1+(1-W){r \over 1-r}}  \;\ge\;
    e^{-(1-W)r/(1-r)}
\ee
whenever $0 \le W \le 1$ and $0 \le r \le 1$.
Therefore, if we set $r_x = r$ for all $x \in X$,
we have
\be
    r_x \prod_{y \neq x}  {1-r_y  \over  1 - W(x,y) r_y}
    \;\ge\;
    r e^{-\Delta_W r/(1-r)}  \;.
  \label{eq.deltaW.star1}
\ee
We then choose $r$ to maximize the right-hand side of \reff{eq.deltaW.star1};
simple calculus yields $r = F(\Delta_W)$ and $\Delta_W r = (1-r)^2$,
so that the right-hand side of \reff{eq.deltaW.star1} is bounded below
by $R(\Delta_W)$.
It follows that if we define $R_x=R(\Delta_W)$ and $r_x=F(\Delta_W)$ for all
$x\in X$ then \reff{eq.dob1a} and so \reff{eq.dob3a} are satisfied.
\qed

\medskip
\par\noindent
{\bf Remarks.}  1. The radius $R(\Delta_W)$ behaves as
\be
    R(\Delta_W)  \;=\;
    \cases{ 1 - 2\Delta_W^{1/2} + {5 \over 2} \Delta_W + O(\Delta_W^{3/2})
               & as $\Delta_W \to 0$  \cr
            \noalign{\vskip 2mm}
            {1 \over e \Delta_W} \left[ 1 - {1 \over \Delta_W}
                 + {3 \over 2\Delta_W} + O(\Delta_W^{-3}) \right]
               & as $\Delta_W \to \infty$  \cr
          }
  \label{eq.Rdelta.1}
\ee
Example~\ref{sec3}.6 (the $r$-ary rooted tree)
shows that this bound is sharp (to leading order) as $\Delta_W \to \infty$.
At the other extreme, the $1 - {\rm const} \times \Delta_W^{1/2}$ behavior
at small $\Delta_W$ is also best possible,
since the two-site lattice gas with $W(x,x) = W(y,y) = 0$
and $W(x,y) = 1-\epsilon$ has $Z_W(w) = 1 + 2w + (1-\epsilon) w^2$
and hence has a root at $w = - 1/(1 + \sqrt{\epsilon})$.
[However, the coefficient 2 rather than 1 in the $\Delta_W^{1/2}$ term
of \reff{eq.Rdelta.1} may not be best possible.]

2. From Proposition~\ref{prop.scrrW.properties}(c) we know that
the set $\scrr(W)$ is log-convex;  it is therefore natural to ask
whether the subset of $\scrr(W)$ produced by \reff{eq.dob1a}
is also log-convex.
(If it weren't, then we could improve Corollary~\ref{cor.dobrushin_1}
by taking the log-convex hull.)
It turns out that the set of vectors satisfying
\reff{eq.dob1a} is indeed always log-convex:
in fact, if $(\r,\R)$ and $(\r',\R')$ are two pairs of vectors
satisfying \reff{eq.dob1a}, and $0 \le \lambda \le 1$,
then $(\r^\lambda \r^{\prime 1-\lambda}, \R^\lambda \R^{\prime 1-\lambda})$
also satisfies \reff{eq.dob1a}.
In the hard-core case this follows from the inequality
$1 - r_y^\lambda (r'_y)^{1-\lambda} \ge (1-r_y)^\lambda (1-r'_y)^{1-\lambda}$,
which is proven by two applications of the weighted arithmetic-geometric
mean inequality as in \reff{eq.arith-geom}.
In the general case it can be shown by a similar argument
(using four applications of the weighted arithmetic-geometric
mean inequality!).

\bigskip

Specializing Corollary~\ref{cor.dobrushin_1}
to the case of a hard-core pair interaction for a graph $G$,
\be
      W(x,y)  \;=\;  \cases{ 0   & if $x=y$ or $xy \in E(G)$  \cr
                             1   & if $x \neq y$ and $xy \notin E(G)$ \cr
                           }
\ee
we have:

\begin{corollary}
    \label{cor.dobrushin_2}
Let $G$ be a finite graph with vertex set $X$,
and let $\R = \{R_x\} _{x \in X} \ge 0$.
Suppose that there exist constants $0 \le r_x < 1$ satisfying
\be
      R_x  \;\le\;  r_x \prod_{y \in \Gamma(x)} (1-r_y)
    \label{eq.dob1b}
\ee
for all $x \in X$.
Then, for all $\w$ satisfying $|\w| \le \R$,
the independent-set polynomial $Z_G$ satisfies
\be
      |Z_G(\w)| \;\ge\; Z_G(-\R) \;\ge\; \prod_{x \in X} (1-r_x)  \;>\; 0
    \label{eq.dob3b}
\ee
and more generally
\be
      \left| {Z_G(\w \, {\bf 1}_{Y \cup Z})
              \over
              Z_G(\w \, {\bf 1}_Z)
             }
      \right|
      \;\ge\; \prod_{x \in Y} (1-r_x)  \;>\; 0  \;.
    \label{eq.dob3b_bis}
\ee

In particular, if $G$ has maximum degree $\Delta$,
then $|Z_G(\w)| \ge [\Delta/(\Delta+1)]^{|X|} > 0$ whenever
$|w_x| \le \Delta^\Delta/(\Delta+1)^{\Delta+1}$ for all $x \in X$.
\end{corollary}

\proof
The last claim is obtained by setting $r_x = 1/(\Delta+1)$
for all $x \in X$.
\qed

\medskip
\par\noindent
{\bf Remark.}  The radius $\Delta^\Delta/(\Delta+1)^{\Delta+1}$
behaves for large $\Delta$ as
\be
    {\Delta^\Delta \over (\Delta+1)^{\Delta+1}}
    \;=\;
    {1 \over e\Delta} \left[ 1 - {1 \over 2\Delta} + {7 \over 24\Delta^2}
                 - {3 \over 16\Delta^3} + O(\Delta^{-4}) \right]
    \;,
  \label{eq.Rdelta.2}
\ee
which agrees with \reff{eq.Rdelta.1} to leading order in $1/\Delta$
but is slightly larger (hence better) at order $1/\Delta^2$.

\bigskip

Combining Corollary~\ref{cor.dobrushin_2} with Theorem~\ref{thm.LLL.1},
we immediately obtain the lopsided Lov\'asz local lemma
(Theorem~\ref{thm.LLL_ESversion}).
Combining Corollary~\ref{cor.dobrushin_1} with Theorem~\ref{thm.softshearer},
we obtain a ``soft-core'' version of the lopsided Lov\'asz local lemma:

\begin{theorem}\label{softlopsided}
Let $(A_x)_{x\in X}$ be a family of events in some probability space, and let
$W \colon\, X\times X\to[0,1]$ be symmetric
and satisfy $W(x,x)=0$ for all $x \in X$.
Suppose that $(A_x)_{x\in X}$ satisfies the weak
dependency conditions \reff{wdc}/\reff{wdc2} with interaction $W$ and
probabilities $(p_x)_{x\in X}$.
Suppose further that $(r_x)_{x\in X}$ are real numbers in $[0,1)$ satisfying
\be
      p_x  \;\le\;  r_x \prod_{y \in \Gamma(x)} (1-r_y)   \;.
\ee
Then
\be\label{softbound1}
  {\mathbb P}(\bigcap_{x\in X}{\overline A_x})
  \;\ge\; \prod_{x\in X}(1-r_x) \;>\; 0  \;,
\ee
and more generally for sets $Y,Z\subseteq X$, we have
\be\label{softbound2}
  {\mathbb P}(\bigcap_{x\in Y}\overline A_x |
                  \bigcap_{x\in Z}\overline A_x)
      \;\ge\;
  \prod_{x\in Y\setminus Z}(1-r_x) \;=\; 0  \;.
\ee
\end{theorem}

Defining the weighted degree $\Delta_W$ as in \reff{weighted.degree},
we obtain the following:

\begin{lemma}\label{softweighted}
Let $(A_x)_{x\in X}$ satisfy the weak dependency conditions
\reff{wdc}/\reff{wdc2} with interaction $W$ and
probabilities $(p_x)_{x\in X}$.
If $p_x < \Delta_W^{\Delta_W}/(\Delta_W+1)^{\Delta_W+1}$
for every $x\in X$, then ${\mathbb P}(\bigcap_{x\in X}\overline{A}_x)>0$.
\end{lemma}

\proof
As in the proof of Corollary \ref{cor.dobrushin_2},
set $r_x = r \equiv 1/(\Delta_W+1)$ for all $x\in X$.
Then check \reff{eq.dob1a}:
\begin{eqnarray}
r_x\prod_{y\ne x}\frac{1-r_y}{1-W(x,y)r_y}
&\le&r_x\prod_{y\ne x}(1-r_y)^{1-W(x,y)}  \nonumber \\
&\leq&r(1-r)^{\Delta_W}  \nonumber \\
&\le&\frac{\Delta_W^{\Delta_W}}{(\Delta_W+1)^{\Delta_W+1}}.
\end{eqnarray}
In the first inequality
we have used the fact that $1-W(x,y)r_y\le(1-r_y)^{W(x,y)}$ for
$0\le W(x,y)\le1$.
\qed

It would
be interesting to see applications of Theorem \ref{softlopsided}
and Lemma \ref{softweighted}.

\subsection{Improved bound}  \label{sec.dob.2}

Let us now attempt to improve the bound of Theorem~\ref{thm.dobrushin}.
Note, first of all, that we need not insist that
the bound \reff{eq.dob2} hold with the {\em same}\/ constant $K_x$
for all $\Lambda \ni x$;
rather, we can use constants $K_{x,\Lambda}$ that depend on $\Lambda$.
Inspection of the inductive argument shows that we can {\em define}\/
the constants $K_{x,\Lambda} \in [0,+\infty]$
as a function of the family $\{R_x\}$ by the recursion
\be
      K_{x,\Lambda}  \;=\;
      \cases{
         \!\!\! \prod\limits_{\begin{scarray}
                   y \in \Lambda \setminus x \\
                   W(x,y) \neq 1 \\
                   R_y > 0
                \end{scarray}}
             \!\!\!
             {\displaystyle
             {1 - W(x,y) K_{y,\Lambda \setminus x} R_y
              \over
              1 - K_{y,\Lambda \setminus x} R_y}
             }
             \quad
         & if $K_{y,\Lambda \setminus x} R_y < 1$ for all terms in the product
           \cr
         \noalign{\vskip 4mm}
         +\infty  & otherwise  \cr
      }
     \label{def.Kxlambda}
\ee
[Note that $K_{x,\{x\}} = 1$ for all $x$,
    because the product \reff{def.Kxlambda} is empty.]
%
It follows immediately by induction that
each $K_{x,\Lambda}$ is an increasing rational function of
$\{R_y\}_{y \in \Lambda \setminus x}$ up to the first pole,
and $+\infty$ thereafter.
More precisely, suppose we define the rational functions
$\widehat{K}_{x,\Lambda}(\R)$
by the recursion \reff{def.Kxlambda}
{\em without}\/ the restrictions $K_{y,\Lambda \setminus x} R_y < 1$;
then it is easy to see that
\be
      K_{x,\Lambda}(\R)  \;=\;
      \cases{ \widehat{K}_{x,\Lambda}(\R)  &
if $\widehat{K}_{x,\Lambda}(\R') < \infty$
                 for $0 \le \R' \le \R$  \cr
              \noalign{\vskip 2pt}
              +\infty  & otherwise  \cr
            }
    \label{Kxlambda.rational}
\ee
Moreover, it is easily proven by induction that
all the partial derivatives of $\widehat{K}_{x,\Lambda}$
are nonnegative at $\R = 0$;
it then follows from Proposition~\ref{prop.pringsheim}
that the Taylor series of $\widehat{K}_{x,\Lambda}(\R)$
about $\mathbf 0$
converges throughout the region where the first case
in \reff{Kxlambda.rational} holds,
and that all the partial derivatives of
$K_{x,\Lambda} = \widehat{K}_{x,\Lambda}$ are nonnegative there.
Finally, it is obvious from
the definition \reff{def.Kxlambda} ff.\ that
$K_{x,\Lambda}$ is an increasing function of $\Lambda$.

Let us now define a graph $G$ with vertex set
$V = \{x \in X \colon\, R_x > 0\}$
and edge set $E = \{x,y \in V \colon\, W(x,y) \neq 1\}$;
and for each $\Lambda \subseteq X$,
let $G_\Lambda$ be the subgraph of $G$
induced by $\Lambda \cap V$.
Then only the connected component of $G_\Lambda$ containing $x$
plays any role in the definition of $K_{x,\Lambda}$:
that is, if $G_\Lambda$ has several connected components
with vertex sets $\Lambda_1,\ldots,\Lambda_k$ and $x \in \Lambda_i$,
then $K_{x,\Lambda} = K_{x,\Lambda_i}$.

Let us now call a pair $(x,\Lambda)$ ``good''
if $K_{x,\Lambda} < \infty$ and $K_{x,\Lambda} R_x < 1$.
It follows immediately from the definition
\reff{def.Kxlambda} ff.\ that if $(x,\Lambda)$ is good,
then $(y,\Lambda\setminus x)$ is also good
whenever $y \in \Lambda\setminus x$ with
$W(x,y) \neq 1$ and $R_y > 0$,
i.e.\ whenever $y$ is a neighbor of $x$ in $G_\Lambda$.
[Indeed, this follows under the weaker hypothesis that
    $K_{x,\Lambda} < \infty$.]

We then have:


\begin{theorem}[improved Dobrushin--Shearer bound]
     \label{thm.Kxlambda}
Let $X$ be a finite set, and let $W$ satisfy
\begin{itemize}
      \item[(a)]  $0 \le W(x,y) \le 1$ for all $x,y \in X$
      \item[(b)]  $W(x,x) = 0$ for all $x \in X$
\end{itemize}
Let $\R = \{R_x\}_{x \in X} \ge 0$.
Define the constants $K_{x,\Lambda} \in [0,+\infty]$
as above.
Suppose that in each connected component of $G_\Lambda$
there exists at least one vertex $x$
for which the pair $(x,\Lambda)$ is good.
Then $Z_\Lambda(\w)$ is nonvanishing in the closed polydisc
$\bar{D}_{\R}$;
and for every good pair $(x,\Lambda)$ and
every $\w \in \bar{D}_{\R}$, we have
\be
     \left|  {\partial\log Z_\Lambda (\w) \over \partial w_x} \right|
     \;\le\;
     {K_{x,\Lambda} \over 1 - K_{x,\Lambda} |w_x|}
     \;.
    \label{eq.dob2x}
\ee
Moreover, if $\w, \w' \in \bar{D}_{\R}$
and $w'_x / w_x \in [0,+\infty]$ for each $x \in \Lambda$,
and in addition the pair $(x,\Lambda)$ is good
whenever $w'_x \neq w_x$, then
\be
     \left|  \log {Z_\Lambda(\w') \over Z_\Lambda(\w)}  \right|
     \;\le\;  \sum_{\begin{scarray}
                       x \in \Lambda \\
                       w'_x \neq w_x
                    \end{scarray}}
     \left|  \log {1 - K_{x,\Lambda} |w'_x|
                   \over
                   1 - K_{x,\Lambda} |w_x|}  \right|,
    \label{eq.dob3x}
\ee
where on the left-hand side we take the standard branch of the log,
i.e.\ $|\imag\log \cdots| \le \pi$.
\end{theorem}

\proof
Note first that \reff{eq.dob2x} for any given $\Lambda$
implies \reff{eq.dob3x} for the same $\Lambda$, by integration.

The proof is by induction on the cardinality of $\Lambda$.
If $\Lambda = \emptyset$ the claims are trivial.
So let us assume that \reff{eq.dob2x}
[and hence also \reff{eq.dob3x}]
holds for all sets of cardinality $< n$
satisfying the stated hypotheses,
and let a set $\Lambda$ of cardinality $n$
satisfying these hypotheses be given.
If $G_\Lambda$ is not connected,
i.e.\ has components with vertex sets $\Lambda_1,\ldots,\Lambda_k$
($k \ge 2$), then $Z_\Lambda(\w)$ factorizes as
\be
      Z_\Lambda(\w)  \;=\;  \prod_{i=1}^k Z_{\Lambda_i}(\w)
      \;;
\ee
moreover, each connected component has cardinality $<n$
and satisfies the hypotheses of the theorem,
so it follows immediately from the inductive hypothesis
that \reff{eq.dob2x} and \reff{eq.dob3x} hold also for $\Lambda$.
We may therefore assume that $G_\Lambda$ is connected.

Let $x$ be any element of $\Lambda$ for which the pair
$(x,\Lambda)$ is good.
Since $G_\Lambda$ is connected,
the vertex $x$ has at least one neighbor $y$
in each connected component of $G_{\Lambda \setminus x}$.
Moreover, as noted previously,
the pair $(y,\Lambda \setminus x)$ is good
whenever $y$ is a neighbor of $x$ in $G_\Lambda$.
Therefore, the inductive hypothesis is applicable to
$\Lambda \setminus x$.
Let us now apply the fundamental identity \reff{eq.dob_basic},
and observe that $W(x,\cdot)\w \in \bar{D}_{\R}$
since $|W(x,y)| \le 1$.
By the inductive hypothesis we have
$Z_{\Lambda \setminus x}(\w) \neq 0$ and
$Z_{\Lambda \setminus x}(W(x,\cdot)\w) \neq 0$;
and from \reff{eq.dob_basic} we have
\be
      {\partial \over \partial w_x} \log Z_\Lambda(\w)
      \;=\;
      {\scrk_{x,\Lambda}(\w)  \over  1 + \scrk_{x,\Lambda}(\w) w_x}
\ee
where
\be
      \scrk_{x,\Lambda}(\w)  \;=\;
         {Z_{\Lambda \setminus x}(W(x,\cdot)\w)  \over
          Z_{\Lambda \setminus x}(\w)}  \;.
\ee
Now each $y \in \Lambda \setminus x$ with $w'_y \neq w_y$
necessarily has $W(x,y) \neq 1$ and $R_y > 0$
(i.e.\ is a neighbor of $x$ in $G_\Lambda$),
so that the pair $(y,\Lambda \setminus x)$ is good.
We may therefore apply the inductive hypothesis \reff{eq.dob3x};
using the fact that $\w' = W(x,\cdot)\w$ satisfies
$w'_y/w_y = W(x,y) \ge 0$, we have
\be
      |\scrk_{x,\Lambda}(\w)|  \;\le\;
      \prod_{\begin{scarray}
                y \in \Lambda \setminus x \\
                W(x,y) \neq 1 \\
                R_y > 0
             \end{scarray}}
          {1 - W(x,y) K_{y,\Lambda \setminus x} |w_y|
           \over
           1 - K_{y,\Lambda \setminus x} |w_y|}
      \;\le\;
      K_{x,\Lambda}
    \label{eq.thm.Kxlambda.laststep}
\ee
since $|w_y| \le R_y$.
This proves \reff{eq.dob2x} for $(x,\Lambda)$,
and hence completes the induction.
\qed

\medskip
\par\noindent
{\bf Remark.}  It is unclear whether the set of vectors $\R$
satisfying the hypotheses of Theorem~\ref{thm.Kxlambda} is log-convex.
If it is not, then the conclusion of Theorem~\ref{thm.Kxlambda}
can be improved by taking the log-convex hull.

\bigskip

As a corollary of Theorem~\ref{thm.Kxlambda},
we can deduce a bound due originally (in the Lov\'asz context)
to Shearer \cite[Theorem 2]{Shearer_85},
which improves the last sentence of Corollary~\ref{cor.dobrushin_2}
by replacing $\Delta$ by $\Delta-1$.
Indeed, we can very slightly improve Shearer's bound
by allowing {\em one}\/ vertex $x_0$
to have a larger radius $R_{x_0}$:

\begin{corollary}
    \label{cor.shearer}
Let $G=(X,E)$ be a finite graph of maximum degree $\Delta \ge 2$,
and fix one vertex $x_0 \in X$.
Suppose that
$|w_{x_0}| \le (\Delta-1)^\Delta/\Delta^\Delta$
and that
$|w_x| \le (\Delta-1)^{\Delta-1}/\Delta^\Delta$ for all $x \neq x_0$.
Then $Z_G(\w) \neq 0$.
\end{corollary}

\proof
Since $Z_G$ factorizes over connected components,
we can assume without loss of generality that $G$ is connected.
(Indeed, if $G$ is disconnected, then
    we can allow one ``$x_0$-like'' vertex in {\em each}\/
    connected component.)
Set $R_{x_0} = (\Delta-1)^\Delta/\Delta^\Delta$
and $R_x = (\Delta-1)^{\Delta-1}/\Delta^\Delta$ for all $x \neq x_0$.

We first claim that if $x_0 \notin \Lambda$,
and $x \in \Lambda$ is a vertex with
at least one neighbor in $X \setminus \Lambda$,
then
\be
      K_{x,\Lambda} \;<\;
      \left( {\Delta \over \Delta-1} \right) ^{\! \Delta-1}
    \label{eq.delta-1}
\ee
(note the strict inequality).
The proof is by induction on $|\Lambda|$,
using the definition \reff{def.Kxlambda}:
it certainly holds if $\Lambda = \{x\}$;
since every $y$ appearing in the product
on the right-hand side of \reff{def.Kxlambda}
has at least one neighbor outside of $\Lambda \setminus x$
(namely, $x$ itself), $K_{y,\Lambda\setminus x}$ satisfies
\reff{eq.delta-1} by the inductive hypothesis
and so $K_{y,\Lambda\setminus x}R_y<1/\Delta$;
and finally, since $x$ has at least one neighbor outside $\Lambda$,
there are at most $\Delta-1$ factors in the product.
Thus
$$K_{x,\Lambda}<\left(\frac{1}{1-1/\Delta}\right)^{\Delta-1}
=\left(\frac{\Delta}{\Delta-1}\right)^{\Delta-1}.$$

It then follows that
\be
      K_{x_0,X} \;<\;
      \left( {\Delta \over \Delta-1} \right) ^{\! \Delta}
      \;,
\ee
since the bound \reff{eq.delta-1}
applies to all the terms $K_{y,\Lambda\setminus x_0}$
appearing on the right-hand side of \reff{def.Kxlambda}.
We therefore have $K_{x_0,X} R_{x_0} < 1$,
and so the pair $(x_0,X)$ is good.
The claim then follows from Theorem~\ref{thm.Kxlambda}.
\qed

Replacing $\Delta^\Delta/(\Delta+1)^{\Delta+1}$
by $(\Delta-1)^{\Delta-1}/\Delta^\Delta$
may seem to be a negligible improvement,
since both quantities have the same leading behavior
$\approx 1/(e\Delta)$ as $\Delta\to\infty$,
and differ only at higher order:
\be
    {(\Delta-1)^{\Delta-1} \over \Delta^\Delta}
    \;=\;
    {1 \over e\Delta} \left[ 1 + {1 \over 2\Delta} + {7 \over 24\Delta^2}
                 + {3 \over 16\Delta^3} + O(\Delta^{-4}) \right]
  \label{eq.Rdelta.3}
\ee
[cf.\ \reff{eq.Rdelta.2}].\footnote{
    The amusing similarity of \reff{eq.Rdelta.2} and \reff{eq.Rdelta.3}
    arises from the fact that
    $-(-\Delta)^{-\Delta}/(-\Delta+1)^{-\Delta+1}
     = (\Delta-1)^{\Delta-1}/\Delta^\Delta$.
}
But Shearer's bound $(\Delta-1)^{\Delta-1}/\Delta^\Delta$
has the great merit of being {\em best possible}\/:
for, as he showed \cite{Shearer_85},
if $G$ is the complete rooted tree
with branching factor $r = \Delta -1$ and depth $n$,
then $Z_G(w)$ has negative real zeros that tend to
$w = - (\Delta-1)^{\Delta-1}/\Delta^\Delta$ as $n \to\infty$
(see Example~\ref{sec3}.6 above).

We remark that Corollary~\ref{cor.shearer} does not appear to extend naturally
to the soft-core case (note that having one neighbor outside $\Lambda$ in the
argument around \reff{eq.delta-1} need not reduce the weighted degree of
a vertex in $\Lambda$ by 1).

\subsection{Optimal bound}  \label{sec.dob.3}

Now let us try to further improve Theorem~\ref{thm.Kxlambda}.
Where in the proof did we lose equality?
This happened only in \reff{eq.thm.Kxlambda.laststep},
where we applied \reff{eq.dob3x}, which in turn arose
from integrating \reff{eq.dob2x}:
the point is that the bound \reff{eq.dob2x}
in general improves as we pass from $\w$ downwards to $\w' = W(x,\cdot)\w$,
but we failed to take advantage of this fact.
The solution is to order (arbitrarily) the vertices
$y_1,\ldots,y_k$ arising in the product \reff{eq.thm.Kxlambda.laststep}
and to write $\scrk_{x,\Lambda}(\w)$ as a telescoping product;
then we will have a sharp bound,
i.e.\ one that becomes equality when $\w = -\R$.

Let us show this first in the special case of a hard-core pair interaction:
setting $\Gamma(x) \cap \Lambda = \{y_1,\ldots,y_k\}$,
we have
\be
      \scrk_{x,\Lambda}(\w)  \;=\;
      {Z_{\Lambda \setminus \Gamma^*(x)}(\w) \over Z_{\Lambda \setminus x}(\w)}
      \;=\;
      \prod\limits_{i=1}^k
      {Z_{\Lambda \setminus x \setminus \{y_1,\ldots,y_i\}}(\w)
       \over
       Z_{\Lambda \setminus x \setminus \{y_1,\ldots,y_{i-1}\}}(\w)
      }
     \;.
\ee
Therefore we can improve \reff{eq.thm.Kxlambda.laststep} by writing
\be
      |\scrk_{x,\Lambda}(\w)|  \;\le\;
      \prod\limits_{i=1}^k
      {1 \over
       1 - K_{y_i, \Lambda \setminus x \setminus \{y_1,\ldots,y_{i-1}\}} 
R_{y_i}
      }
      \;,
\ee
and so we can replace the definition \reff{def.Kxlambda} by
\be
      K_{x,\Lambda}^{\rm opt}  \;=\;
      \prod\limits_{i=1}^k
      {1
       \over
      1 - K_{y_i, \Lambda \setminus x \setminus 
\{y_1,\ldots,y_{i-1}\}}^{\rm opt}
          R_{y_i}
      }
     \label{def.Kxlambda.opt}
\ee
if
$K_{y_i, \Lambda \setminus x \setminus \{y_1,\ldots,y_{i-1}\}}^{\rm opt}
    R_{y_i} < 1$
for all terms in the product,
and $K_{x,\Lambda}^{\rm opt} = +\infty$ otherwise
(note that $K_{x,\{x\}}^{\rm opt} = 1$).
Moreover, if we use $K_{x,\Lambda}^{\rm opt}$,
then all the bounds in the proof become {\em equality}\/ when $\w = -\R$.

In the general case of a ``soft'' interaction $W(x,y)$,
things become slightly more complicated,
since the vectors $\w' = W(x,\cdot)\w$ have their components
depressed but not set to zero.
To handle this case, we need to define the numbers
$K_{x,\Lambda}^{\rm opt}$ explicitly as functions of
a vector $\R_{\neq x}$.
We begin by writing $\scrk_{x,\Lambda}(w)$ as a telescoping product
as in \reff{eq.scrk.telescoping1}:
\be
      \scrk_{x,\Lambda}(\w)  \;=\;
         \prod\limits_{i=1}^k
         {Z_{\Lambda \setminus x}(\wtilde^{(i)})
          \over
          Z_{\Lambda \setminus x}(\wtilde^{(i-1)})
         }
    \label{eq.scrk.telescoping1_bis}
\ee
where the vectors $\wtilde^{(i)}$ are defined by
\be
      (\wtilde^{(i)})_y  \;=\;
      \cases{ W(x,y) w_y  & if $y=y_j$ for some $j \le i$  \cr
              w_y         & otherwise \cr
            }
    \label{def.wtildei}
\ee
Therefore we can improve \reff{eq.thm.Kxlambda.laststep} by writing
\be
      |\scrk_{x,\Lambda}(\w)|  \;\le\;
      \prod\limits_{i=1}^k
      {1 - W(x,y_i) K_{y_i,\Lambda \setminus x}(\Rtilde^{(i-1)}_{\neq x}) 
R_{y_i}
       \over
       1 - K_{y_i,\Lambda \setminus x}(\Rtilde^{(i-1)}_{\neq x}) R_{y_i}
      }
\ee
[where $\Rtilde^{(i)}_{\neq x}$ is defined by the obvious analogue
    of \reff{def.wtildei}],
and so we could have replaced the definition \reff{def.Kxlambda} by
\be
      K_{x,\Lambda}^{\rm opt}  \;=\;
      \prod\limits_{i=1}^k
      {1 - W(x,y_i)
           K^{\rm opt}_{y_i,\Lambda \setminus x}(\Rtilde^{(i-1)}_{\neq x}) 
R_{y_i}
       \over
       1 - K^{\rm opt}_{y_i,\Lambda \setminus x}(\Rtilde^{(i-1)}_{\neq x}) 
R_{y_i}
      }
     \label{def.Kxlambda.opt2}
\ee
if
$K_{y_i, \Lambda \setminus x}^{\rm opt}(\Rtilde^{(i-1)}_{\neq x}) R_{y_i} < 1$
for all terms in the product,
and $K_{x,\Lambda}^{\rm opt} = +\infty$ otherwise.

Another way of looking at all this is:
We want to choose the constants $K_{x,\Lambda}$ so that
the bound \reff{eq.dob2x} holds.
By \reff{eq.logZLambda.KxLambda} and \reff{eq.prop.scrk},
the optimal choice is manifestly
\be
      K_{x,\Lambda}^{\rm opt}  \;=\;
      \sup\limits_{0 \le \R' \le \R}  \scrk_{x,\Lambda}(-\R')   \;,
    \label{def.Kxlambda.opt3}
\ee
where $\scrk_{x,\Lambda}(\w)$ was defined in \reff{def.scrk}.
We now claim that \reff{def.Kxlambda.opt3} is identical to
\reff{def.Kxlambda.opt2} et seq.
Indeed, the recursion \reff{eq.scrk.telescoping2}
shows that the rational function $\scrk_{x,\Lambda}(-\R)$
is identical to what one obtains from the recursion \reff{def.Kxlambda.opt2}
if one {\em omits}\/ the condition following the equation;
and the monotonicity of \reff{def.Kxlambda.opt2} up to the first pole
guarantees that implementing the condition following the equation
is equivalent to taking the supremum over all $\R'$ satisfying
$0 \le \R' \le \R$.

In summary, this second improvement of Theorem~\ref{thm.dobrushin}
is {\em optimal}\/ in the sense that it is equivalent to
calculating the {\em exact}\/ $\scrk_{x,\Lambda}(\w)$
and hence [by \reff{eq.reconstructZ}] the exact $Z_\Lambda(\w)$.
So this ``optimally improved Dobrushin--Shearer theorem''
is not usually going to be useful in practice;
but it does give insight into what has been lost in
Theorems~\ref{thm.dobrushin} and \ref{thm.Kxlambda}.

\section{Tree interpretation}   \label{sec.tree_interpretation}

Further insight into Theorem~\ref{thm.Kxlambda}
and its ``optimal'' improvement \`a la
\reff{def.Kxlambda.opt2}/\reff{def.Kxlambda.opt3}
can be obtained by considering the tree structure
underlying the recursions \reff{def.Kxlambda} and \reff{def.Kxlambda.opt2}.

\subsection{Tree interpretation of Theorem~\ref{thm.Kxlambda}}
      \label{sec.tree_interpretation.1}

Let us first consider \reff{def.Kxlambda},
which provides a recursive definition of the functions $K_{x,\Lambda}(\R)$
[or $\widehat{K}_{x,\Lambda}(\R)$ if we ignore
    the conditions $K_{y,\Lambda \setminus x} R_y < 1$
    following \reff{def.Kxlambda}].
This recursion can be encoded as a tree:
at the top (root) of the tree is the pair $(x,\Lambda)$;
immediately underneath it are all the pairs
$(y,\Lambda \setminus x)$ for which $W(x,y) \neq 1$
[that is, for which $xy$ is an edge of the support graph $G=G_W$
    restricted to $\Lambda$];
and so on.
Another way of saying this is as follows:

\begin{definition}[see also \protect\cite{Lovasz_86}]
      \label{def.SAW-tree}
Let $G$ be a simple loopless graph, and let $x_1 \in V(G)$.
Then the {\em path-tree}\/ or {\em self-avoiding-walk tree}\/ (SAW-tree)
of $G$ rooted at $x_1$
--- let us call it ${\rm SAW}(G,x_1)$ ---
is the graph whose vertex set is the set of paths in $G$ starting at $x_1$
[i.e.\ paths $\overline{x_1 x_2 \cdots x_k}$ in $G$ with $k \ge 1$]
and whose edges connect $P$ to $P'$ whenever $P'$ is a
one-step extension of $P$
[i.e.\ whenever $P' = \overline{x_1 \cdots x_k}$ with $k \ge 2$
    and $P = \overline{x_1 \cdots x_{k-1}}$].
The root of ${\rm SAW}(G,x_1)$ is the zero-step path $\overline{x_1}$.
[Note that if $G$ happens to be a tree, then ${\rm SAW}(G,x_1)$
is isomorphic to $G$ for any $x_1 \in V(G)$.]
\end{definition}

\noindent
We then associate a path $\overline{x_1 \cdots x_k}$
with the pair $(x_k,\Lambda)$ where
$\Lambda = V(G) \setminus \{x_1,\ldots,x_{k-1}\}$.
Note that a given pair $(x_k,\Lambda)$ can correspond
to many paths $\overline{x_1 \cdots x_k}$,
which arise from different orderings of the set
$\{x_2,\ldots,x_{k-1}\} = V(G) \setminus \Lambda \setminus x_1$.

Given a nonnegative vector $\R = \{R_x\}_{x \in V(G)}$,
let us now assign to the vertices of ${\rm SAW}(G,x_1)$
the fugacities
\be
      \widehat{w}_{\overline{x_1 \cdots x_k}}  \;=\;  -R_{x_k}   \;,
    \label{def.what.SAW}
\ee
and to the edges of ${\rm SAW}(G,x_1)$ the weights
\be
      \widehat{W}(\overline{x_1 \cdots x_{k-1}}, \overline{x_1 \cdots x_k})
      \;=\;
      W(x_{k-1},x_k)   \;.
    \label{def.What.SAW}
\ee
We can then apply Algorithm T from Section~\ref{sec.tree_algorithm}
to calculate the effective fugacities
$\widehat{w}_{\overline{x_1 \cdots x_k}}^{\rm eff}$
on the tree ${\rm SAW}(G,x_1)$, working upwards from the leaves:

\begin{proposition}
      \label{prop.tree_interpretation.1}
In the foregoing set-up, we have
\be
      \widehat{w}_{\overline{x_1 \cdots x_k}}^{\rm eff}
      \;=\;
      -R_{x_k}  \widehat{K}_{x_k, V(G) \setminus \{x_1,\ldots,x_{k-1}\}}(\R)
    \label{eq.prop.tree_interpretation.1}
\ee
where the rational functions $\widehat{K}_{x,\Lambda}(\R)$
are defined by the recursion \reff{def.Kxlambda}
ignoring the conditions $K_{y,\Lambda \setminus x} R_y < 1$.
\end{proposition}

\proof
The proof runs inductively up from the leaves.

If $\overline{x_1 \cdots x_k}$ is a leaf (i.e.\ a maximal SAW),
then $\widehat{w}_{\overline{x_1 \cdots x_k}}^{\rm eff} =
         \widehat{w}_{\overline{x_1 \cdots x_k}} = -R_{x_k}$
by Algorithm T and the definition \reff{def.what.SAW};
moreover, $\widehat{K}_{x_k, V(G) \setminus \{x_1,\ldots,x_{k-1}\}} \equiv 1$
because $x_k$ is an isolated vertex in $G \setminus \{x_1,\ldots,x_{k-1}\}$
[since $\overline{x_1 \cdots x_k}$ is a maximal SAW],
so that the product \reff{def.Kxlambda} is empty.

Let $P = \overline{x_1 \cdots x_k}$ be a non-leaf,
and suppose that \reff{eq.prop.tree_interpretation.1}
holds for all one-step extensions $P' = \overline{x_1 \cdots x_k x_{k+1}}$.
Then Algorithm T gives
\begin{eqnarray}
      \widehat{w}_{\overline{x_1 \cdots x_k}}^{\rm eff}
      & = &
\widehat{w}_{\overline{x_1 \cdots x_k}}
      \prod_{x_{k+1} \in V(G) \setminus \{x_1,\ldots,x_k\}}
      {1 \,+\, W(x_k,x_{k+1})
               \widehat{w}_{\overline{x_1 \cdots x_k x_{k+1}}}^{\rm eff}
       \over
       1 \,+\, \widehat{w}_{\overline{x_1 \cdots x_k x_{k+1}}}^{\rm eff}
      }
          \nonumber \\[2mm]
      & = &
-R_{x_k}
      \prod_{x_{k+1} \in V(G) \setminus \{x_1,\ldots,x_k\}}
      {1 \,-\, W(x_k,x_{k+1}) R_{x_{k+1}}
               \widehat{K}_{x_{k+1}, V(G) \setminus \{x_1,\ldots,x_k\}}
       \over
       1 \,-\, R_{x_{k+1}}
               \widehat{K}_{x_{k+1}, V(G) \setminus \{x_1,\ldots,x_k\}}
      }
          \nonumber \\[2mm]
      & = &
-R_{x_k} \widehat{K}_{x_k, V(G) \setminus \{x_1,\ldots,x_{k-1}\}},
\end{eqnarray}
where the second equality uses the inductive hypothesis,
and the last equality follows immediately from \reff{def.Kxlambda}.
\qed

Now let us apply Theorem~\ref{thm.algorithm.tree} to the tree
${\rm SAW}(G,x_1)$ with
the fugacities $\what$ defined by \reff{def.what.SAW}
and edge weights $\widehat{W}$ defined by \reff{def.What.SAW}.
We conclude that the vector $\Rhat$ defined by
\be
      \widehat{R}_{\overline{x_1 \cdots x_k}}  \;=\;  R_{x_k}
\ee
lies in $\scrr(\widehat{W})$ if and only if
$R_x \widehat{K}_{x,\Lambda} < 1$
for all $\Lambda \subseteq V(G)$ and all $x \in \Lambda$.
[By~\reff{eq.prop.tree_interpretation.1},
 this latter condition is sufficient to have $\scrr(\widehat{W})$;
 indeed, only certain sets $\Lambda \subseteq V(G)$ actually arise.
 Conversely, given any $\Lambda \subseteq V(G)$ and any $x \in \Lambda$,
 we can choose $x_1 = x$ and restrict to the subtree of ${\rm SAW}(G,x_1)$
 consisting of paths inside $\Lambda$.
 Using the monotonicity of $\scrr(\widehat{W})$
 to restrict to this subtree, we conclude by Algorithm T that
 $\widehat{w}_{\overline{x_1}}^{\rm eff} \in (-1,0]$
 and hence by \reff{eq.prop.tree_interpretation.1}
 that $R_x \widehat{K}_{x,\Lambda} < 1$.]
On the other hand, by Theorem~\ref{thm.Kxlambda}
this is a {\em sufficient}\/ condition to have $\R \in \scrr(W)$.

In summary, the bounds produced by Theorem~\ref{thm.Kxlambda}
for the lattice gas on $G$ with edge weights $W$
correspond to solving exactly (via Algorithm T)
the lattice gas on the tree ${\rm SAW}(G,x_1)$
with edge weights $\widehat{W}$ [for any $x_1 \in V(G)$].
And this produces a {\em lower}\/ bound on the set $\scrr(W)$.

\subsection{Tree interpretation of the ``optimal'' bound: hard-core case}
      \label{sec.tree_interpretation.2}

Next let us consider the ``optimal'' recursion \reff{def.Kxlambda.opt2}
in the special case of a hard-core pair interaction,
i.e.\ the independent-set polynomial for the graph $G$.

For each vertex $\overline{x_1 \cdots x_k}$ of the SAW-tree,
let us choose (in any way we like) an ordering of the children
$\overline{x_1 \cdots x_k x_{k+1}}$.  We then define the
{\em pruned SAW-tree}\/ corresponding to this ordering to be
the subtree of ${\rm SAW}(G,x_1)$ whose vertex set consists of
those paths $\overline{x_1 \cdots x_k}$ satisfying the rule that
``you can't use an elder sibling of a vertex of $G$ that you have
     previously used''.
More precisely, for each path $\overline{x_1 \cdots x_i}$,
let us define the set $S(\overline{x_1 \cdots x_i})$
of ``spurned vertices at step $i$\/'' to be the set of all $x \in V(G)$
such that $\overline{x_1 \cdots x_{i-1} x}$ is a path that
precedes $\overline{x_1 \cdots x_{i-1} x_i}$ in the ordering of
children of $\overline{x_1 \cdots x_{i-1}}$.
A path $\overline{x_1 \cdots x_k}$ then belongs to the pruned SAW-tree
if and only if for all $2 \le i < j \le k$ we have
$x_j \notin S(\overline{x_1 \cdots x_i})$.

Let us remark that the pruned SAW-tree is in general much smaller
than the full SAW-tree.  For example, if $G=K_n$, then the SAW-tree
has $(n-1)!$ leaves and $(n-1)! \sum_{i=0}^{n-1} 1/i!$ vertices,
while the pruned SAW-tree (for any choice of ordering)
has $2^{n-2}$ leaves and $2^{n-1}$ vertices.

Having defined the pruned SAW-tree,
we then identify a path $\overline{x_1 \cdots x_k}$
in the pruned SAW-tree with the pair $(x_k,\Lambda)$ where
\be
      \Lambda  \,\equiv\, \Lambda(x_1,\ldots,x_k)  \;=\;
      V(G) \setminus \{x_1,\ldots,x_{k-1}\} \setminus
               \bigcup\limits_{i=2}^k S(\overline{x_1 \cdots x_i})
      \;,
\ee
that is, $V(G)$ minus the vertices already visited or already spurned.
[Note that this differs from the definition
    $\Lambda = V(G) \setminus \{x_1,\ldots,x_{k-1}\}$
    used in the preceding subsection.]

Given a vector $\w = \{w_x\}_{x \in V(G)}$ of fugacities on $G$,
let us now assign to the vertices of the pruned SAW-tree the fugacities
\be
      \widehat{w}_{\overline{x_1 \cdots x_k}}  \;=\;  w_{x_k}
    \label{def.what.prunedSAW}
\ee
and, as before, assign to the edges the weights
\be
      \widehat{W}(\overline{x_1 \cdots x_{k-1}}, \overline{x_1 \cdots x_k})
      \;=\;
      W(x_{k-1},x_k)
    \label{def.What.prunedSAW}
\ee
(which, in the hard-core case currently under consideration,
    takes the value 0 for each edge of the pruned SAW-tree).
We can then apply Algorithm T from Section~\ref{sec.tree_algorithm}
to the pruned SAW-tree
to calculate the effective fugacities
$\widehat{w}_{\overline{x_1 \cdots x_k}}^{\rm eff}$:

\begin{proposition}
      \label{prop.tree_interpretation.2}
For all $\overline{x_1 \cdots x_k}$ belonging to the pruned SAW-tree, we have
\be
      \widehat{w}_{\overline{x_1 \cdots x_k}}^{\rm eff}
      \;=\;
      w_{x_k}  \scrk_{x_k,\Lambda(x_1,\ldots,x_k)}(\w)
      \;.
    \label{eq.prop.tree_interpretation.2}
\ee
\end{proposition}

\proof
The proof runs inductively up from the leaves.

If $\overline{x_1 \cdots x_k}$ is a leaf (i.e.\ a maximal {\em pruned}\/ SAW),
then $\widehat{w}_{\overline{x_1 \cdots x_k}}^{\rm eff} =
         \widehat{w}_{\overline{x_1 \cdots x_k}} = w_{x_k}$
by Algorithm T and the definition \reff{def.what.prunedSAW};
moreover, $\scrk_{x_k,\Lambda(x_1,\ldots,x_k)} \equiv 1$
because $x_k$ is an isolated vertex in the subgraph of $G$
induced by $\Lambda(x_1,\ldots,x_k)$
[since $\overline{x_1 \cdots x_k}$ is a maximal pruned SAW],
so that all factors in the product \reff{eq.scrk.telescoping2_G} are 1.

Let $P = \overline{x_1 \cdots x_k}$ be a non-leaf,
and suppose that \reff{eq.prop.tree_interpretation.2}
holds for all one-step extensions $P' = \overline{x_1 \cdots x_k x_{k+1}}$
that are pruned SAWs.
These one-step extensions are children of $\overline{x_1 \cdots x_k}$,
hence are ordered:  let the $i$th such walk in order be
$\overline{x_1 \cdots x_k y_i}$ ($1 \le i \le l$ for some $l \ge 1$).
Note that $\{y_1,\ldots,y_l\} = \Gamma(x_k) \cap \Lambda(x_1,\ldots,x_k)$
and that
\be
      \Lambda(x_1,\ldots,x_k,y_i)  \;=\;
      \Lambda(x_1,\ldots,x_k) \setminus \{x_k,y_1,\ldots,y_{i-1}\}
      \;.
    \label{eq.Lambdas}
\ee
Then Algorithm T gives
\begin{eqnarray}
      \widehat{w}_{\overline{x_1 \cdots x_k}}^{\rm eff}
      & = &
\widehat{w}_{\overline{x_1 \cdots x_k}} \,
      \prod\limits_{i=1}^l
      {1 \,+\, W(x_k,y_i)
               \widehat{w}_{\overline{x_1 \cdots x_k y_i}}^{\rm eff}
       \over
       1 \,+\, \widehat{w}_{\overline{x_1 \cdots x_k y_i}}^{\rm eff}
      }
          \nonumber \\[2mm]
      & = &
\widehat{w}_{\overline{x_1 \cdots x_k}} \,
      \prod\limits_{i=1}^l
      {1
       \over
       1 \,+\, w_{y_i} \scrk_{y_i,\Lambda(x_1,\ldots,x_k,y_i)}(\w)
      }
          \nonumber \\[2mm]
      & = &
\widehat{w}_{\overline{x_1 \cdots x_k}} \,
      \prod\limits_{i=1}^l
      {1
       \over
       1 \,+\, w_{y_i} \scrk_{y_i,
                \Lambda(x_1,\ldots,x_k) \setminus 
\{x_k,y_1,\ldots,y_{i-1}\}}(\w)
      }
          \nonumber \\[2mm]
      & = &
\widehat{w}_{\overline{x_1 \cdots x_k}} \,
      \scrk_{x_k,\Lambda(x_1,\ldots,x_k)}(\w)
\end{eqnarray}
where the second equality uses the inductive hypothesis
and the fact that $W(x_k,y_i) = 0$,
the third equality uses \reff{eq.Lambdas},
and the final equality uses \reff{eq.scrk.telescoping2_G}.
\qed

Now let us apply Theorem~\ref{thm.algorithm.tree} to the
pruned SAW-tree with
the fugacities $\what$ defined by \reff{def.what.prunedSAW}
with $\w = -\R$,
and edge weights $\widehat{W}$ defined by \reff{def.What.prunedSAW}.
We conclude that the vector $\Rhat$ defined by
\be
      \widehat{R}_{\overline{x_1 \cdots x_k}}  \;=\;  R_{x_k}
\ee
lies in $\scrr(\widehat{W})$ if and only if
$R_x \scrk_{x,\Lambda}(-\R) < 1$
for all $\Lambda \subseteq V(G)$ and all $x \in \Lambda$;
and by the discussion surrounding
\reff{def.Kxlambda.opt2}/\reff{def.Kxlambda.opt3},
this is a {\em necessary and sufficient}\/ condition to have $\R \in \scrr(W)$.

Therefore, we have shown that the ``optimal''  Dobrushin--Shearer bound
\`a la \reff{def.Kxlambda.opt2}/\reff{def.Kxlambda.opt3}
for the independent-set polynomial (= hard-core lattice gas) on $G$
corresponds to computing exactly (via Algorithm T)
the independent-set polynomial for the {\em pruned}\/ SAW-tree
[for any $x_1 \in V(G)$ and any choice of orderings of children].
And this produces an {\em exact computation}\/ of the set $\scrr(W)$.

\subsection{Tree interpretation of the ``optimal'' bound: general case}
      \label{sec.tree_interpretation.3}

Finally, let us consider the ``optimal'' recursion \reff{def.Kxlambda.opt2}
in the general case of a soft-core pair interaction $W$.
Here we work again on the full SAW-tree ${\rm SAW}(G,x_1)$,
where $G=G_W$ is the support graph of $W$;
``pruning'' will be replaced by a ``soft suppression of spurned vertices''.
As before, we begin by choosing (in any way we like)
an ordering on the children of each vertex in the SAW-tree,
and we use this ordering to define the set
$S(\overline{x_1 \cdots x_i})$ of ``spurned vertices at step $i$\/''.
Then, given a vector $\w = \{w_x\}_{x \in V(G)}$ of fugacities on $G$,
we assign to the vertices of the SAW-tree the modified fugacities
\be
      \widehat{w}_{\overline{x_1 \cdots x_k}}  \;=\;
      w_{x_k} \!\!\!\!\!\!\prod\limits_{\begin{scarray}
                               2 \le i \le k  \\
                               x_k \in S(\overline{x_1 \cdots x_i})
                            \end{scarray}}
               \!\!\!\! W(x_{i-1},x_k)
      \;.
    \label{def.what.softprunedSAW}
\ee
[Note that the product in \reff{def.what.softprunedSAW}
could equally well be written $2 \le i < k$,
since $x_k \notin S(\overline{x_1 \cdots x_k})$.]
As before, we assign to the edges the weights
\be
      \widehat{W}(\overline{x_1 \cdots x_{k-1}}, \overline{x_1 \cdots x_k})
      \;=\;
      W(x_{k-1},x_k)   \;.
    \label{def.What.softprunedSAW}
\ee
We can then apply Algorithm T to calculate the effective fugacities
$\widehat{w}_{\overline{x_1 \cdots x_k}}^{\rm eff}$
on the tree ${\rm SAW}(G,x_1)$.  We obtain:

\begin{proposition}
      \label{prop.tree_interpretation.3}
In the foregoing set-up, we have
\be
      \widehat{w}_{\overline{x_1 \cdots x_k}}^{\rm eff}
      \;=\;
      \widetilde{w}_{x_k}^{[x_1 \cdots x_k]}
        \scrk_{x_k, V(G) \setminus \{x_1,\ldots,x_{k-1}\}}
           (\widetilde{\w}^{[x_1 \cdots x_k]})
    \label{eq.prop.tree_interpretation.3}
\ee
where $\widetilde{\w}^{[x_1 \cdots x_k]}$ is defined by
\be
      \widetilde{w}_y^{[x_1 \cdots x_k]}  \;=\;
      w_y  \!\!\!\!\!\!\prod\limits_{\begin{scarray}
                            2 \le i \le k  \\
                            y \in S(\overline{x_1 \cdots x_i})
                         \end{scarray}}
            \!\!\!\! W(x_{i-1},y)
      \;.
    \label{def.tree_interpretation.wtilde}
\ee
\end{proposition}

\proof
The proof runs inductively up from the leaves.

If $\overline{x_1 \cdots x_k}$ is a leaf (i.e.\ a maximal SAW),
then
\be
      \widehat{w}_{\overline{x_1 \cdots x_k}}^{\rm eff}  \;=\;
      \widehat{w}_{\overline{x_1 \cdots x_k}}  \;=\;
      w_{x_k} \!\!\!\!\!\!\prod\limits_{\begin{scarray}
                               2 \le i \le k  \\
                               x_k \in S(\overline{x_1 \cdots x_i})
                            \end{scarray}}
               \!\!\!\! W(x_{i-1},x_k)
       \;=\; \widetilde{w}_{x_k}^{[x_1 \cdots x_k]}
      \;,
\ee
where the first equality uses Algorithm T
and the subsequent equalities are the definitions
\reff{def.what.softprunedSAW} and \reff{def.tree_interpretation.wtilde}.
On the other hand,
$\scrk_{x_k, V(G) \setminus \{x_1,\ldots,x_{k-1}\}} \equiv 1$
because $\overline{x_1 \cdots x_k}$ is a maximal SAW,
so that the product \reff{eq.scrk.telescoping2_G} is empty.

Let $P = \overline{x_1 \cdots x_k}$ be a non-leaf,
and suppose that \reff{eq.prop.tree_interpretation.3}
holds for all one-step extensions $P' = \overline{x_1 \cdots x_k x_{k+1}}$.
These one-step extensions are children of $\overline{x_1 \cdots x_k}$,
hence are ordered:  let the $i$th such walk in order be
$\overline{x_1 \cdots x_k y_i}$ ($1 \le i \le l$ for some $l \ge 1$).
Note that $\{y_1,\ldots,y_l\} = \Gamma(x_k) \cap \{x_1,\ldots,x_{k-1}\}$.
Then Algorithm T gives
\begin{eqnarray}
      \widehat{w}_{\overline{x_1 \cdots x_k}}^{\rm eff}
      & = &
\widehat{w}_{\overline{x_1 \cdots x_k}}
      \prod\limits_{i=1}^l
      {1 \,+\, W(x_k,y_i)
               \widehat{w}_{\overline{x_1 \cdots x_k y_i}}^{\rm eff}
       \over
       1 \,+\, \widehat{w}_{\overline{x_1 \cdots x_k y_i}}^{\rm eff}
      }
          \nonumber \\[2mm]
      & = &
\widetilde{w}_{x_k}^{[x_1 \cdots x_k]}
      \prod\limits_{i=1}^l
      {1 \,+\, W(x_k,y_i) \widetilde{w}_{y_i}^{[x_1 \cdots x_k y_i]}
               \scrk_{y_i, V(G) \setminus \{x_1,\ldots,x_k\}}
                  (\widetilde{\w}^{[x_1 \cdots x_k y_i]})
       \over
       1 \,+\, \widetilde{w}_{y_i}^{[x_1 \cdots x_k y_i]}
               \scrk_{y_i, V(G) \setminus \{x_1,\ldots,x_k\}}
                  (\widetilde{\w}^{[x_1 \cdots x_k y_i]})
      }
      \;. \qquad
     \label{eq.star4}
\end{eqnarray}
Note now that from \reff{def.tree_interpretation.wtilde} we have
\be
      \widetilde{w}_{y}^{[x_1 \cdots x_k y_i]}
      \;=\;
      \cases{ W(x_k,y) \widetilde{w}_{y}^{[x_1 \cdots x_k]}   &
if $y=y_j$ for some $j<i$  \cr
              \noalign{\vskip 2pt}
              \widetilde{w}_{y}^{[x_1 \cdots x_k]}    & otherwise \cr
            }
     \label{eq.star5}
\ee
Therefore, the product on the right-hand side of \reff{eq.star4}
is identical to \reff{eq.scrk.telescoping2} if we set
$x=x_k$ and $\Lambda = V(G) \setminus \{x_1,\ldots,x_{k-1}\}$
and $\w$ is replaced by $\widetilde{\w}^{[x_1 \cdots x_k]}$,
for then \reff{eq.star5} becomes precisely the vector
$\wtilde^{(i-1)}$ defined in \reff{def.wtildei_sec3}.
Therefore, \reff{eq.star4} equals the right-hand side of
\reff{eq.prop.tree_interpretation.3}, as claimed.
\qed

\section{Unfolding}  \label{sec.unravel}

In Section~\ref{sec.tree_interpretation.2} we showed that the
``optimal''  Dobrushin--Shearer bound
\`a la \reff{def.Kxlambda.opt2}/\reff{def.Kxlambda.opt3}
can be interpreted, in the special case of the independent-set polynomial
(= hard-core lattice gas) for a graph $G$,
as an {\em exact computation}\/ of the set $\scrr(G)$
based on computing exactly (via Algorithm T)
the independent-set polynomial of the pruned SAW-tree of $G$.
In this section we would like to show how this tree bound
can be understood as arising from the repeated application
of a single ``unfolding'' step.


\begin{figure}[t]
\setlength{\unitlength}{1cm}
\begin{center}
\begin{tabular}{c@{\qquad\qquad}c}
\begin{picture}(2,2)(0,0)
      \put(1,1){\circle{2}}
      \put(0.7,1){\line(1,0){0.6}}
      \put(0.7,1){\circle*{0.15}}
      \put(1.3,1){\circle*{0.15}}
      \put(0.4,0.95){\footnotesize\it x}
      \put(1.4,0.95){\footnotesize\it y}
\end{picture}
&
\begin{picture}(5,2)(0,0)
      \put(1,1){\circle{2}}
      \put(4,1){\circle{2}}
      \put(1.8,1){\line(1,0){1.4}}
      \put(1.8,1){\circle*{0.15}}
      \put(3.2,1){\circle*{0.15}}
      \put(1.5,0.95){\footnotesize\it x}
      \put(3.3,0.95){\footnotesize\it y}
      \put(0.2,1.5){\footnotesize\it G${}'$ $\!\equiv\!$ G $\!\backslash\!$ y}
      \put(3.2,1.5){\footnotesize\it G${}''$ $\!\equiv\!$ G $\!\backslash\!$ x}
\end{picture}
\\
$G$  & $\widehat{G}^{xy}$
\end{tabular}
\end{center}
\caption{
      The graphs $G$ and $\widehat{G}^{xy}$.
}
     \label{fig1}
\end{figure}
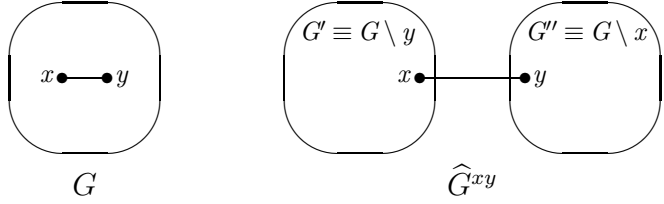


Let $G=(V,E)$ be a graph,
and let us select a pair of adjacent vertices $x,y$.
As previously, $G \setminus x$ (resp.\ $G \setminus y$)
denotes the graph obtained from $G$
by deleting the vertex $x$ (resp.\ $y$) and all edges incident with it.
We then define $\widehat{G}^{xy}$ to be the graph
obtained from the disjoint union of
$G' \equiv G \setminus y$ and $G'' \equiv G \setminus x$
by adjoining an extra edge connecting
the vertex $x$ in $G'$ to the vertex $y$ in $G''$ (Figure~\ref{fig1}).
To each vertex $z$ in
$G \setminus x \setminus y \equiv (G \setminus x) \setminus y$,
there corresponds a vertex $z'$ in $G'$ and a vertex $z''$ in $G''$.
Given a vector $\w$ with index set $V(G)$,
we define the corresponding ``diagonal'' vector $\what$
with index set $V(\widehat{G}^{xy})$ by
\begin{subeqnarray}
      \widehat{w}_x  & = & w_x    \\
      \widehat{w}_y  & = & w_y    \\
      \widehat{w}_{z'}   & = & w_z    \\
      \widehat{w}_{z''}  & = & w_z
\end{subeqnarray}
By considering the occupation of sites $x$ and $y$,
it is easily seen that
\be
      Z_G(\w)  \;=\;  Z_{G \setminus x \setminus y}(\w)  \,+\,
                      w_x Z_{G \setminus \Gamma^*(x)}(\w)  \,+\,
                      w_y Z_{G \setminus \Gamma^*(y)}(\w)
\ee
and
\begin{subeqnarray}
      Z_{\widehat{G}^{xy}}(\what)  & = &
Z_{G \setminus x \setminus y}(\w)^2 \,+\,
         w_x Z_{G \setminus x \setminus y}(\w)
             Z_{G \setminus \Gamma^*(x)}(\w)  \,+\,
         w_y Z_{G \setminus x \setminus y}(\w)
             Z_{G \setminus \Gamma^*(y)}(\w)  \nonumber \\ \\
      & = & Z_{G \setminus x \setminus y}(\w) \, Z_G(\w)   \;.
        \slabel{eq.unravel.b}
\end{subeqnarray}
We have the following fundamental result:

\begin{theorem}
      \label{thm.unravelling}
\quad
$\R \in \scrr(G)$ if and only if $\Rhat \in \scrr(\widehat{G}^{xy})$.
In other words, $\scrr(G)$ is (isomorphic to) the
``diagonal cross section'' of $\scrr(\widehat{G}^{xy})$.
\end{theorem}

\proof
Suppose first that $\R \in \scrr(G)$.
Then we also have $\R \in \scrr(G \setminus x \setminus y)$
by the monotonicity statement at the end of
Proposition~\ref{prop.scrrW.properties}.
Therefore, if $-\R \le \w \le 0$ we have
$Z_G(\w) > 0$ and $Z_{G \setminus x \setminus y}(\w) > 0$,
and hence deduce by \reff{eq.unravel.b} that
$Z_{\widehat{G}^{xy}}(\what) > 0$.
In particular, $Z_{\widehat{G}^{xy}}(-\lambda\Rhat) > 0$
for $0 \le \lambda \le 1$.
Applying Theorem~\ref{thm2.fund}(a)$\implies$(b)
to the line segment connecting 0 to $-\Rhat$,
we conclude that $Z_{\widehat{G}^{xy}}(\w') > 0$
whenever $-\Rhat \le \w' \le 0$
(whether or not $\w'$ lies ``on the diagonal'').
Hence $\Rhat \in \scrr(\widehat{G}^{xy})$.

Conversely, suppose that $\Rhat \in \scrr(\widehat{G}^{xy})$.
If $-\R \le \w \le 0$ we have $-\Rhat \le \what \le 0$
and hence $Z_{\widehat{G}^{xy}}(\what) > 0$;
by \reff{eq.unravel.b} this implies $Z_G(\w) \neq 0$.
\qed

Assume now that $G$ is connected.  Let $\widetilde{G}^{xy}$
be the component of $\widehat{G}^{xy}$ containing $x$ and $y$.
Any other component of $\widehat{G}^{xy}$ must be contained
either in $G' \setminus x$ or in $G'' \setminus y$;
either way, it corresponds to some component of $G \setminus x \setminus y$.
Now, for any component $H$ of $G \setminus x \setminus y$,
there are three possibilities:  either $H$ is adjacent to $x$ in $G$,
or $H$ is adjacent to $y$ in $G$, or both.
In the first case, the copy $H'$ of $H$ in $G'$ is contained in
$\widetilde{G}^{xy}$, while the copy $H''$ of $H$ in $G''$
is disjoint from $\widetilde{G}^{xy}$;
in the second case, the reverse holds;
in the third case, both copies of $H$ are contained in $\widetilde{G}^{xy}$.
Therefore, any component of $\widehat{G}^{xy}$ other than
$\widetilde{G}^{xy}$ has a mirror image contained in $\widetilde{G}^{xy}$,
and hence is isomorphic to a subgraph of $\widetilde{G}^{xy}$
(with the same weights when we are ``on-diagonal'').

Given a vector $\w$ with index set $V(G)$,
let us define the vector $\wtilde$ with index set $\widetilde{G}^{xy}$
by restricting $\what$ from $V(\widehat{G}^{xy})$ to $V(\widetilde{G}^{xy})$.

\begin{lemma}
      \label{lemma.restrictions}
Let $H$ be a subgraph of $G$,
and let $\R = \{R_x\}_{x \in V(G)} \in \scrr(G)$.
Then $\R_H \equiv \{R_x\}_{x \in V(H)} \in \scrr(H)$.
\end{lemma}

\proof
Define first $\R \, {\bf 1}_{V(H)}$ [with index set $V(G)$] by
\be
      (\R \, {\bf 1}_{V(H)})_x  \;=\;
      \cases{R_x  & if $x \in V(H)$ \cr
             0    & if $x \notin V(H)$ \cr
            }
\ee
Since $\R \, {\bf 1}_{V(H)} \le \R$, we have
$\R \, {\bf 1}_{V(H)} \in \scrr(G)$ because $\scrr(G)$ is a down-set.
Equivalently, $\R_H \in \scrr(G[V(H)])$,
where $G[V(H)]$ is the induced subgraph of $G$ with vertex set $V(H)$.
But $\scrr(H) \supseteq \scrr(G[V(H)])$
by the monotonicity statement at the end of
Proposition~\ref{prop.scrrW.properties}.
\qed

\begin{corollary}
      \label{cor.unravelling}
Assume that $G$ is connected.
Then $\R \in \scrr(G)$ if and only if $\Rtilde \in \scrr(\widetilde{G}^{xy})$.
\end{corollary}

\proof
If $\R \in \scrr(G)$, then $\Rhat \in \scrr(\widehat{G}^{xy})$
by Theorem~\ref{thm.unravelling}.
Then $\Rtilde \in \scrr(\widetilde{G}^{xy})$
by Lemma~\ref{lemma.restrictions},
because $\widetilde{G}^{xy}$ is a subgraph of $\widehat{G}^{xy}$.

Conversely, suppose that $\Rtilde \in \scrr(\widetilde{G}^{xy})$.
Let $H_1,\ldots,H_k$ be the components of $\widehat{G}^{xy}$
other than $\widetilde{G}^{xy}$.  Then
\be
      Z_{\widehat{G}^{xy}}(\what)  \;=\;
      Z_{\widetilde{G}^{xy}}(\wtilde)
      \prod\limits_{i=1}^k Z_{H_i}(\what \restrict H_i)
    \label{eq.cor.unravelling}
\ee
If $-\R \le \w \le 0$, then $-\Rtilde \le \wtilde \le 0$
and hence $Z_{\widetilde{G}^{xy}}(\wtilde) > 0$.
But since each component of $H_i$ is isomorphic to a subgraph of
$\widetilde{G}^{xy}$ (with the same weights in the mirror copy),
Lemma~\ref{lemma.restrictions} implies that
$Z_{H_i}(\what \restrict H_i) > 0$ as well.
Hence $Z_{\widehat{G}^{xy}}(\what) > 0$ by \reff{eq.cor.unravelling},
so that $Z_G(\w) \neq 0$ by \reff{eq.unravel.b}.
This shows that $\R \in \scrr(G)$.
\qed

\medskip
\noindent
{\bf Remark.}  It would be interesting to know whether there is
a version of unfolding in the case of soft-core interaction.

\section{Infinite graphs}  \label{sec.infinite}

In this section we discuss briefly the repulsive lattice gas
on a countably infinite graph.
We begin by deriving some general properties valid on an arbitrary
countably infinite graph (Section~\ref{sec.infinite.general}).
Next we discuss two cases of special interest:
trees (Section~\ref{sec.infinite.trees})
and regular lattices (Section~\ref{sec.infinite.regular}).
Finally we show how quantitative bounds on $\scrr(W)$ can be obtained,
using as an example the square lattice $\Z^2$ (Section~\ref{sec.infinite.Z2}).

\subsection{General properties}  \label{sec.infinite.general}

Let $X$ be a countably infinite set,
and let $W \colon\, X \times X \to [0,1]$ be symmetric.
For every nonempty finite subset $\Lambda \subset X$
we can consider the partition function $Z_\Lambda(\w)$ defined
for $\w \in \C^\Lambda$ in the obvious way,
i.e.\ by considering the lattice gas on $\Lambda$
with interaction $W \restrict \Lambda$.
Then, to each such $\Lambda$ there corresponds a set
$\scrr(W \restrict \Lambda) \subseteq [0,\infty)^\Lambda$.
{}From the Remark after Proposition~\ref{prop.scrrW.properties},
we observe that if $\Lambda' \subseteq \Lambda$
and $\R \in [0,\infty)^{\Lambda'}$, then
\be
    \R \in \scrr(W \restrict \Lambda')
    \;\;\Longleftrightarrow\;\;
    (\R, {\bf 0}) \in \scrr(W \restrict \Lambda)
  \label{eq.infinite.contain}
\ee
(here ${\bf 0}$ has index set $\Lambda \setminus \Lambda'$).

Now let us define, for each finite $\Lambda \subset X$,
the set
\be
    \scrr_\Lambda(W)  \;=\;
    \scrr(W \restrict \Lambda) \,\times\, [0,\infty)^{X \setminus \Lambda}
\ee
in the infinite-dimensional space $[0,\infty)^X$.
Otherwise put, a vector $\R \in [0,\infty)^X$
belongs to $\scrr_\Lambda(W)$ if and only if
$\R \restrict \Lambda$ lies in $\scrr(W \restrict \Lambda)$.
Note that
\be
    \partial\scrr_\Lambda(W)  \;=\;
    \partial\scrr(W \restrict \Lambda) \,\times\,
       [0,\infty)^{X \setminus \Lambda}
    \;.
\ee
Clearly $\scrr_\Lambda(W)$ is open (in the product topology)
and is a down-set.
Moreover, it follows immediately from \reff{eq.infinite.contain}
and the fact that $\scrr(W \restrict \Lambda)$ is a down-set
that if $\Lambda' \subseteq \Lambda$, then
$\scrr_{\Lambda'}(W) \supseteq \scrr_\Lambda(W)$.
In other words, the $\{\scrr_\Lambda(W)\}$
form a decreasing family of sets in $[0,\infty)^X$
[when $\Lambda$ runs over the collection of finite subsets of $X$
  ordered by inclusion].
We define the limiting set
\be
    \scrr(W)  \;=\;  \bigcap\limits_\Lambda  \scrr_\Lambda(W)
    \;.
  \label{def.scrrW.infinite}
\ee
Note that, because $\{\scrr_\Lambda(W)\}$ is a decreasing family,
we also have
\be
    \scrr(W)  \;=\;  \bigcap\limits_{n=1}^\infty  \scrr_{\Lambda_n}(W)
\ee
for any increasing sequence $\Lambda_1 \subseteq \Lambda_2 \subseteq \ldots$
whose union is all of $X$.

\medskip
\par\noindent
{\bf Remark.}  If $W(x,x) < 1$ for all $x \in X$,
then it follows from Proposition~\ref{prop.scrr.bounded}
that there exist constants $C_x < \infty$
such that $\scrr(W) \subseteq \prod_{x \in X} [0,C_x]$,
so that $\scrr(W)$ is relatively compact in the product topology.

\bigskip

In view of the fact that each set $\scrr_\Lambda(W)$ is open,
it is perhaps surprising that the limiting set $\scrr(W)$
is ``almost'' closed:

\begin{theorem}
  \label{thm.infinite.closed}
Let $X$ be countably infinite, and let $\R \in \overline{\scrr(W)}$
(where the closure is taken in the product topology).
If every component of $G_W[\supp\R]$ is infinite, then $\R \in \scrr(W)$.
\end{theorem}

\begin{corollary}
  \label{cor.infinite.closed}
Let $X$ be countably infinite, let $\r \in [0,\infty)^X$,
and suppose that every component of $G_W[\supp\r]$ is infinite.
Then $\{\lambda \ge 0 \colon\, \lambda\r \in \scrr(W)\}$
is a {\em closed}\/ interval of $[0,\infty)\;$
(which may reduce to $\{0\}$ or to all of $[0,\infty)\;$).
\end{corollary}

\medskip
\par\noindent
{\bf Remark.}  If $W(x,y) < 1$ for some $x,y \in \supp\r$,
then it follows from Proposition~\ref{prop.scrr.bounded}
that the interval $\{\lambda \ge 0 \colon\, \lambda\r \in \scrr(W)\}$
is bounded, i.e.\ not all of $[0,\infty)$.

\medskip

\proofof{Theorem~\ref{thm.infinite.closed}}
By hypothesis we have
\be
    \R \;\in\; \overline{\bigcap\limits_\Lambda \scrr_\Lambda(W)}
    \;\subseteq\; \bigcap\limits_\Lambda \overline{\scrr_\Lambda(W)}
    \;.
  \label{eq.thm.infinite.star1}
\ee
If $\R \in \scrr_\Lambda(W)$ for all $\Lambda$, we are done.
So assume that $\R \in \partial\scrr_\Lambda(W)$ for some $\Lambda$,
i.e.\ $\R \restrict \Lambda \in \partial\scrr(W \restrict \Lambda)$
and in particular $Z_\Lambda(-\R) = 0$
by Proposition~\ref{prop.closure_scrr}(b).
Let $\Lambda_1,\ldots,\Lambda_k$ be the vertex sets
of the components of $G_W[\supp \R \cap \Lambda]$.
Since $Z_\Lambda(\w) = \prod_{i=1}^k Z_{\Lambda_i}(\w)$
whenever $\supp\w \subseteq \supp\R$,
there must be at least one index $i$ for which
$Z_{\Lambda_i}(-\R) = 0$,
so that $\R \restrict \Lambda_i \in \partial\scrr(W \restrict \Lambda_i)$
[recalling that
  $\R \restrict \Lambda_i \in \overline{\scrr(W \restrict \Lambda_i)}$
  by \reff{eq.thm.infinite.star1}].
Now choose a vertex $x \in (\supp \R) \setminus \Lambda_i$
that is adjacent to $\Lambda_i$ in $G_W$
[this is possible since each component of $G_W[\supp\R]$ is infinite];
and define $\widetilde{\Lambda}_i = \Lambda_i \cup \{x\}$.
By \reff{eq.thm.infinite.star1} we have
$\R \restrict \widetilde{\Lambda}_i \in
  \overline{\scrr(W \restrict \widetilde{\Lambda}_i)}$;
and hence by Corollary~\ref{cor.strict_monotonicity} we have
$(\R \restrict \Lambda_i, 0) \in \scrr(W \restrict \widetilde{\Lambda}_i)$
[where 0 corresponds to the $x$th entry],
i.e.\ $\R \restrict \Lambda_i \in \scrr(W \restrict \Lambda_i)$.
But this contradicts
$\R \restrict \Lambda_i \in \partial\scrr(W \restrict \Lambda_i)$
since $\scrr(W \restrict \Lambda_i)$ is open.
\qed

\subsection{Infinite trees}  \label{sec.infinite.trees}

In Example~\ref{sec3}.6 we considered the complete $r$-ary rooted tree
and showed, following Shearer \cite{Shearer_85},
that there are negative real roots of the
univariate polynomial $Z_G(w)$ that tend to $w_\infty = -r^r / (r+1)^{r+1}$
as the number of levels tends to infinity.
Let us now use the ``homogenization'' ideas of Section~\ref{sec3.homo}
to extend this result to trees that are not complete.

\begin{proposition}
    \label{prop.infinite_tree}
Let $G$ be an infinite tree with (arbitrarily chosen) root vertex $x_0$,
and let $X_i = \{x \colon\, {\rm dist}(x,x_0) = i\}$.
Let $\bar{b} = \limsup\limits_{i\to\infty} |X_i|^{1/i}$.
If $R > \bar{b}^{\bar{b}} / (\bar{b}+1)^{\bar{b}+1}$,
then any vector $\R \ge 0$ satisfying
$\left( \prod\limits_{x \in X_i} R_x \right) ^{\! 1/|X_i|} \ge R$
for all $i$ does not lie in $\scrr(G)$.
In particular, $R {\bf 1} \notin \scrr(G)$.
\end{proposition}

\proof
Let $\R\ge\mathbf 0$ be a vector satisfying
$\widetilde{R}_i \equiv (\prod_{x\in X_i}R_x)^{1/|X_i|}\ge R$
for all $i$, and suppose that $\R\in\scrr(G)$.  Then by definition we have
$\R\in\scrr(G_D)$ for every $D$, where $G_D=G[\bigcup_{i\le D}X_i]$ is the
subtree consisting of the first $D+1$ levels of $G$.

Let us fix $D$, and apply Algorithm T of Section~\ref{sec.tree_algorithm}
to the tree $G_D$ with $\w = -\R$.
For $0\le i\le D$, define $\widetilde{p}_i$ as in \reff{ptildedef}.
Let $b_i=|X_{i+1}|/|X_i|$ and define $\widehat{p}_i$ by the recursion
\reff{precursiondef}.  Then it follows from Proposition \ref{prop.tree.homo1}
[or the discussion after \reff{precursiondef}] that
$\widetilde{p}_i\ge \widehat{p}_i$ for $0 \le i\leq D$,
since by hypothesis $\R\in\scrr(G_D)$.
By monotonicity of \reff{precursiondef} in the $\widetilde{R}_i$,
 it follows that the sequence
$(q_i^{(D)})_{i=1}^\infty$
defined by the recursion
\be\label{qrecdef}
q_i^{(D)}  \;=\;  \frac{R}{(1-q_{i+1}^{(D)})^{b_i}},
\ee
with initial condition $q_i^{(D)}=0$ for $i\ge D+1$ satisfies
$0\le q_i^{(D)}\le\widetilde{p}_i$ for $0 \le i \le D$.
In particular, we have $0\le q_i^{(D)}<1$ for every $i$.

Now let $D\to\infty$.  Monotonicity of \reff{qrecdef} in $q_{i+1}^{(D)}$
implies that, for each $i$, $q_i^{(D)}$ is increasing in $D$.
Since $q_i^{(D)}$ is bounded above by 1,
taking $q_i=\lim_{D\to\infty}q_i^{(D)}$ gives a sequence
$(q_i)_{i=0}^\infty$ satisfying
\be\label{qrecdef2}
q_i \;=\; \frac{R}{(1-q_{i+1})^{b_i}}
\ee
for every $i$.  Furthermore, $0\le q_i\le1$ for every $i$, and so 
\reff{qrecdef2}
implies that $0\le q_i<1$ for $i\ge1$.

The proposition is then a consequence of the following lemma.  \qed

\begin{lemma}
 \label{lemma8.4}
Let $(b_i)_{i=1}^\infty$ be a sequence of positive real numbers,
let $\bar{b} = \limsup\limits_{i\to\infty} (b_1 b_2 \cdots b_{i-1})^{1/i}$,
and let $R \ge 0$.
Suppose that there exists a sequence $(q_i)_{i=1}^\infty$
satisfying $0 \le q_i < 1$ and
\be
   q_i  \;\ge\;  \frac{R}{(1-q_{i+1})^{b_i}}
\ee
for all $i \ge 1$.  Then $R \le \bar{b}^{\bar{b}} / (\bar{b}+1)^{\bar{b}+1}$.
\end{lemma}

As preparation for proving Lemma~\ref{lemma8.4},
let us prove an analogous result for finite sequences.

\begin{lemma}
 \label{lemma8.5}
Let $k \ge 1$, let $b_0,b_1,\ldots,b_{k-1} > 0$,
and define $\bar{b} = \left( \prod\limits_{i=0}^k b_i \right) ^{\! 1/k}$.
Let $R \ge 0$.
Suppose that there exist $q_0,q_1,\ldots,q_k \in [0,1)$ satisfying
\be
   q_i  \;\ge\;  \frac{R}{(1-q_{i+1})^{b_i}}
   \qquad\hbox{for $i=0,1,\ldots,k-1$}
\ee
and
\be
   q_0 \;\le\; q_k  \;.
\ee
Then $R \le \bar{b}^{\bar{b}} / (\bar{b}+1)^{\bar{b}+1}$.
\end{lemma}

\proof
When $k=1$, the lemma is proved by straightforward calculus.
So let us treat the case $k > 1$.
Define weights
\be
   \gamma_i  \;=\;  \left( \prod\limits_{j=0}^{i-1} b_j \right)
                    \Big/ \bar{b}^i
\ee
for $i=0,1,\ldots,k$ and note that $\gamma_0 = \gamma_k = 1$.
Let $\Gamma = \sum_{i=1}^k \gamma_i$.
Now define
\begin{subeqnarray}
   \widetilde{q}_0  & = &
        \left( \prod\limits_{i=0}^{k-1} q_i^{\gamma_i} \right) ^{\! 1/\Gamma}
      \\[2mm]
   \widetilde{q}_1  & = &
        \left( \prod\limits_{i=1}^{k}   q_i^{\gamma_i} \right) ^{\! 1/\Gamma}
\end{subeqnarray}
The argument at \reff{def.gammai.sec3}--\reff{ineq.twotilde.3} shows that
\be
   \widetilde{q}_0  \;\ge\;  \frac{R}{(1-\widetilde{q}_1)^{\bar{b}}}
   \;.
\ee
If $q_0 \le q_k$, then $\widetilde{q}_0 \le \widetilde{q}_1$;
it then follows from the $k=1$ case of the lemma that
$R \le \bar{b}^{\bar{b}} / (\bar{b}+1)^{\bar{b}+1}$.
\qed

\proofof{Lemma~\ref{lemma8.4}}
If $R > \bar{b}^{\bar{b}} / (\bar{b}+1)^{\bar{b}+1}$,
let us choose $\epsilon > 0$ and $\bar{c} < \bar{b}$ such that
\be
   {R \over 1+\epsilon}  \;>\;
   {\bar{c}^{\bar{c}} \over (\bar{c}+1)^{\bar{c}+1}}  \;>\;
   {\bar{b}^{\bar{b}} \over (\bar{b}+1)^{\bar{b}+1}}  \;.
\ee
Since $\limsup_{i\to\infty} (b_1 b_2 \cdots b_{i-1})^{1/i} > \bar{c}$,
we can find integers $1 < j_1 < j_2 < \ldots$ with
\be
   \prod\limits_{i= j_t + 1}^{j_{t+1}} b_i
   \;>\;
   \bar{c}^{j_{t+1} - j_t}
 \label{eq.lemma8.4.star}
\ee
for $t=1,2,\ldots\;$.
Since the sequence $(q_{j_t})_{t=1}^\infty$
lies in the interval $[R,1)$ and $R > 0$,
there must exist $t$ such that
$q_{j_t} \le (1+\epsilon) q_{j_{t+1}}$.
To simplify the notation, let us set $k = j_{t+1} - j_t$
and $\widehat{q}_i = q_{j_t + i}$ for $i=0,\ldots,k$.
By hypothesis we have
\be
   \widehat{q}_i  \;\ge\;  \frac{R}{(1-\widehat{q}_{i+1})^{b_i}}
   \qquad\hbox{for $i=0,1,\ldots,k-1$}
\ee
and $\widehat{q}_0 \le (1+\epsilon)\widehat{q}_k$.
It follows that if we set $S = R/(1+\epsilon)$,
$\widehat{\widehat{q}}_i = \widehat{q}_i$ for $i=1,\ldots,k$,
and $\widehat{\widehat{q}}_0 = \widehat{q}_0/(1+\epsilon)$,
we have
\be
   \widehat{\widehat{q}}_i  \;\ge\;
   \frac{S}{(1-\widehat{\widehat{q}}_{i+1})^{b_i}}
   \qquad\hbox{for $i=0,1,\ldots,k-1$}
\ee
and $\widehat{\widehat{q}}_0 \le \widehat{\widehat{q}}_k$.
Lemma~\ref{lemma8.5} together with \reff{eq.lemma8.4.star}
then imply that $S \le {\bar{c}^{\bar{c}} / (\bar{c}+1)^{\bar{c}+1}}$,
which is a contradiction.
\qed

\subsection{Regular lattices}  \label{sec.infinite.regular}

The most important situation in statistical mechanics is
that of a model defined on a regular lattice
(which we take for simplicity to be $\Z^d$)
with a translation-invariant interaction \cite{Simon_93}.
In this case one expects to be able to prove that the
free energy per site (or ``pressure'')
$F_\Lambda \equiv |\Lambda|^{-1} \log Z_\Lambda$
converges to an infinite-volume limit
\be
    F_\infty \;\equiv\;
    \lim_{\Lambda \nearrow \infty} F_\Lambda
    \;,
\ee
where $\Lambda \nearrow \infty$ denotes convergence in the
F\o{}lner--van Hove sense, i.e.\ $|\Lambda| \to \infty$
in such a way that the surface-to-volume ratio tends to zero
(see \cite[Section 2.4.1 and Appendix A.3.1]{vanEnter_93}
  for a variety of equivalent conditions).
Indeed, there are several standard arguments for proving such convergence:
\begin{itemize}
    \item[1)]  {\em Almost additivity.}\/
        A large volume $\Lambda$ is subdivided into
        smaller (but still large) cubes $\Lambda_i$ separated by
        wide ``corridors'',
        and $|\log Z_\Lambda - \sum_i \log Z_{\Lambda_i}|$
        is shown to be suitably small \cite[Section I.2]{Israel_79}.
        This method applies to all models with bounded interaction energies,
        and yields F\o{}lner--van Hove convergence
        \cite[Theorems I.2.3--I.2.5]{Israel_79}.
    \item[2)]  {\em Superadditivity.}\/
        In some cases it can be shown that
        $\log Z_{\Lambda \cup \Lambda'} \ge \log Z_\Lambda + \log Z_{\Lambda'}$
        whenever $\Lambda$ and $\Lambda'$ are disjoint.
        This occurs, in particular, for systems with certain symmetries,
        for ferromagnets, and for systems with negative interaction energies.
        Moreover, arbitrary models with bounded interaction energies
        can be reduced to the case of negative interaction energies
        simply by adding suitable constants to the interaction terms
        \cite[Theorem II.2.4 and Examples 1--4 following it;
          see also Section II.4]{Simon_93}.
        However, this method does not yield F\o{}lner--van Hove convergence,
        but only a slightly weaker convergence in which
        $|\Lambda|/{\rm diam}(\Lambda)^d$ must be bounded below away from zero
        \cite[Appendix A.3.3]{vanEnter_93}.
        Moreover, the limiting free energy $F_\infty$ could be $+\infty$;
        a separate argument is needed to exclude this possibility.
    \item[3)]  {\em Subadditivity.}\/
        In some cases it can be shown that
        $\log Z_{\Lambda \cup \Lambda'} \le \log Z_\Lambda + \log Z_{\Lambda'}$
        whenever $\Lambda$ and $\Lambda'$ are disjoint.
        This occurs, in particular,
        for systems with positive interaction energies.
        Moreover, arbitrary models with bounded interaction energies
        can be reduced to the case of positive interaction energies
        by adding suitable constants to the interaction terms.
        Like superadditivity,
        this method does not yield F\o{}lner--van Hove convergence,
        but only the slightly weaker convergence in which
        $|\Lambda|/{\rm diam}(\Lambda)^d$ is bounded below away from zero.
        Moreover, the limiting free energy $F_\infty$ could be $-\infty$;
        a separate argument is needed to exclude this possibility.
\end{itemize}

{\bf Remark.}
Although we shall concentrate here on the case of a regular lattice
(namely, $\Z^d$),
arguments of the foregoing types can usually be generalized to handle
arbitrary quasi-transitive amenable infinite graphs
\cite{Jonasson_99,Procacci_03};
we expect that the same should be true for our results.

\bigskip

Let us therefore consider the lattice gas on
$X = \Z^d$ with a translation-invariant interaction
$W(x,y) = W(x-y)$ satisfying $0 \le W(x,y) \le 1$.
Let us place the same fugacity $w \ge 0$ at each site.
[More generally, we could consider a periodic or quasiperiodic fugacity
  $\w = \{w_x\}_{x \in \Z^d}$;
  but let us stick to a constant fugacity for simplicity.]
Under these assumptions we can prove that the infinite-volume limit
of the free energy exists.
Indeed, the subadditivity argument applies immediately to our model,
thanks to the hypothesis of {\em repulsive}\/ interactions;
moreover, since $Z_\Lambda \ge 1$ for all $\Lambda$
(thanks to the contribution of the empty configuration),
it follows that $F_\infty \ge 0$
and in particular that $F_\infty \neq -\infty$.
But this method does not yield F\o{}lner--van Hove convergence.

To prove F\o{}lner--van Hove convergence, we use the
almost-additivity argument.
The standard theorems \cite[Theorems I.2.3--I.2.5]{Israel_79}
do not apply to our model, because the interaction energies
can be unbounded, indeed for either of two reasons:
\begin{itemize}
    \item[(a)] the infinite interaction energy when $W(x,y) = 0$
      for some $x \neq y$;  and/or
    \item[(b)] the unboundedness of the interaction energies
      in the absence of hard-core self-repulsion
      [i.e.\ for $W(x,x) \neq 0$],
      arising from the fact that the occupation number $n_x$
      at a site can be arbitrary large.
\end{itemize}
We can nevertheless make slight adaptations in the standard
almost-additivity argument so as to make the proof go through.
We assume that the total interaction of any site with the
rest of the world is finite, i.e.
\be
    \sum_{y \in \Z^d}  [1-W(x,y)]  \;<\; \infty  \;.
  \label{eq.W.finite}
\ee
In particular, there is a number $R<\infty$ such that $W(x,y) > 0$
whenever $|x-y| > R$.

Let us begin with the case of hard-core self-repulsion,
so that we need only deal with problem (a).
This is handled by the following lemma:

\begin{lemma}
   \label{lemma.decomp}
Consider any repulsive lattice gas with hard-core self-repulsion
and fugacities $w_x \ge 0$.
Suppose that $\Lambda = \bigcup\limits_{i=1}^n \Lambda_i \,\cup\, \Lambda_0$
(disjoint union).  Then
\be
    \left( \prod\limits_{i=1}^n Z_{\Lambda_i} \right)
    \left( \prod\limits_{\begin{scarray}
                             1 \le i < j \le n \\
                             x \in \Lambda_i \\
                             y \in \Lambda_j
                         \end{scarray}}
               W(x,y) \right)
    \;\le\;
    Z_\Lambda
    \;\le\;
    \left( \prod\limits_{i=1}^n Z_{\Lambda_i} \right)
    \left( \prod\limits_{x \in \Lambda_0} (1+w_x) \right)
    \;.
\ee
\end{lemma}

\proof
We make use of the definition
\be
      Z_\Lambda  \;=\;  \sum_{X' \subseteq \Lambda}  \;
              \prod_{x \in X'} w_x
              \prod_{\{x,y\} \subseteq X'}   W(x,y)
      \;.
\ee
The lower bound is on $Z_\Lambda$ is obtained by considering
only those configurations in which $\Lambda_0$ is empty
(i.e.\ $X' \cap \Lambda_0 = \emptyset$)
and writing
\be
    \prod_{\{x,y\} \subseteq X'}   W(x,y)   \;\ge\;
    \left( \prod_{i=1}^n \prod_{\{x,y\} \subseteq X' \cap X_i} W(x,y) \right)
    \left( \prod_{1 \le i < j \le n}
            \prod\limits_{\begin{scarray}
                             x \in \Lambda_i \\
                             y \in \Lambda_j
                         \end{scarray}}
               W(x,y) \right)
    \;,
\ee
which is valid since $0 \le W(x,y) \le 1$.
The upper bound on $Z_\Lambda$ is obtained by writing
\be
    \prod_{\{x,y\} \subseteq X'}   W(x,y)   \;\le\;
    \prod_{i=1}^n \prod_{\{x,y\} \subseteq X' \cap X_i} W(x,y)
    \;.
\ee
\qed

We then have:

\begin{theorem}[infinite-volume limit, hard-core case]
   \label{thm.sec8.infvol.1}
Consider a translation-invariant repulsive lattice gas on $\Z^d$,
with hard-core self-repulsion, satisfying
$\sum_{y \in \Z^d}  [1-W(x,y)] < \infty$,
with the same fugacity $w \ge 0$ at each site.
Then $F_\infty(w) \equiv
       \lim_{\Lambda \nearrow \infty} |\Lambda|^{-1} \log Z_\Lambda(w)$
exists in F\o{}lner--van Hove sense and satisfies
$0 \le F_\infty(w) \le \log(1+w)$.
Moreover, $F_\infty(w)$ is an increasing and convex function of $\log w$;
in particular, it is continuous on $0 \le w < \infty$.
\end{theorem}

\proof
Fix integers $a,c > 0$ and consider the paving of $\Z^d$
by disjoint cubes of side $a+c$ with corners located at $(a+c)\Z^d$,
i.e.\ cubes $C_\n = [0,a+c)^d + (a+c)\n$ for $\n \in \Z^d$.
Consider also the subcubes $C'_\n \subset C_\n$
of side $a$ with the same lowermost corner,
i.e.\ $C'_\n = [0,a)^d + (a+c)\n$.
For any finite subset $\Lambda \subset \Z^d$,
let $\scrc_\Lambda$ be the collection of all cubes $C_\n$
that are contained in $\Lambda$,
and let $\scrc'_\Lambda = \{ C'_\n \colon\, C_\n \in \scrc_\Lambda \}$.
By Lemma~\ref{lemma.decomp},
\be
    \left| \log Z_\Lambda \,-\,
           \sum\limits_{C' \in \scrc'_\Lambda} \log Z_{C'}
    \right|
    \;\le\;
    \alpha(c) |\Lambda| \,+\,
    \beta \left| \Lambda \setminus \bigcup\limits_{C' \in \scrc'_\Lambda} C'
          \right|
    \;,
  \label{eq.infvol.1}
\ee
where $\alpha(c) = - \sum\limits_{\begin{scarray}
                                      x \in \Z^d \\
                                      |x| > c
                                   \end{scarray}}  \log W(x)$
and $\beta = \log(1+w)$.
By translation invariance, $\log Z_{C'} = \log Z_{C'_{\bf 0}}$
for all $C' \in \scrc'_\Lambda$.
Moreover, $|\scrc'_\Lambda| = |\scrc_\Lambda|$ by construction.
Dividing \reff{eq.infvol.1} by $|\Lambda|$, we get
\be
    \left| {1 \over |\Lambda|} \log Z_\Lambda  \,-\,
           {a^d |\scrc'_\Lambda| \over |\Lambda|} \, {1 \over a^d}
           \log Z_{C'_{\bf 0}}
    \right|
    \;\le\;
    \alpha(c) \,+\, \beta \left( 1 - {a^d |\scrc_\Lambda| \over |\Lambda|}
                          \right)
  \label{eq.infvol.2}
\ee
and hence
\begin{subeqnarray}
    |F_\Lambda - F_{C'_{\bf 0}}|
    & \le &
    \alpha(c) \,+\, \left( 1 - {a^d |\scrc_\Lambda| \over |\Lambda|}
                    \right)
                    \left( \beta + {1 \over a^d} |\log Z_{C'_{\bf 0}}| \right)
       \\[2mm]
    & \le &
    \alpha(c) \,+\, 2\beta \left( 1 - {a^d |\scrc_\Lambda| \over |\Lambda|}
                           \right)
    \slabel{eq.infvol.3b}
\end{subeqnarray}
since $1 \le Z_{C'_{\bf 0}} \le (1+w)^{a^d}$.
Taking \reff{eq.infvol.3b} for two finite subsets
$\Lambda, \widetilde{\Lambda} \subset \Z^d$ and subtracting, we get
\be
   |F_\Lambda - F_{\widetilde{\Lambda}}|  \;\le\;
   2\alpha(c) \,+\, 2\beta \left( 1 - {a^d |\scrc_\Lambda| \over |\Lambda|}
                           \right)
              \,+\, 2\beta \left( 1 - {a^d |\scrc_{\widetilde{\Lambda}}|
                                       \over |{\widetilde{\Lambda}}|}
                           \right)
    \;.
    \label{eq.infvol.4}
\ee

Now let $(\Lambda_n) \nearrow \infty$ in F\o{}lner--van Hove sense;
this implies that the fraction of volume of $\Lambda_n$
contained in the cubes $\scrc_{\Lambda_n}$ tends to 1,
i.e.\ $\lim_{n\to\infty} (a+c)^d |\scrc_{\Lambda_n}|/|\Lambda_n| = 1$.
Therefore,
\be
    \limsup\limits_{m,n\to\infty} |F_{\Lambda_m} - F_{\Lambda_n}|
    \;\le\;
    2\alpha(c) \,+\, 4\beta \left[ 1 - \left( {a \over a+c} \right) {\! d}
                            \right]
    \;.
   \label{eq.infvol.5}
\ee
Now, given any $\epsilon > 0$, we can [by \reff{eq.W.finite}]
choose $c$ large enough so that $\alpha(c) \le \epsilon$;
and we can then choose $a$ large enough so that
$1 - [a/(a+c)]^d \le \epsilon$.  Hence
\be
    \limsup\limits_{m,n\to\infty} |F_{\Lambda_m} - F_{\Lambda_n}|
    \;\le\;
    2\epsilon + 4\beta\epsilon
    \;.
   \label{eq.infvol.6}
\ee
Since $\epsilon$ is arbitrary, we have
$\limsup_{m,n\to\infty} |F_{\Lambda_m} - F_{\Lambda_n}| = 0$,
so that $(|\Lambda_n|^{-1} \log Z_{\Lambda_n})_{n=1}^\infty$
is a Cauchy sequence and hence converges.
(The limit is the same for all F\o{}lner--van Hove sequences,
since interleaving two F\o{}lner--van Hove sequences yields another one.)

The bounds $0 \le F_\infty(w) \le \log(1+w)$
follow immediately from $1 \le Z_\Lambda \le (1+w)^{|\Lambda|}$.
Finally, each $F_\Lambda(w)$ is manifestly increasing,
and by Lemma~\ref{lemma.sec2.convexity} it is a convex function of $\log w$;
and these properties are preserved under pointwise limits.
\qed

Now let us turn to problem (b), arising from the unboundedness of
the occupation number $n_x$ when $W(x,x) \neq 0$.
We begin with a well-known fact about positive correlation
of increasing functions on a totally ordered space:

\begin{lemma}
   \label{lemma.corrineq.1}
Let $\mu \not\equiv 0$ be a nonnegative measure on a totally ordered
space $\Omega$, and let $f$ and $g$ be increasing functions on $\Omega$.
Then (provided the functions concerned are integrable):
\begin{itemize}
    \item[(a)] $ \left( \int fg \, d\mu \right)
                    \left( \int {\bf 1} \, d\mu \right)
                    \;\ge\;
                    \left( \int f \, d\mu \right)
                    \left( \int g \, d\mu \right)
                 \;.$
    \item[(b)] ${\displaystyle
                    {\int f e^{\alpha g} \, d\mu
                     \over
                     \int e^{\alpha g} \, d\mu
                    }}$
        is an increasing function of $\alpha \in \RR$.
\end{itemize}
\end{lemma}

\proof
(a)  We have
\be
    \left( {\textstyle\int} fg \, d\mu \right)
    \left( {\textstyle\int} {\bf 1} \, d\mu \right)
    \,-\,
    \left( {\textstyle\int} f \, d\mu \right)
    \left( {\textstyle\int} g \, d\mu \right)
    \;=\;
    {1 \over 2}
    \int [f(x)-f(y)] \, [g(x)-g(y)] \, d\mu(x) \, d\mu(y)
    \;,
\ee
and the integrand on the right-hand side is nonnegative
both when $x \le y$ and when $x \ge y$.

(b) We have
\be
    {d \over d\alpha}
    \left( {\int f e^{\alpha g} \, d\mu
            \over
            \int e^{\alpha g} \, d\mu
           }
    \right)
    \;=\;
    {(\int fg e^{\alpha g} \, d\mu)(\int e^{\alpha g} \, d\mu)
     \,-\,
     (\int f e^{\alpha g} \, d\mu)(\int g e^{\alpha g} \, d\mu)
     \over
     (\int e^{\alpha g} \, d\mu)^2
    }
    \;.
\ee
Now apply part (a) with $\mu$ replaced by $e^{\alpha g} \mu$.
\qed

Now let $\Lambda$ be a finite set;
for each $x \in \Lambda$, let $\Omega_x$ be a totally ordered space;
and let $\Omega = \prod_{x \in X} \Omega_x$.
For each $x \in \Lambda$, let $\mu_x \not\equiv 0$ be a
nonnegative measure on $\Omega_x$
and let $F_x$ be a nonnegative {\em decreasing}\/ function on $\Omega_x$.
Finally, let $H$ be an increasing function on $\Omega$
and let $\alpha \ge 0$.
Define
\be
    Z_\Lambda^{(\{F_x\},\alpha,H)}  \;=\;
    \int \left( \prod\limits_{x \in \Lambda} F_x(\varphi_x) \right)
         \exp[-\alpha H(\varphi)]
         \prod\limits_{x \in \Lambda} d\mu_x(\varphi_x)
    \;.
\ee
\begin{lemma}
   \label{lemma.corrineq.2}
Under the above hypotheses,
\be
    Z_\Lambda^{(\{F_x\},\alpha,H)}  \;\ge\;
    \left( \prod\limits_{x \in \Lambda}
           {\int F_x \, d\mu_x  \over  \int {\bf 1} \, d\mu_x}
    \right)
    \,\times\,
    Z_\Lambda^{(\{{\bf 1}\},\alpha,H)}
    \;.
  \label{eq.corrineq.2}
\ee
\end{lemma}

\proof
Choose one site $z \in \Lambda$,
and let us study the integral over $\varphi_z$ with
$\varphi_{\neq z}$ held fixed.
By Lemma~\ref{lemma.corrineq.1}(b), the quantity
\be
    {\int F_z(\varphi_z) \,
          \exp[-\alpha H(\varphi_z,\varphi_{\neq z})]
          \, d\mu_z(\varphi_z) 
     \over
     \int \exp[-\alpha H(\varphi_z,\varphi_{\neq z})]
          \, d\mu_z(\varphi_z)
    }
\ee
is an increasing function of $\alpha \in \RR$,
so its value at $\alpha \ge 0$ is bounded below by its value at $\alpha=0$,
which is
\be
    {\int F_z(\varphi_z) \, d\mu_z(\varphi_z)
     \over
     \int d\mu_z(\varphi_z)
    }
    \;.
\ee
Therefore,
\be
    \int  F_z(\varphi_z) \,
          \exp[-\alpha H(\varphi_z,\varphi_{\neq z})] \,
          d\mu_z(\varphi_z)
    \;\ge\;
    \left( {\int F_z \, d\mu_z  \over  \int {\bf 1} \, d\mu_z}
    \right)
    \,\times\,
    \int \exp[-\alpha H(\varphi_z,\varphi_{\neq z})] \,
         d\mu_z(\varphi_z)
    \;.
\ee
Now multiply both sides by $\prod_{x \in \Lambda \setminus z} F_x(\varphi_x)$
and integrate with respect to
$\prod_{x \in \Lambda \setminus z} d\mu_x(\varphi_x)$;
we obtain
\be
    Z_\Lambda^{(\{F_x\},\alpha,H)}  \;\ge\;
    \left( {\int F_z \, d\mu_z  \over  \int {\bf 1} \, d\mu_z}
    \right)
    \,\times\,
    Z_\Lambda^{(\{{\bf 1}_x, F_{\neq x}\},\alpha,H)}
    \;.
\ee

Applying the same argument successively to each site in $\Lambda$,
we obtain \reff{eq.corrineq.2}.
\qed

Let us now specialize these results to the repulsive lattice gas,
by taking $\Omega_x = \N$, $d\mu_x(n_x) = w_x^{n_x}/n_x!$,
\be
    H(\n)  \;=\;
    \sum\limits_{x \in \Lambda} [-\log W(x,x)] {n_x (n_x -1) \over 2}
    \,+\,
    \sum\limits_{\{x,y\} \subseteq \Lambda} [-\log W(x,y)] n_x n_y
    \;,
\ee
and $F_x(n_x) = {\bf 1}(n_x \le K_x)$
for arbitrarily chosen positive constants $K_x$.
We then have
\be
    {Z_\Lambda^{(\K)} \over Z_\Lambda}
    \;\ge\;
    \prod\limits_{x \in \Lambda} [1-\gamma(K_x,w_x)]
    \;,
  \label{eq.cutoffZ.bound}
\ee
where $Z_\Lambda^{(\K)}$ denotes the sum \reff{defZ_2}
restricted to the configurations satisfying $n_x \le K_x$ for all $x$,
and
\be
    \gamma(K,w)  \;=\;  e^{-w} \sum\limits_{n=K}^\infty {w^n \over n!}
    \;.
\ee
We can now prove a variant of Lemma~\ref{lemma.decomp}:

\begin{lemma}
    \label{lemma.decomp.2}
Consider any repulsive lattice gas with fugacities $w_x \ge 0$,
and let $(K_x)_{x \in \Lambda}$ be arbitrary positive constants.
Suppose that $\Lambda = \bigcup\limits_{i=1}^n \Lambda_i \,\cup\, \Lambda_0$
(disjoint union).  Then
\begin{eqnarray}
   & &
    \left( \prod\limits_{i=1}^n Z_{\Lambda_i} \right)
    \left( \prod\limits_{\begin{scarray}
                             1 \le i \le n \\
                             x \in \Lambda_i
                         \end{scarray}}
               [1-\gamma(K_x,w_x)] \right)
    \left( \prod\limits_{\begin{scarray}
                             1 \le i < j \le n \\
                             x \in \Lambda_i \\
                             y \in \Lambda_j
                         \end{scarray}}
               W(x,y)^{K_x K_y} \right)
       \nonumber \\[4mm]
    & &  \hspace*{5cm}
    \;\le\;
    Z_\Lambda
    \;\le\;
    \left( \prod\limits_{i=1}^n Z_{\Lambda_i} \right)
    \left( \prod\limits_{x \in \Lambda_0} e^{w_x} \right)
    \;.
\end{eqnarray}
\end{lemma}

\proof
The upper bound is proved exactly as in Lemma~\ref{lemma.decomp}.
For the lower bound, we begin from the trivial fact
that $Z_\Lambda \ge Z_\Lambda^{(\K)}$.
Now consider only those configurations satisfying $\n \le \K$
and, in addition, $n_x = 0$ for all $x \in \Lambda_0$;
for such configurations we have
\be
    \prod_{\{x,y\} \subseteq \Lambda \setminus \Lambda_0}   W(x,y)^{n_x n_y}
    \;\ge\;
    \left( \prod_{i=1}^n \prod_{\{x,y\} \subseteq X_i} W(x,y)^{n_x n_y} \right)
    \left( \prod_{1 \le i < j \le n}
            \prod\limits_{\begin{scarray}
                             x \in \Lambda_i \\
                             y \in \Lambda_j
                         \end{scarray}}
               W(x,y)^{K_x K_y} \right)
    \;.
\ee
It follows that
\be
    Z_\Lambda^{(\K)}  \;\ge\;
    \left( \prod_{i=1}^n  Z_{\Lambda_i}^{(\K \restrict \Lambda_i)} \right)
    \left( \prod_{1 \le i < j \le n}
            \prod\limits_{\begin{scarray}
                             x \in \Lambda_i \\
                             y \in \Lambda_j
                         \end{scarray}}
               W(x,y)^{K_x K_y} \right)
    \;.
\ee
Now use \reff{eq.cutoffZ.bound} for each set $\Lambda_i$.
\qed

\begin{theorem}[infinite-volume limit, general case]
   \label{thm.sec8.infvol.2}
Consider a translation-invariant repulsive lattice gas on $\Z^d$,
satisfying $\sum_{y \in \Z^d}  [1-W(x,y)] < \infty$,
with the same fugacity $w \ge 0$ at each site.
Then $F_\infty(w) \equiv
       \lim_{\Lambda \nearrow \infty} |\Lambda|^{-1} \log Z_\Lambda(w)$
exists in F\o{}lner--van Hove sense and satisfies
$0 \le F_\infty(w) \le w$.
Moreover, $F_\infty(w)$ is an increasing and convex function of $\log w$;
in particular, it is continuous on $0 \le w < \infty$.
\end{theorem}

\proof
We begin by defining families of cubes $\scrc_\Lambda$ and $\scrc'_\Lambda$
as in the proof of Theorem~\ref{thm.sec8.infvol.1}.
By Lemma~\ref{lemma.decomp.2} with $K_x = K$ for all $x$, we have
\be
    \left| \log Z_\Lambda \,-\,
           \sum\limits_{C' \in \scrc'_\Lambda} \log Z_{C'}
    \right|
    \;\le\;
    \alpha(c) K^2 |\Lambda| \,+\,
    w \left| \Lambda \setminus \bigcup\limits_{C' \in \scrc'_\Lambda} C'
          \right|
    \,+\, |\Lambda| \, |\log[1-\gamma(K,w)]|
  \label{eq.infvol.bis.1}
\ee
where $\alpha(c) = - \sum\limits_{\begin{scarray}
                                      x \in \Z^d \\
                                      |x| > c
                                   \end{scarray}}  \log W(x)$.
By the same arguments as before, we have
\be
   |F_\Lambda - F_{C'_{\bf 0}}|
    \;\le\;
    \alpha(c) K^2 \,+\, 2w \left( 1 - {a^d |\scrc_\Lambda| \over |\Lambda|}
                           \right)
      \,+\, |\log[1-\gamma(K,w)]|
    \;.
  \label{eq.infvol.bis.2}
\ee
Taking \reff{eq.infvol.bis.2} for two finite subsets
$\Lambda, \widetilde{\Lambda} \subset \Z^d$ and subtracting, we get
\be
   |F_\Lambda - F_{\widetilde{\Lambda}}|  \;\le\;
   2\alpha(c) K^2 \,+\, 2w \left( 1 - {a^d |\scrc_\Lambda| \over |\Lambda|}
                           \right)
              \,+\, 2w \left( 1 - {a^d |\scrc_{\widetilde{\Lambda}}|
                                   \over |{\widetilde{\Lambda}}|}
                           \right)
      \,+\, 2 |\log[1-\gamma(K,w)]|
    \;.
    \label{eq.infvol.bis.3}
\ee

Now let $(\Lambda_n) \nearrow \infty$ in F\o{}lner--van Hove sense,
so that
$\lim_{n\to\infty} (a+c)^d |\scrc_{\Lambda_n}|/|\Lambda_n| = 1$.
It follows that
\be
    \limsup\limits_{m,n\to\infty} |F_{\Lambda_m} - F_{\Lambda_n}|
    \;\le\;
    2\alpha(c) K^2 \,+\, 4w \left[ 1 - \left( {a \over a+c} \right) {\! d}
                            \right]
      \,+\, 2 |\log[1-\gamma(K,w)]|
    \;.
   \label{eq.infvol.bis.4}
\ee
Now, given any $\epsilon > 0$, we first choose $K$ large enough
so that $|\log[1-\gamma(K,w)]| \le \epsilon$;
then we choose $c$ large enough so that $\alpha(c) K^2 \le \epsilon$;
and finally we choose $a$ large enough so that
$1 - [a/(a+c)]^d \le \epsilon$.  Therefore,
\be
    \limsup\limits_{m,n\to\infty} |F_{\Lambda_m} - F_{\Lambda_n}|
    \;\le\;
    4\epsilon + 4\epsilon w
    \;.
   \label{eq.infvol.bis.5}
\ee
The remainder of the argument is exactly as in
Theorem~\ref{thm.sec8.infvol.1}.
\qed


\medskip
\par\noindent
{\bf Historical remark.}
Results like Theorem~\ref{thm.sec8.infvol.2}
have been proven in the vastly more general context
of ``unbounded spin systems''
by Lebowitz and Presutti \cite{Lebowitz_76},
based on superstability estimates due to Ruelle \cite{Ruelle_76}.
See also \cite{Kunsch_81,Bellissard_82} for related work.
We think that it is nevertheless useful to give an elementary
and self-contained proof for our special case.

\bigskip

Let us now return to the case of hard-core self-repulsion,
and consider the convergence of the finite-volume free energies $F_\Lambda(w)$
for {\em complex}\/ fugacities $w$.
By hypothesis \reff{eq.W.finite} we have
\be
    \Delta_W \;\equiv\;  \sum\limits_{y \neq 0} [1-W(0,y)]
      \;<\; \infty  \;.
\ee
Therefore, by Corollary~\ref{cor.dobrushin_1},
all of the partition functions $Z_\Lambda(w)$
are nonvanishing in the disc $|w| < R(\Delta_W)$
where $R(\Delta_W)$ is defined by \reff{def.Rdelta}.
Moreover, they satisfy the trivial bounds
$|Z_\Lambda(w)| \le e^{|\Lambda| \, |w|}$.
It follows that the free energies
$F_\Lambda(w) \equiv |\Lambda|^{-1} \log Z_\Lambda(w)$
are analytic in the disc $|w| < R(\Delta_W)$
and satisfy there $\real F_\Lambda(w) \le |w| \le R(\Delta_W)$.
Finally, Theorem~\ref{thm.sec8.infvol.2} shows that
the $F_\Lambda(w)$ converge to a limit when $w$ lies in the
{\em real}\/ interval $[0, R(\Delta_W))$.
These three facts are sufficient to imply
the convergence of $F_\Lambda(w)$ to an analytic limit $F_\infty(w)$
everywhere in the disc $|w| < R(\Delta_W)$,
using the following standard result on normal families
of analytic functions:

\begin{proposition}[exp log Vitali]
    \label{prop.explogVitali}
Let $D$ be a domain in $\C$, let $S$ be a subset of $D$
having at least one accumulation point in $D$,
let $M < \infty$, and let $(f_n)_{n=1}^\infty$ be analytic functions in $D$
satisfying:
\begin{itemize}
    \item[(a)]  $\real f_n(z) \le M$ for all $n$ and all $z \in D$; and
    \item[(b)]  $\lim\limits_{n \to\infty} f_n(z)$ exists (and is finite)
        for all $z \in S$.
\end{itemize}
Then there exists an analytic function $f_\infty$ on $D$
such that $f_n(z) \to f_\infty(z)$ uniformly for $z$ in compact subsets
of $D$.
\end{proposition}

\noindent
For a proof of Proposition~\ref{prop.explogVitali},
see e.g.\ \cite[p.~343]{Simon_74}.

\begin{corollary}
    \label{cor.explogVitali}
Consider a translation-invariant repulsive lattice gas on $\Z^d$,
with hard-core self-repulsion, with
\be
    \sum\limits_{y \neq 0} [1-W(0,y)]
    \;\equiv\;  \Delta_W  \;<\; \infty  \;.
\ee
Then:
\begin{itemize}
    \item[(a)]  Each function
        $F_\Lambda(w) \equiv |\Lambda|^{-1} \log Z_\Lambda(w)$
        is analytic in the disc $|w| < R(\Delta_W)$
        and satisfies there $\real F_\Lambda(w) \le |w|$.
    \item[(b)]  There exists an analytic function $F_\infty(w)$
        on the disc $|w| < R(\Delta_W)$ such that
        $\lim_{\Lambda \nearrow \infty} F_\Lambda(w) = F_\infty(w)$
        uniformly for $w$ in compact subsets of $|w| < R(\Delta_W)$.
\end{itemize}
\end{corollary}

\medskip
\par\noindent
{\bf Remark.}  Proposition~\ref{prop.explogVitali} is a very special case
of a much more general ``Vitali--Porter--type'' theorem
for normal families of analytic functions
\cite[Lemma 3.5]{Sokal_chromatic_roots},
in which condition (a) can be weakened to
``there exists a nonempty disc $\Delta \subset \C$
   such that $f_n(z) \notin \Delta$ for all $z \in D$''
\cite[Example~2.3.9]{Schiff_93}
or even to
``there exist $w_1,w_2 \in \C$ with $w_1 \neq w_2$ such that
   $f_n(z) \notin \{w_1,w_2\}$ for all $z \in D$''
\cite[Section 2.7]{Schiff_93}.
Moreover, it is sufficient for this hypothesis to hold locally in $D$.
Detailed accounts of these results can be found in
\cite{Montel_27,Schiff_93}.

\subsection{Quantitative bounds for the lattice $\Z^d$}
    \label{sec.infinite.Z2}

By Corollary~\ref{cor.infinite.closed},
the set $\{\lambda \ge 0 \colon\, \lambda{\bf 1} \in \scrr(W)\}$
is a {\em closed}\/ interval $[0,\lambda_c]$
provided that each component of $G_W$ is infinite.
In this section we sketch briefly some methods for finding
reasonably sharp upper and lower bounds on $\lambda_c$.
We shall illustrate these examples with reference to the
hard-core lattice gas on $\Z^d$ with nearest-neighbor edges,
paying particular attention to the case of the square lattice $\Z^2$.

\subsubsection{Upper bounds on $\lambda_c$}  \label{sec.infinite.Z2.upper}

Let $G$ be any finite subgraph of $\Z^d$,
and let $w_\star = -\lambda_\star$ be the negative real root of $Z_G$
of smallest magnitude.  Then, by the definition
of $\scrr(W)$ for an infinite graph [cf.\ \reff{def.scrrW.infinite}]
together with the monotonicity of $\scrr(G)$ in $G$
[Proposition~\ref{prop.scrrW.properties}],
we can conclude immediately that $\lambda_c(\Z^d) \le \lambda_\star$.
Moreover, these bounds converge to the exact value $\lambda_c$
if we take any increasing sequence $G_1 \subseteq G_2 \subseteq \ldots$
whose union is all of $\Z^d$.

More generally, as a consequence of the discussion in
Section~\ref{sec.tree_interpretation.2},
we can take any finite subtree $T$ of the pruned SAW-tree for $\Z^d$,
and let $w_\star = -\lambda_\star$ be the negative real root of $Z_T$
of smallest magnitude;  we again have $\lambda_c(\Z^d) \le \lambda_\star$.

\bigskip

{\bf Example \thesection.1.}
Clearly $\Z^d$ contains finite paths of all lengths.
By Example~\ref{sec3}.2
(or alternatively Example~\ref{sec3}.6 with $r=1$),
these give bounds $\lambda_\star$ tending to $1/4$
as the path length tends to infinity.
Thus $\lambda_c(\Z^d)\le1/4$.

\bigskip

{\bf Example \thesection.2.}
Now consider a finite subgraph $G \subseteq \Z^d$ consisting of
a long path $P$ along one axis (the {\em spine}\/)
and, radiating from each vertex of $P$,
$2d-2$ disjoint long paths perpendicular to $P$ (the {\em antennae}\/).
Let us assign fugacity $w\le 0$ to every vertex of $G$.
Using Algorithm T on the antennae, we must first
iterate the recursion $w_{\rm eff} \mapsto w/(1+w_{\rm eff})$.
If the antennae were infinitely long, we would approach the fixed point
$w' = -{1 \over 2} + \sqrt{w + {1 \over 4}}$
[provided that $-1/4\le w\le0$; otherwise we are outside $\scrr(G)$].
By taking the antennae sufficiently long,
we can get as close to this value as we wish.
We are now left with a ``caterpillar'' consisting of
the spine (with fugacity $w$ on each vertex)
along with $2d-2$ pendant vertices (each with fugacity $w'$)
attached to each vertex of the spine.
Applying Algorithm T to this graph, we get the recursion
$w_{\rm eff} \mapsto w/[(1+w_{\rm eff})(1+w')^{2d-2}]
 \equiv w_\star/(1+w_{\rm eff})$,
where
\be
   w_\star  \;\equiv\;  {w \over (1+w')^{2d-2}}
     \;=\;
     {w  \over
      \left({1 \over 2} + \sqrt{w + {1 \over 4}} \, \right)^{\! 2d-2}}
   \;.
\ee
We require $-{1 \over 4} \le w_\star \le 0$
in order to stay within $\scrr(G)$.
For $d=2$ this yields the bound $\lambda_c(\Z^2) \le 4/25$.
For $d \to\infty$ it yields
$\lambda_c(\Z^d) \le (\log d)/(2d) + O((\log\log d)/d)$.

\bigskip

{\bf Example \thesection.3.}
To get the correct bound $\lambda_c(\Z^d) = O(1/d)$,
one can argue using a subtree of the pruned SAW-tree.
Note first that any walk using only steps in the positive coordinate
directions is guaranteed to be self-avoiding.
Moreover, if we define the pruning such that at each vertex
all steps in the positive coordinate directions are preferred to all steps
in the negative coordinate directions, then every walk using only
positive coordinate steps appears as a vertex in the pruned
SAW-tree of $\Z^d$.
Thus, the complete $d$-ary rooted tree (consisting of these walks)
is a subtree of the pruned SAW-tree of $\Z^d$,
and so by Example~\ref{sec3}.6 we have
\be
    \lambda_c(\Z^d) \;\le\;  {d^d \over (d+1)^{d+1}}  \;\sim\;  {1 \over ed}
    \;.
 \label{eq8.treebound}
\ee

\bigskip

{\bf Example \thesection.4.}
Asymptotically correct upper bounds on $\lambda_c(\Z^d)$
can be obtained by using large cylinders,
for which $\lambda_c$ can be computed by
the transfer-matrix method.\footnote{
    See e.g.\ \cite{Todo_99} for a brief discussion of transfer matrices
    for the hard-core lattice gas.
}
Let us illustrate the method for $\Z^2$.
Consider the strip $S_L=\{(x,y) \in \Z^2 \colon\, 0\le x<L\}$.
Since $S_1 \subseteq S_2 \subseteq \ldots \subseteq \Z^2$,
we have $\lambda_c(S_1) \ge \lambda_c(S_2) \ge \ldots \ge \lambda_c(\Z^2)$.
[In particular, $\lim_{L\to\infty} \lambda_c(S_L)$ exists.]
On the other hand, since each finite subgraph of $\Z^2$
is contained (modulo translation) in $S_L$ for all sufficiently large $L$,
we have $\lim_{L\to\infty} \lambda_c(S_L) \le \lambda_c(\Z^2)$.
It follows that $\lambda_c(S_L) \downarrow \lambda_c(\Z^2)$ as $L\to\infty$.

An analogous argument can also be made using strips with periodic
boundary conditions, which are more convenient for computation \cite{Todo_99}.
To see this, let $\widetilde{S}_L$ be the strip $S_L$
with an extra edge added from $(L-1,y)$ to $(0,y)$ for each $y$.
As before, each finite subgraph of $\Z^2$ is contained (modulo translation)
in $\widetilde{S}_L$ for all sufficiently large $L$,
so that $\limsup_{L\to\infty} \lambda_c(\widetilde{S}_L) \le \lambda_c(\Z^2)$.
On the other hand, the pruned SAW-tree of $\widetilde{S}_L$
(for any given choice of ordering)
is a subtree of the pruned SAW-tree of $\Z^2$
(provided we make an appropriately compatible choice of ordering):
just map each path in $\widetilde{S}_L$ to its ``universal cover'' in $\Z^2$,
i.e.\ to the path in $\Z^2$ obtained by making the same sequence of
north, south, east or west steps.
Thus $\lambda_c(\widetilde{S}_L) \ge \lambda_c(\Z^2)$ for every $L$
(though we do not necessarily have monotonicity in $L$).
It follows that $\lambda_c(S_L) \to \lambda_c(\Z^2)$ as $L\to\infty$.

\bigskip

{\bf Example \thesection.5.}
Let us also remark that $\lambda_c(\Z^d)$
is {\em strictly}\/ decreasing in $d$;
indeed, we can derive an inequality bounding $\lambda_c(\Z^{d+1})$ above
in terms of $\lambda_c(\Z^d)$.
To see this, note that $\Z^{d+1}$ contains a copy of the graph $G_d$
obtained from $\Z^d$ by attaching two semi-infinite paths (``antennae'')
to every vertex.
If we place fugacity $w\le0$ on every vertex of $G_d$
and integrate out the antennae as in Example~\thesection.2,
we are left with a copy of $\Z^d$ with effective fugacities
\be
   w_{\rm eff}  \;=\;  {w \over (1+w')^2}  \;=\;
       {w \over  w + {1 \over 2} + \sqrt{w+{1 \over 4}}}  \;.
\ee
If $w \in (-\lambda_c(\Z^{d+1}),0]$,
then $w_{\rm eff}$ must be in $(-\lambda_c(\Z^d),0]$.
It follows that
\be
   \lambda_c(\Z^d)  \;\ge\;
   {\lambda_c(\Z^{d+1})
    \over
    {1 \over 2} - \lambda_c(\Z^{d+1}) + \sqrt{{1 \over 4} - \lambda_c(\Z^{d+1})}
   }
   \;,
\ee
or equivalently
\be
   \lambda_c(\Z^{d+1})   \;\le\;
   {\lambda_c(\Z^d)  \over  [1+\lambda_c(\Z^d)]^2}
   \;,
  \label{eq8.d+1}
\ee
which is the desired bound.

Using \reff{eq8.d+1} together with the initial condition
$\lambda_c(\Z^1) = 1/4$,
it is easy to show by induction that
$\lambda_c(\Z^d) \le 1/(2d+2)$.
However, this bound is less sharp than the bound \reff{eq8.treebound}
obtained in Example~\thesection.3.

\subsubsection{Lower bounds on $\lambda_c$}  \label{sec.infinite.Z2.lower}

We now turn to proving lower bounds on $\lambda_c$.
Corollary \ref{cor.dobrushin_2} gives
\be
\lambda_c(\Z^d) \;\ge\; \frac{\Delta^\Delta}{(\Delta+1)^{\Delta+1}}
         \;=\;  \frac{(2d)^{2d}}{(2d+1)^{2d+1}} \;\sim\;
         \frac{1}{2ed}   \;,
  \label{bound2a}
\ee
while Corollary \ref{cor.shearer} gives the slightly better bound
\be
\lambda_c(\Z^2) \;\ge\; \frac{(\Delta-1)^{\Delta-1}}{\Delta^\Delta}
         \;=\;  \frac{(2d-1)^{2d-1}}{(2d)^{2d}} \;\sim\;
         \frac{1}{2ed}   \;.
  \label{bound2b}
\ee
For $d=2$ the latter bound yields $\lambda_c(\Z^2) \ge 27/256$.
Either of these bounds shows,
when combined with Example~\thesection.3,
that $\lambda_c(\Z^d) = \Theta(1/d)$ as $d \to\infty$.

To get quantitatively better bounds, we can take
a supertree of the pruned SAW-tree of $\Z^d$ and calculate exactly for it.
In particular, if we choose a supertree that is eventually periodic,
the computation reduces to finding the fixed point of a recursion.

\bigskip

{\bf Example \thesection.6.}
The pruned SAW-tree of $\Z^d$ is obviously contained in
the $2d$-branching tree, for which the computation has been performed
in Example~\ref{sec3}.6.  This yields the bound \reff{bound2a}.
[This argument obviously works for any infinite graph of
 maximum degree $\Delta$, not just $\Z^d$.]

\bigskip

{\bf Example \thesection.7.}
A smaller supertree of the pruned SAW-tree is obtained by giving
the root $2d$ children and every other vertex $2d-1$ children.
Applying Algorithm T, the recursion (until we reach children of the root) is
$w' \mapsto w/(1+w')^{\Delta-1}$.
We end up with the bound \reff{bound2b}.
[The argument again works for any infinite graph of maximum degree $\Delta$.]

\bigskip

{\bf Example \thesection.8.}
Another possibility (not the only one)
is to take a large finite piece $T$ of the pruned SAW-tree of $\Z^2$,
and then repeat it periodically (attaching to each leaf of $T$ a copy
of $T$ starting at the root, and repeating this).
It is easily seen that this gives a supertree of the
pruned SAW-tree.
We therefore use the recursion
$w_{\rm eff,leaves} \mapsto w_{\rm eff,root}$
and demand that there exist an attractive fixed point in $(-1,0]$.
We have various choices about how to order the edges to define the pruning.
For instance, we can choose an ordering of the edges at each vertex
in a translation-invariant way, find the first $k$ levels
of the pruned SAW-tree, and then repeat these periodically.
Alternatively, we can order the vertices arbitrarily
at the root of the SAW-tree, and
at all other vertices of the SAW-tree order the edges according to the
angle the path turns through (e.g.\ $0, +\pi/2, -\pi/2$)
in taking the step.  This defines a pruned SAW-tree, and
we can take the first $k$ levels and repeat periodically as
before.  We conjecture that either method  gives
bounds converging to $\lambda_c(\Z^2)$ as $k$ grows, but we do not
know how to prove this.

\bigskip

\noindent
{\bf Remark.}
In Examples~\thesection.2 and \thesection.7,
we have proven for $\Z^2$ the rigorous bounds
\be
   0.105468\ldots \;=\; {27 \over 256} \;\le\; \lambda_c(\Z^2)
     \;\le\; {4 \over 25}  \;=\;  0.16  \;.
\ee
By extensions of those arguments and with a little calculation,
these bounds can be narrowed further;
it would be interesting to see how far one can go.
It is worth noting that Todo \cite{Todo_99} has given
the extraordinarily precise numerical estimate
\be
   \lambda_c(\Z^2) \;=\; 0.119\,338\,881\,88(1)  \;,
\ee
obtained by using transfer matrices and the phenomenological-renormalization
method (a variant of finite-size scaling).
Furthermore, his computations up to $L=38$
show \cite{Todo_private} that\footnote{
   What Todo \protect\cite{Todo_private} actually computed is the number
   $\lambda_L^\times$ for which
   \begin{itemize}
      \item[(a)] the transfer matrix has a unique eigenvalue of
           largest modulus for $w \in (-\lambda_L^\times,0]$, and
      \item[(b)] the transfer matrix has two dominant eigenvalues of
           largest modulus for $w = -\lambda_L^\times$.
   \end{itemize}
   Let us show that $\lambda_L^\times = \lambda_c(\widetilde{S}_L)$.
   Let $\widetilde{S}_L^{(n)}$ be the cylinder of width $L$ and length $n$.
   By monotonicity, $\lambda_c(\widetilde{S}_L^{(n)})$ decreases in $n$
   and hence has a limit as $n \to\infty$;
   it is easy to see that this limit is $\lambda_c(\widetilde{S}_L)$.
   Now, the Beraha--Kahane--Weiss theorem
   \protect\cite{BKW_78,Sokal_chromatic_roots} tells us that
   \begin{itemize}
      \item[(a\textprime)] for every $\epsilon > 0$, there exist $\delta > 0$
         and $n_0 < \infty$ such that $Z_{\widetilde{S}_L^{(n)}}$
         has no (real or complex) zeros within a distance $\delta$
         from the interval $[-\lambda_L^\times + \epsilon, 0]$
         when $n \ge n_0$; and
      \item[(b\textprime)] there exist (possibly complex) zeros of
         $Z_{\widetilde{S}_L^{(n)}}$ tending to $-\lambda_L^\times$
         as $n \to\infty$.
   \end{itemize}
   Since $-\lambda_c(\widetilde{S}_L^{(n)})$ is the closest zero
   to the origin of $Z_{\widetilde{S}_L^{(n)}}$,
   it follows easily from (a\textprime) and (b\textprime)
   that $\lambda_L^\times = \lambda_c(\widetilde{S}_L)$.
}
\be
   \lambda_c(\widetilde{S}_{38}) \;=\; 0.119\,365(1)  \;,
\ee
which by Example~\thesection.4 provides an upper bound on
$\lambda_c(\Z^2)$.
See also Guttmann \cite{Guttmann_87} for an earlier
and only slightly less precise estimate of $\lambda_c(\Z^2)$,
obtained by series analysis.

\section*{Acknowledgments}

We wish to thank Keith Ball for informing us of the important work
of Shearer \cite{Shearer_85},
and Joel Lebowitz for informing us of the work of Ginibre \cite{Ginibre_70}.
We also thank Keith Ball, Pierre Leroux and Joel Spencer
for helpful discussions,
and Synge Todo for communicating to us some of his unpublished
numerical results.

One of us (Sokal) wishes to thank
the Department of Mathematics at University College London
and the Theoretical Physics Group at Imperial College
for hospitality during his stays in London.

This research was supported in part by
U.S.\ National Science Foundation grants PHY--9900769 and PHY--0099393.

\clearpage


\end{document}